\documentclass[oneside,pdflatex,sn-mathphys-num]{sn-jnl}

\usepackage{graphicx}%
\usepackage{subcaption}
\usepackage{multirow}%
\usepackage{amsmath,amssymb,amsfonts}%
\usepackage{amsthm}%
\usepackage{mathrsfs}%
\usepackage[title]{appendix}%
\usepackage{xcolor}%
\usepackage{textcomp}%
\usepackage{manyfoot}%
\usepackage{booktabs}%
\usepackage{algorithm}%
\usepackage{algorithmicx}%
\usepackage{algpseudocode}%
\usepackage{listings}%
\usepackage{booktabs}
\usepackage{tabularx}
\usepackage{ragged2e}

\usepackage[displaymath,mathlines]{lineno}

\theoremstyle{thmstyleone}%

\theoremstyle{thmstyletwo}%

\theoremstyle{thmstylethree}%

\begin{document}

\title[Article Title]{Accurate de novo sequencing of the modified proteome with OmniNovo}

\author[1,2]{\fnm{Yuhan} \sur{Chen}}
\author[2,3]{\fnm{Shang} \sur{Qu}}
\author[2]{\fnm{Zhiqiang} \sur{Gao}}
\author[2,4]{\fnm{Yuejin} \sur{Yang}}
\author[5]{\fnm{Xiang} \sur{Zhang}}
\author[2,4]{\fnm{Sheng} \sur{Xu}}
\author[2,6,7]{\fnm{Xinjie} \sur{Mao}}
\author[6]{\fnm{Liujia} \sur{Qian}}
\author[2,8]{\fnm{Jiaqi} \sur{Wei}}
\author[2]{\fnm{Zijie} \sur{Qiu}}
\author[9]{\fnm{Chenyu} \sur{You}}
\author[2]{\fnm{Lei} \sur{Bai}}
\author*[2,3]{\fnm{Ning} \sur{Ding}}
\author*[6]{\fnm{Tiannan} \sur{Guo}}
\author*[2,3]{\fnm{Bowen} \sur{Zhou}}
\author*[4]{\fnm{Siqi} \sur{Sun}}

\affil[1]{\orgdiv{Shanghai Research Institute for Intelligent Autonomous Systems}, \orgname{Tongji University}, \orgaddress{\city{Shanghai}, \country{China}}}
\affil[2]{\orgname{Shanghai Artificial Intelligence Laboratory}, \orgaddress{\city{Shanghai}, \country{China}}}
\affil[3]{\orgdiv{Department of Electronic Engineering}, \orgname{Tsinghua University}, \orgaddress{\city{Beijing}, \country{China}}}
\affil[4]{\orgdiv{Research Institute of lntelligent Complex Systems}, \orgname{Fudan University}, \orgaddress{\city{Shanghai}, \country{China}}}
\affil[5]{\orgdiv{Department of computer science}, \orgname{University of British Columbia}, \orgaddress{\city{Vancouver}, \country{Canada}}}
\affil[6]{\orgdiv{School of Medicine}, \orgname{Westlake University}, \orgaddress{\city{Hangzhou}, \state{Zhejiang}, \country{China}}}
\affil[7]{\orgname{Shanghai Innovation Institute}, \orgaddress{\city{Shanghai}, \country{China}}}
\affil[8]{\orgdiv{Department of computer science}, \orgname{Zhejiang University}, \orgaddress{\city{Hangzhou}, \state{Zhejiang}, \country{China}}}
\affil[9]{\orgdiv{Department of computer science}, \orgname{Stony Brook University}, \orgaddress{\city{New York}, \country{America}}}

\abstract{
Post-translational modifications (PTMs) serve as a dynamic chemical language regulating protein function, yet current proteomic methods remain blind to a vast portion of the modified proteome. 
Standard database search algorithms suffer from a combinatorial explosion of search spaces, limiting the identification of uncharacterized or complex modifications. 
Here we introduce OmniNovo, a unified deep learning framework for reference-free sequencing of unmodified and modified peptides directly from tandem mass spectra. 
Unlike existing tools restricted to specific modification types, OmniNovo learns universal fragmentation rules to decipher diverse PTMs within a single coherent model. By integrating a mass-constrained decoding algorithm with rigorous false discovery rate estimation, OmniNovo achieves state-of-the-art accuracy, identifying 51\% more peptides than standard approaches at a 1\% false discovery rate.
Crucially, the model generalizes to biological sites unseen during training, illuminating the dark matter of the proteome and enabling unbiased comprehensive analysis of cellular regulation.
}

\maketitle

\section*{Introduction}\label{sec1}

Protein sequencing forms the foundation of modern proteomics, enabling the unbiased characterization of proteoforms and the discovery of novel biological mechanisms\cite{aebersold2003mass,smith2013proteoform}. While the primary amino acid sequence provides the structural blueprint, post-translational modifications (PTMs) constitute a dynamic chemical language that regulates protein function, stability, and interaction\cite{nesvizhskii2014proteogenomics,chen2013characterization}. Common modifications such as phosphorylation and methylation are pivotal in cellular signaling, transcriptional control, and cell cycle regulation\cite{kouzarides2007chromatin,walsh2005protein,choudhary2010decoding,seet2006reading, pagel2015current,hunter2007age}. 
However, the biological impact of these events is often determined by a precise molecular logic—specifically, the exact residue modified and the combinatorial presence of multiple PTMs on a single peptide \cite{walsh2005protein, beltrao2012systematic, nishi2011phosphorylation}
Consequently, the ability to accurately sequence and localize diverse co-existing PTMs is not merely a technical convenience but a prerequisite for mapping the functional landscape of the cell\cite{walsh2005protein,smith2013proteoform}.

Despite the ubiquity of mass spectrometry for high-throughput sequencing, identifying modified peptides remains a fundamental bottleneck. The conventional approach relies on matching experimental spectra against static reference databases\cite{perkins1999probability,eng1994approach,cox2008maxquant,zhang2012peaks,bateman2014maximizing,rost2014openswath,chi2018comprehensive}. This strategy is inherently biased: it is effectively blind to the ``dark matter'' of the proteome, failing to identify non-canonical peptides, unexpected mutations, or modifications not explicitly pre-defined in the search space\cite{azari2019ga,beslic2023comprehensive,karunratanakul2019uncovering,hettich2013metaproteomics}. Furthermore, attempting to capture the full spectrum of biological variation triggers a combinatorial explosion; including even a modest number of variable modifications expands the search space exponentially, rendering database search computationally prohibitive and statistically underpowered\cite{chick2015mass,kong2017msfragger}.

To overcome these reference-dependent limitations, deep learning-based \textit{de novo} sequencing has emerged as a powerful alternative\cite{ma2003peaks,frank2005pepnovo}. Recent transformer-based models \cite{karunratanakul2019uncovering,bittremieux2024deep,tran2017novo,qiao2021computationally,qiao2021computationally,yang2024introducing,petrovskiy2024powernovo,yang2019pnovo,yilmaz2022novo,yilmaz2024sequence,zhang2025pi,eloff2025instanovo}, such as Casanovo\cite{yilmaz2022novo}, InstaNovo\cite{eloff2025instanovo} and $\pi$-PrimeNovo\cite{zhang2025pi}, have leveraged large-scale training data to achieve high accuracy on unmodified peptides. However, extending these successes to the modified proteome has proven difficult. Current solutions are fragmented: tools like InstaNovo-P\cite{lauridsen2025instanovo} are restricted to single modification types (e.g., phosphorylation) or rely on fine-tuning separate models for specific PTM classes\cite{zhang2025pi,petrovskiy2024powernovo}. This ``one-model-per-PTM'' paradigm is scalable neither computationally nor biologically, as it typically models modifications as fixed residue-PTM pairs. Such rigid representations prevent models from learning generalizable spectral rules, limiting their ability to resolve complex peptides where diverse modifications coexist.

Here we introduce OmniNovo, a unified deep learning framework engineered to decipher the comprehensive language of modified peptides directly from mass spectra. Unlike prior approaches, OmniNovo integrates eleven distinct PTM classes—ranging from abundant acetylation to complex ubiquitination—into a single, coherent model without sacrificing performance on unmodified peptides. A core architectural innovation is the representation of PTMs as independent tokens within the model's vocabulary, decoupling the chemical modification from amino acid identity. This allows OmniNovo to learn universal fragmentation signatures that generalize across different residues. To ensure these predictions are physically rigorous, we introduce a precise-mass-and-modification-control (PMMC) decoding unit, which dynamically enforces precursor mass tolerance and biochemical constraints during sequence generation.

We demonstrate that OmniNovo establishes a new performance standard for the field. It achieves state-of-the-art recall on diverse benchmarks, identifying 69\% of unmodified peptides on the nine-species dataset\cite{tran2017novo} and 63\% of modified peptides on our curated PTMBench, surpassing existing fine-tuned models. Crucially, by combining this accuracy with a robust semi-supervised FDR control strategy, OmniNovo identifies 51\% more peptides than MaxQuant\cite{cox2008maxquant} at a 1\% false discovery rate (FDR). Furthermore, the model exhibits strong zero-shot generalization: when tested on biological sites never seen during training, OmniNovo delivers a 65.8\% accuracy improvement over baseline settings. This capability suggests the model has learned the underlying physics of peptide fragmentation rather than merely memorizing training examples, providing a reliable tool for illuminating the dark matter of the proteome.

\section*{Results}\label{sec2}

\subsection*{A unified deep learning framework trained on a comprehensive modified proteome atlas}\label{sec2.1}

\begin{figure*}[t]
    \centering
    \includegraphics[width=\textwidth]{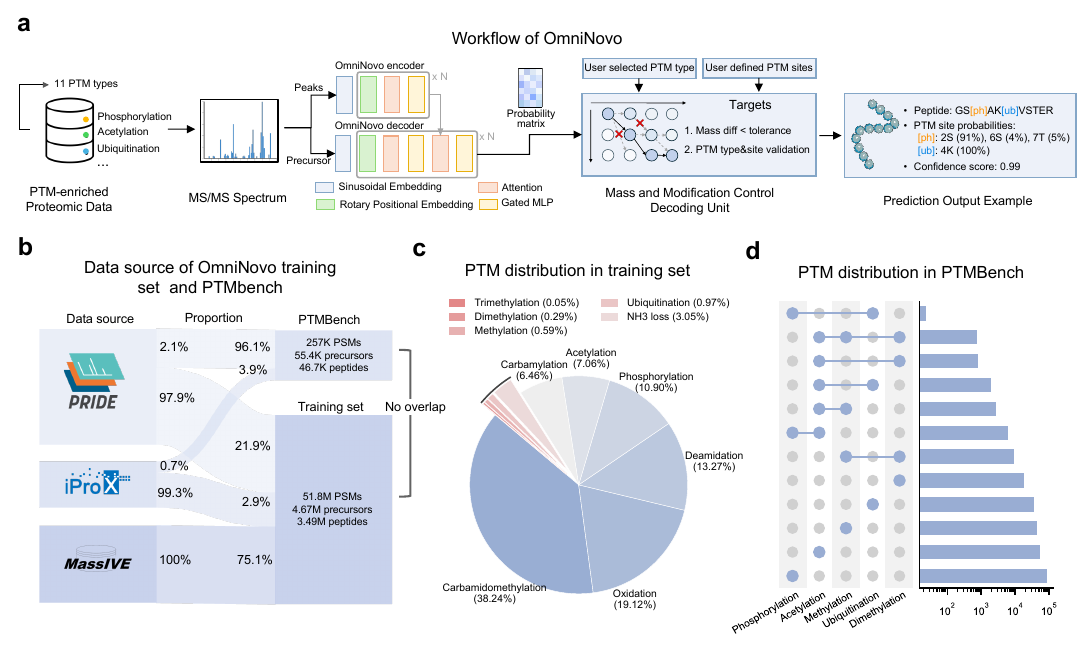}
    \caption{Overview of OmniNovo workflow and training data. \textbf{a}, OmniNovo takes the peaks from each MS/MS spectrum as input and generate the predicted peptide sequence. Modification tokens have been added in the output embedding layer as individual tokens. OmniNovo uses the precise mass and modification control decoding method to optimize peptide generation to meet mass tolerance and user-defined modification rules. \textbf{b}, The data source of OmniNovo training set and PTMbench. \textbf{c}, Percent of each kind of modification in the training set. \textbf{d}, Distribution of PSMs with and without modifications in peptides obtained from database search in each project used to construct our dataset. \textbf{d}, Percent of each kind of modification in the training set. }
    \label{fig:1}
\end{figure*}

Accurate de novo sequencing of post-translationally modified peptides is a longstanding challenge in proteomics, limited primarily by the lack of large-scale, unified data resources. To fill this gap, we constructed the largest diverse proteomic training corpus to date for PTM-aware modeling. We integrated 22 large-scale mass spectrometry projects from the PRIDE\cite{perez2025pride}, iProX\cite{ma2019iprox}, and MassIVE\cite{wang2018assembling} repositories to compile a dataset of 51.8 million PSMs (Fig.\ref{fig:1}b). After strict filtering, this corpus contains 4.7 million unique precursors spanning eleven common PTM types (Fig.\ref{fig:1}c). It captures the full chemical diversity of the cellular proteome, ranging from common artifacts like carbamidomethylation to low-frequency regulatory events such as phosphorylation, methylation, and ubiquitination. This dataset provides an comprehensive quantitative atlas of fragmentation patterns, establishing a robust foundation for learning the universal spectral signatures of PTMs. 

Based on this resource, we developed OmniNovo, a deep learning frameworkdesigned to sequence modified peptides directly from mass spectra. The model uses a non-autoregressive transformer that directly translates tandem mass spectra into amino acid sequences (Fig.\ref{fig:1}a). Key to our approach is the representation of PTMs as independent tokens within the model’s vocabulary, unlike previous approaches that relied on fixed residue-modification pairs. This allows OmniNovo to learn generalizable spectral principles rather than the statistical patterns of specific residue types, avoiding the combinatorial complexity that limits prior models.To ensure biological validity, we introduce the PMMC decoding unit. 
This module restricts the decoding process via a knapsack-like optimization, strictly enforcing both precursor mass tolerance and plausible PTM placement.

To validate the model, we established PTMBench, a comprehensive benchmarking system designed for reproducible and rigorous assessment (Fig.\ref{fig:1}d). It consists of 257,000 high-quality PSMs from 25 projects, including five projects held out entirely for cross-experiment, zero-shot validation to test model generalization. The system tests models across a range of scenarios, including peptides with single versus multiple co-occurring modifications. PTMBench thus represents the standardized benchmark for de novo sequencing of modified peptides, offering the field a robust tool for evaluation.

\subsection*{OmniNovo achieves state-of-the-art accuracy on unmodified and modified peptides}\label{sec2.2}

\begin{figure*}[h]
    \centering
    \includegraphics[width=\textwidth]{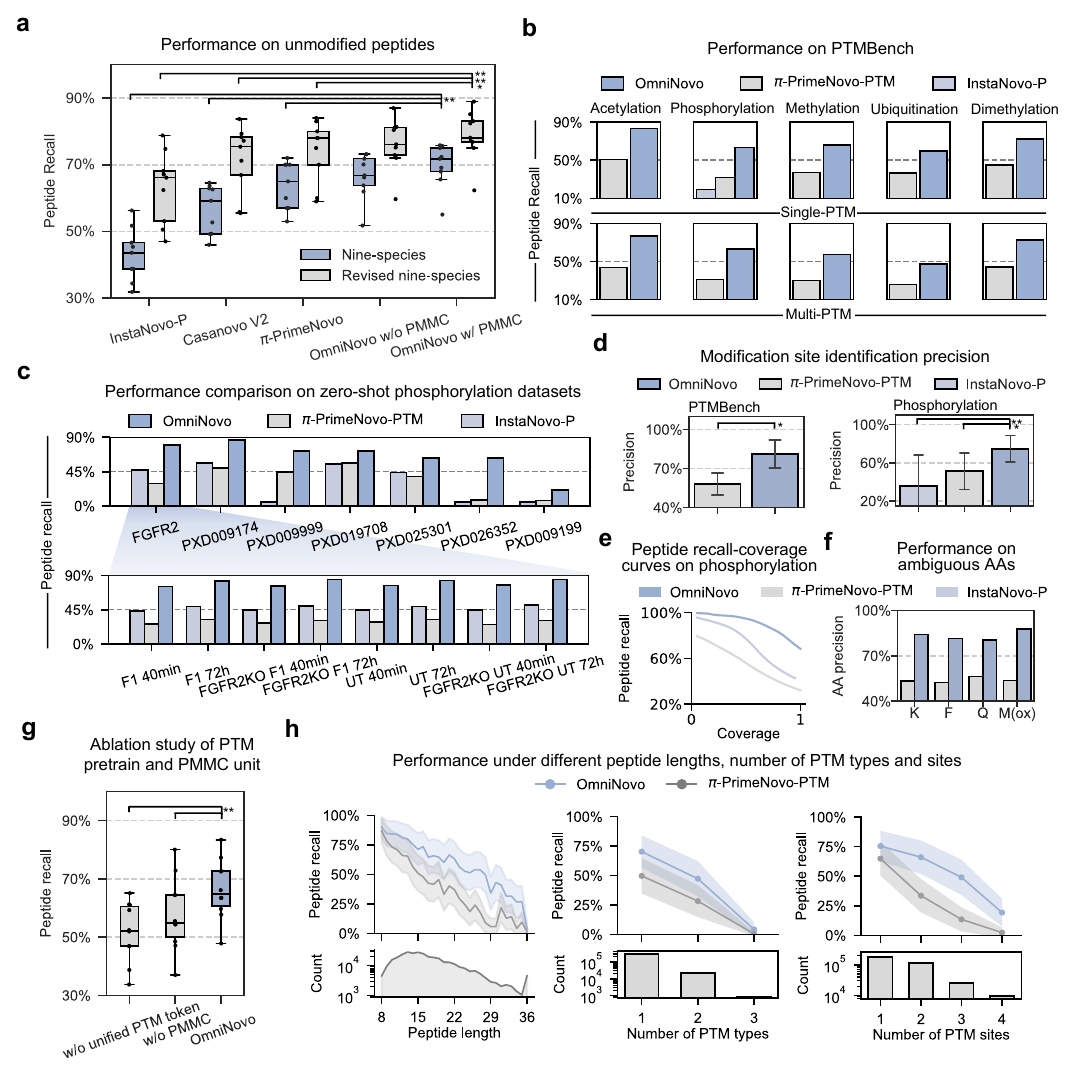}
    \caption{
    \textbf{OmniNovo achieves state-of-the-art performance and demonstrates robustness on both unmodified and post-translationally modified peptides.}
    \textbf{a}, Peptide recall of OmniNovo, an ablated variant (without PMMC), and other leading methods on unmodified peptides from the nine-species and revised nine-species benchmarks. Statistical significance was calculated using single-sided Wilcoxon signed-rank test. * $p<0.05$; ** $p<0.01$.
    \textbf{b}, Peptide recall on PTMBench. Results are shown separately for peptides containing only the specified PTM (top) and for peptides that include the specified PTM along with other modifications (bottom).
    \textbf{c}, Head-to-head peptide recall comparison on seven zero-shot phosphorylation datasets, benchmarking OmniNovo against $\pi$-PrimeNovo-PTM and the phosphorylation-specialized model InstaNovo-P. A detailed comparison is also shown for subsets of the FGFR2 dataset generated under different experimental conditions.
    \textbf{d}, Precision of modification site on PTMBench and phosphorylation datasets.
    \textbf{e}, Peptide recall versus prediction coverage for OmniNovo and baselines on phosphorylation datasets.
    \textbf{f}, Performance on amino acids with similar masses.
    \textbf{g}, Ablation study of OmniNovo performance on modified peptides, evaluating the contribution of PTM pretraining and the PMMC unit.
    \textbf{h}, Peptide recall on modified peptides, comparing OmniNovo and $\pi$-PrimeNovo-PTM. Performance is stratified by the number of peptide length, distinct PTM types per peptide and by the total number of PTM sites per peptide. The number of peptides in each category is shown below.
}
    \label{fig:2}
    
\end{figure*}

OmniNovo sets a new standard for canonical peptide sequencing. When evaluated on the standard nine-species benchmark\cite{tran2017novo}, the model consistently outperformed existing methods, achieving a peptide recall of 69.0\% on the standard dataset and 77.2\% on the revised version\cite{yilmaz2024sequence}, as detailed in Fig.\ref{fig:2}a. 
These results represent a statistically significant improvement over InstaNovo-P\cite{lauridsen2025instanovo}, Casanovo V2\cite{yilmaz2024sequence}, and $\pi$-PrimeNovo\cite{zhang2025pi}, confirming that OmniNovo provides a superior foundation for sequencing unmodified peptides before extending its capabilities to the more complex modified proteome.

Building on this foundation, we assessed the model’s ability to decode PTMs using PTMBench. A key limitation of prior approaches is their reliance on PTM-specific fine-tuning, which we replicated by training a specialized $\pi$-PrimeNovo-PTM baseline. OmniNovo surpasses this by an average of 28.7\% across five major modification classes (Fig.\ref{fig:2}b). We further validated generalization via zero-shot testing on seven real-world phosphorylation datasets from PRIDE spanning diverse human tissues. OmniNovo consistently outperformed competing methods, including the phosphorylation-specialized InstaNovo-P, achieving peptide recall improvements of 27.7\% across seven zero-shot scenarios (Fig.\ref{fig:2}c). This robustness is further demonstrated by the peptide recall-coverage curves, where OmniNovo maintains significantly higher recall than baselines across the entire range of prediction confidence (Fig.\ref{fig:2}e).

Beyond identifying peptide sequences, we evaluated the model's precision in localizing modification sites and resolving mass-ambiguous residues. As shown in Fig.\ref{fig:2}d, OmniNovo achieved significantly higher site-level precision compared to baselines. Furthermore, the model effectively distinguishes between amino acids with near-identical masses (Fig.\ref{fig:2}f), suggesting it relies on fine-grained fragmentation patterns rather than precursor mass matching alone.
Finally, stress tests revealed that OmniNovo maintains stable performance even as peptide complexity increases. It significantly outperformed baselines when analyzing long peptides or those containing multiple distinct PTM types and sites (Fig.\ref{fig:2}h). This robustness is supported by the model’s architectural design, specifically the unified modeling strategy and the PMMC unit, both of which were essential in ablation studies (Fig.\ref{fig:2}g).

\subsection*{OmniNovo refines database-driven identifications through improved spectral interpretation}\label{sec2.3}

\begin{figure*}[h]
    \centering
    \includegraphics[width=\textwidth]{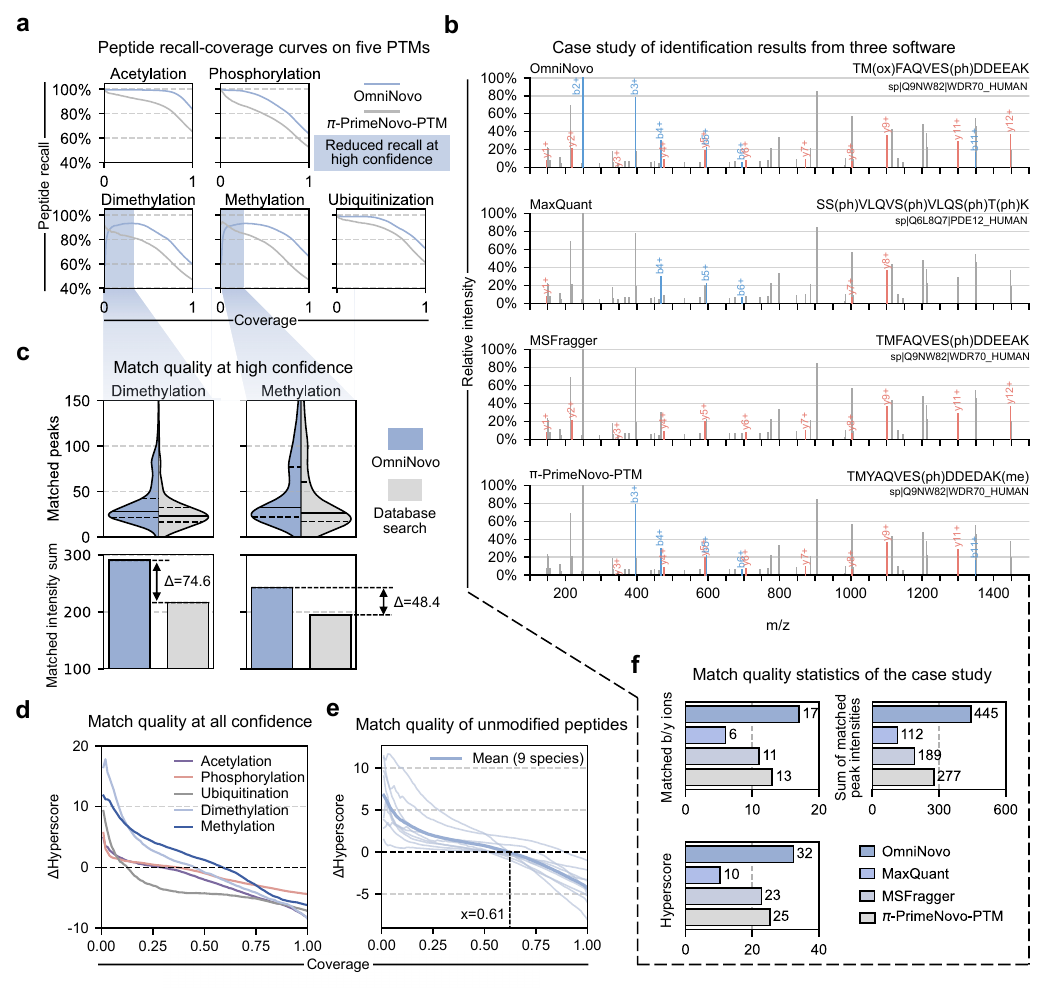}
    \caption{
    Spectral match quality for OmniNovo predictions compared to database search labels.
    \textbf{a}, Peptide recall versus prediction coverage for OmniNovo and $\pi$-PrimeNovo-PTM across five PTM types. Coverage is the fraction of spectra considered, ranked by model confidence. The shaded region indicates an unexpected recall drop for methylation and dimethylation.
    \textbf{b}, An experimental spectrum annotated with conflicting peptide identifications from OmniNovo, MaxQuant, MSFragger and $\pi$-PrimeNovo-PTM. Matched theoretical b/y ions are indicated for each prediction.
    \textbf{c}, Spectral matching metrics for conflicting predictions within the high-confidence methylation and dimethylation region from \textbf{a}. Distributions compare the number of matched peaks and the sum of matched peak intensities between OmniNovo predictions and database search labels. Peak intensities were pre-normalized to a [1, 100] scale for the sum calculation.
    \textbf{d}, Difference in hyperscore ($\Delta$Hyperscore) between OmniNovo predictions and database labels versus prediction coverage on PTMBench. It includes all spectra with conflicting sequence identifications.
    \textbf{e}, $\Delta$Hyperscore versus prediction coverage for conflicting predictions on the revised nine-species benchmark. The average trend is bolded, with results for individual species in the background. The x-intercept indicates the coverage at which the average $\Delta$Hyperscore is zero.
    \textbf{f}, Quantitative comparison of spectral matching metrics (number of matched b/y ions, sum of matched peak intensities, and hyperscore) for the OmniNovo, MaxQuant,MSFragger, and $\pi$-PrimeNovo-PTM predictions from the case study in \textbf{b}.
}
    \label{fig:3}
\end{figure*}

To assess the model's reliability, we examined the correlation between peptide recall and prediction confidence across various PTM types in the PTMBench dataset.
As shown in Fig.\ref{fig:3}a, OmniNovo generally demonstrated superior recall-coverage characteristics compared to the fine-tuned $\pi$-PrimeNovo-PTM baseline.
However, a distinct divergence emerged in the high-confidence regime for methylation and dimethylation, where OmniNovo appeared to exhibit a counter-intuitive drop in recall relative to database labels. We hypothesized that this discrepancy is due to the fundamental difference in training paradigms: while the baseline relies on database matching, OmniNovo learns intrinsic spectral patterns. Consequently, in high-confidence scenarios, OmniNovo may identify sequences that are physically more accurate than the database search results used as ground truth.

To validate this, we scrutinized the spectral evidence for cases where OmniNovo’s high-confidence predictions conflicted with database labels. 
A representative case study, detailed in Fig.\ref{fig:3}b and quantified in Fig.\ref{fig:3}f, reveals that the sequence identified by OmniNovo aligns significantly better with the experimental spectrum than candidates proposed by standard tools like MaxQuant\cite{cox2008maxquant}, MSFragger\cite{kong2017msfragger}, as well as the baseline model $\pi$-PrimeNovo-PTM (see Supplementary Note.5 for additional examples).
Specifically, OmniNovo annotated a higher number of matching ions and captured a greater proportion of the total peak intensity. This superior spectral alignment was reflected in the hyperscore metric, a measure of the similarity between the peptide and the spectrum, where OmniNovo achieved a score of 32 compared to scores ranging from 10 to 25 for other methods.
These results suggest that OmniNovo does not merely hallucinate sequences but provides a more exhaustive and statistically robust interpretation of the raw data, effectively capturing complex spectral patterns that confound traditional database search algorithms, which is pivotal for constructing a more reliable and comprehensive atlas of the post-translational modification landscape.

We then confirmed that this refining capability is systematic rather than anecdotal. By aggregating all high-confidence conflicting predictions for methylation and dimethylation, we observed a consistent trend: OmniNovo’s predictions yielded distributions shifted towards higher matched peak counts and intensities compared to the corresponding database labels (Fig.\ref{fig:3}c). We extended this analysis to all five PTM types using the $\Delta$Hyperscore metric, which measures the score difference between the model’s prediction and the database label. As illustrated in Fig.\ref{fig:3}d, the mean $\Delta$Hyperscore remains consistently positive in the high-confidence region across all modifications. This indicates that when OmniNovo expresses high certainty, its predictions systematically offer a better statistical explanation of the underlying spectra than the assigned database labels.

Finally, we demonstrated that this capacity to refine proteomic identification extends beyond modified peptides. When applied to the revised nine-species benchmark, OmniNovo continued to assign sequences with higher hyperscores than the database ground truth in the top 61\% of conflicting cases (Fig.\ref{fig:3}e). These findings establish that de novo sequencing with OmniNovo has matured from a supplementary method into a critical validation tool capable of refining database-driven annotations. By uncovering high-fidelity peptides that are misidentified by conventional workflows, OmniNovo offers a new avenue for constructing a more accurate and comprehensive atlas of the proteome.

\subsection*{OmniNovo expands the landscape of modified peptides with validated false discovery control}\label{sec2.4}

To enable robust error estimation without a target-decoy database, we employed a database search-assisted calibration method. As shown in Fig.\ref{fig:4}a, high-confidence PSMs identified by a standard database search at 1\% FDR served as ``anchors'' to calibrate the confidence scoring. For these anchor spectra, we compared the hyperscore of the database assignment ($S_{Search}$) against the OmniNovo prediction ($S_{Prediction}$). A prediction is flagged as a ``proxy false discovery'' if the database search yields a significantly better spectral match ($S_{search} > S_{pred}$). Subsequently, the OmniNovo confidence threshold was established at the point where 99\% of these anchors scored higher in OmniNovo than in the database search. This semi-supervised strategy ensures that our \textit{de novo} discoveries meet the same spectral quality standards as established methods.

We validated this approach using the FGFR2 dataset\cite{lauridsen2025instanovo}, comparing OmniNovo against MaxQuant and InstaNovo-P, with all methods filtered at a 1\% FDR. As shown in Fig.\ref{fig:4}b, OmniNovo detected 51\% more peptides than the database search. Notably, it outperformed the phosphorylation-specialized model, InstaNovo-P, by over an order of magnitude. This performance gap is likely due to OmniNovo's ability to generalize across diverse fragmentation patterns without being restricted to specific PTM motifs.
Crucially, OmniNovo recovered 51.3\% of the peptides found by the database search, demonstrating a substantial overlap that validates the reliability of the \textit{de novo} predictions while highlighting its capacity to capture the ``dark matter'' of the proteome missed by standard workflows.

We next assessed the spectral quality of these identifications. 
For both the complete datasets and the phosphorylated subsets, the hyperscore distribution for OmniNovo is shifted towards higher values compared to both database search and InstaNovo-P (Fig.\ref{fig:4}c). This indicates that the increase in identification numbers does not come at the cost of precision; rather, at a controlled 1\% FDR, OmniNovo yields spectral matches of equal or superior quality.
To rigorously stress-test our FDR control, we conducted a paired entrapment-based FDP estimation\cite{wen2025assessment}(Fig.\ref{fig:4}d). 
This method artificially expands the search space (up to 100$\times$) to challenge the scoring specificity. 
OmniNovo demonstrated exceptional robustness, consistently maintaining estimated FDPs substantially below the nominal thresholds across the entire range.
Specifically, at the critical 1\% nominal FDR, FDPs remained negligible under both 1$\times$ (0.019\%) and 100$\times$ entrapment settings (0.03\%). 
This exceptionally low error rate confirms that OmniNovo achieves its gains through fundamental improvements in discrimination power rather than loose filtering. 
In contrast, MaxQuant exhibited significant susceptibility to this expanded search space, with FDP inflating from 1.49\% (1$\times$) to 29.22\% (100$\times$).

We further validated the physical basis of our scoring by analyzing the relationship between OmniNovo's confidence score and the mean hyperscore of PSMs (Fig.\ref{fig:4}e). A strong positive correlation was observed across all peptide lengths, confirming that the model's confidence serves as a reliable proxy for PSM quality.

Finally, we evaluated the generalizability of OmniNovo across diverse datasets and conditions (Fig.\ref{fig:4}f).
On subsets of the FGFR2 dataset representing different experimental conditions, OmniNovo consistently identified more peptides than the database search while retaining a high proportion of shared identifications. 
This advantage was even more pronounced on external, modification-enriched datasets evaluated in a zero-shot manner. For instance, on the ubiquitin-enriched PXD024309 dataset\cite{sewduth2023spatial}, OmniNovo achieved a threefold increase in identified peptides. This substantial gain is attributable to OmniNovo's deep learning architecture, which generates full peptide sequences directly from spectra without referencing a database. This approach effectively bypasses the combinatorial search space explosion that forces standard engines to miss peptides containing multiple co-occurring PTMs, establishing OmniNovo as a robust method for discovering novel modified peptides.

\subsection*{OmniNovo enables precise PTM site localization and cross-residue generalization}\label{sec2.5}

Accurate localization of PTMs is critical yet computationally challenging, particularly for peptides with multiple potential modification sites. We evaluated OmniNovo’s ability to resolve these ambiguities using a representative peptide containing multiple serine and threonine residues complicated by dual phosphorylation and carbamidomethylation (Fig. \ref{fig:5}a-c). OmniNovo successfully disentangles the positional isomers, generating a residue-wise probability distribution that correctly assigns phosphorylation to T3 and S5 with a high confidence score of 0.89 (Fig. \ref{fig:5}a-b). Crucially, this assignment is strictly aligned with physical spectral evidence, supported by a comprehensive series of backbone fragment ions (b2–b4 and y3–y9) that unambiguously define the modification sites. In contrast, alternative isomers (e.g., phosphorylation at S1 or S10) are assigned lower scores (0.73 and 0.64), as OmniNovo effectively utilizes refuting evidence—specifically the absence of critical expected ions (Fig. \ref{fig:5}c)—to discriminate based on signal integrity rather than mere guessing.

We further assessed the robustness of this localization capability on PTMBench (Fig.\ref{fig:5}d). OmniNovo demonstrates high localization accuracy at strict coverage levels, confirming that the confidence score is a well-calibrated indicator of reliability. Notably, the model maintains high performance even for hyper-modified peptides containing four or more PTMs, a regime where traditional search engines often struggle. Simultaneously, the backbone sequence accuracy remains robust at an average of 74.08\%, ensuring that PTM site precision does not come at the cost of peptide identification correctness.

Finally, a transformative feature of OmniNovo is its ability to generalize PTM discovery to unobserved residues, facilitated by the PMMC unit and unified PTM token modeling. To test this, we designed a strict zero-shot experiment by training the model exclusively on phosphorylated Tyrosine and Threonine and then testing its ability to identify phosphorylated Serine. This target represents a residue the model had never seen modified during training, as shown in the top panel of Fig. \ref{fig:5}e. Even under these conditions, the model achieved substantial recall. Furthermore, when identifying phosphorylation on Threonine after training only on Serine and Tyrosine, OmniNovo correctly assigned the modification site with a +65.8\% improvement over random guessing as illustrated in the bottom panel of Fig. \ref{fig:5}e. This result suggests that OmniNovo learns the transferable biochemical signatures of modifications such as mass shifts and fragmentation patterns rather than memorizing residue-specific associations, highlighting its potential for discovering novel PTM events on non-canonical amino acids.

\section*{Discussion}\label{sec3}

The accurate identification of PTMs remains a major challenge in proteomics, limiting our understanding of cellular regulation. By combining the sequencing of unmodified and modified peptides into a single deep learning framework, OmniNovo addresses a key gap in current methods. Our results show that OmniNovo not only improves recall and spectral matching compared to existing \textit{de novo} models but also provides a reliable way to estimate false discovery rates. This makes \textit{de novo} sequencing a practical, standalone tool for analyzing the modified proteome, reducing the reliance on restrictive database search spaces.

A key advantage of OmniNovo is how it handles modifications. Instead of treating residue-PTM pairs as fixed units, OmniNovo represents PTMs as separate tokens in the model’s vocabulary. This design separates the chemical identity of the amino acid from the modification, allowing the model to learn general fragmentation rules—such as neutral losses and specific ion patterns—rather than just memorizing specific sequences. This approach is supported by our zero-shot tests, where the model correctly identified modifications on residue types it had not seen during training. This suggests that OmniNovo has learned the physical principles of peptide fragmentation, enabling it to interpret modifications in new biological contexts.

These capabilities allow us to look closer at what is considered ``ground truth'' in proteomics. In many high-confidence cases where OmniNovo disagreed with database search results, our model provided a better explanation of the experimental spectra, showing higher hyperscores and more complete fragment ion series. This indicates that public datasets contain many spectra that are chemically interpretable but are currently missed or mislabeled by standard methods. OmniNovo thus acts as a valuable independent validation tool, helping to refine existing annotations and identify peptides that are typically overlooked.

However, the flexibility of deep learning requires careful statistical control. A main issue with \textit{de novo} sequencing has been the lack of robust error estimation. We address this with the PMMC unit, which applies strict physical and biochemical rules during decoding to prevent the generation of impossible sequences. Additionally, our semi-supervised FDR strategy aligns the model's confidence scores with established database search engines. This allowed OmniNovo to identify significantly more peptides than MaxQuant at a 1\% FDR, with entrapment experiments confirming that the error rates remain low. This demonstrates that deep learning can provide the reliability needed for downstream biological analysis.

While OmniNovo is a significant step forward, there are limitations. The current model covers 11 major PTM types, but the proteome contains many rare modifications. The token-based structure of our model makes it straightforward to add these rarer modifications as more training data becomes available. Also, while our FDR strategy works well for standard samples, future work is needed to handle samples where database searching fails completely, such as in non-model organisms. Looking ahead, OmniNovo has potential beyond standard PTM analysis. Its ability to sequence peptides without a reference genome is particularly useful for applications where databases are incomplete, such as identifying neoantigens in cancer or analyzing complex microbiomes. By decoding peptide sequences directly from spectra, OmniNovo offers researchers a way to discover new biomarkers and regulatory mechanisms that are difficult to find with traditional pipelines.

\bibliography{sn-bibliography}

\begin{figure*}[h]
    \centering
    \includegraphics[width=\textwidth]{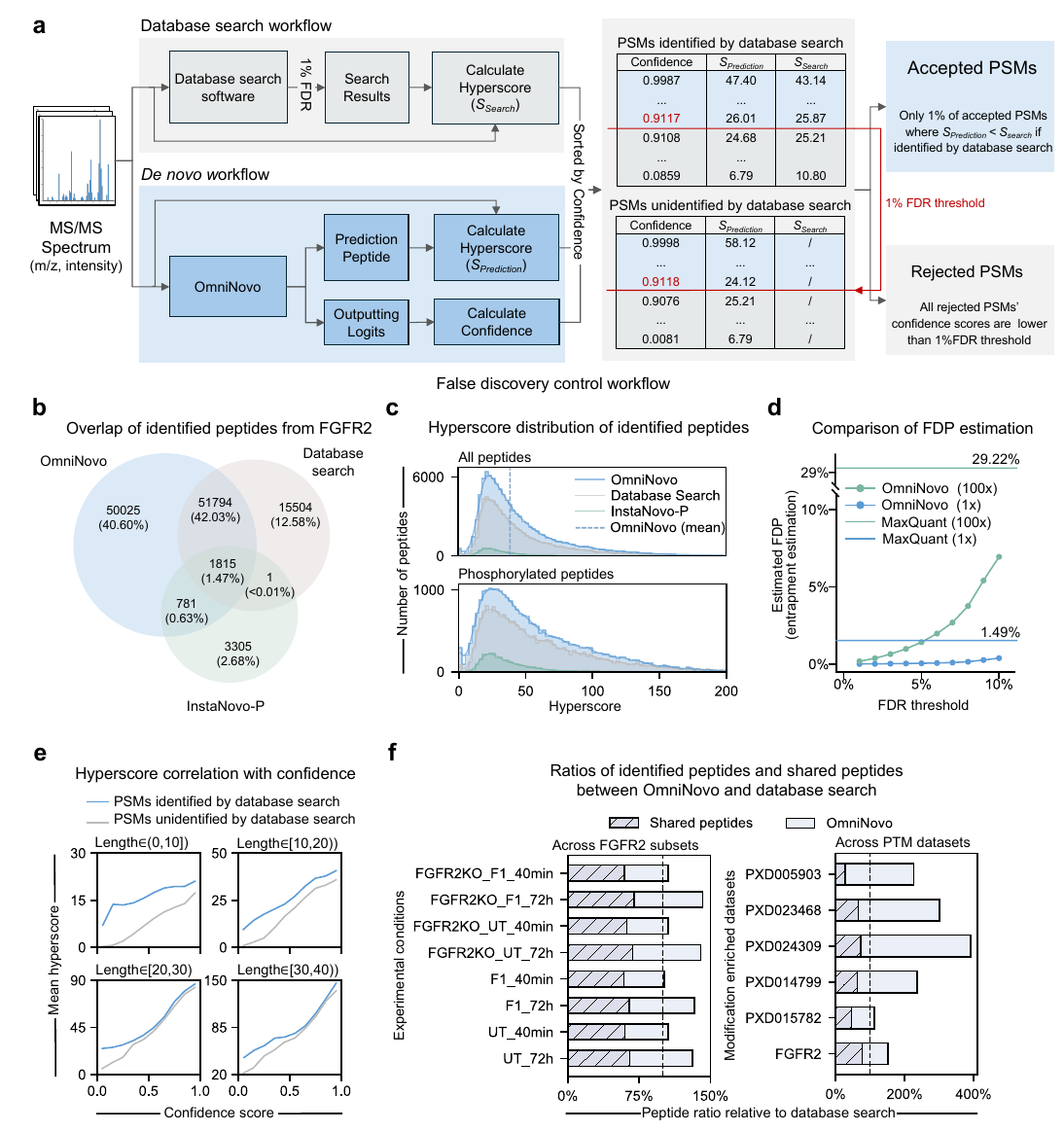}
    \caption{
    \textbf{Performance evaluation and false discovery rate control of OmniNovo.}
    \textbf{a}, Workflow for FDR control. PSMs from the \textit{de novo} workflow are sorted by OmniNovo confidence, and a reference database search is used to establish a confidence threshold corresponding to a 1\% FDR.
    \textbf{b}, Overlap of unique peptides identified by OmniNovo, database search, and InstaNovo-P from the FGFR2 dataset at 1\% FDR.
    \textbf{c}, Distributions for all identified peptides (top) and phosphorylated peptides (bottom). Each unique peptide is represented by the hyperscore of its best‑scoring PSM.
    \textbf{d}, Comparison of estimated false discovery proportions on the FGFR2 dataset as a function of the FDR threshold. FDPs were estimated for OmniNovo and MaxQuant using either one (1×) or one hundred (100×) entrapment sequences per PSM.
    \textbf{e}, Relationship between mean hyperscore and OmniNovo confidence. PSMs are stratified by peptide length and by whether they were identified in the database search.
    \textbf{f}, Ratios of identified peptide counts between OmniNovo and database search at 1\% FDR. Left: comparisons across FGFR2 subsets under different experimental conditions. Right: comparisons across multiple modification-enriched public datasets. Ratios represent the numbers of total and shared peptides identified by OmniNovo relative to those identified by database search.
}
    \label{fig:4}
\end{figure*}

\begin{figure*}[h]
    \centering
    \includegraphics[width=\textwidth]{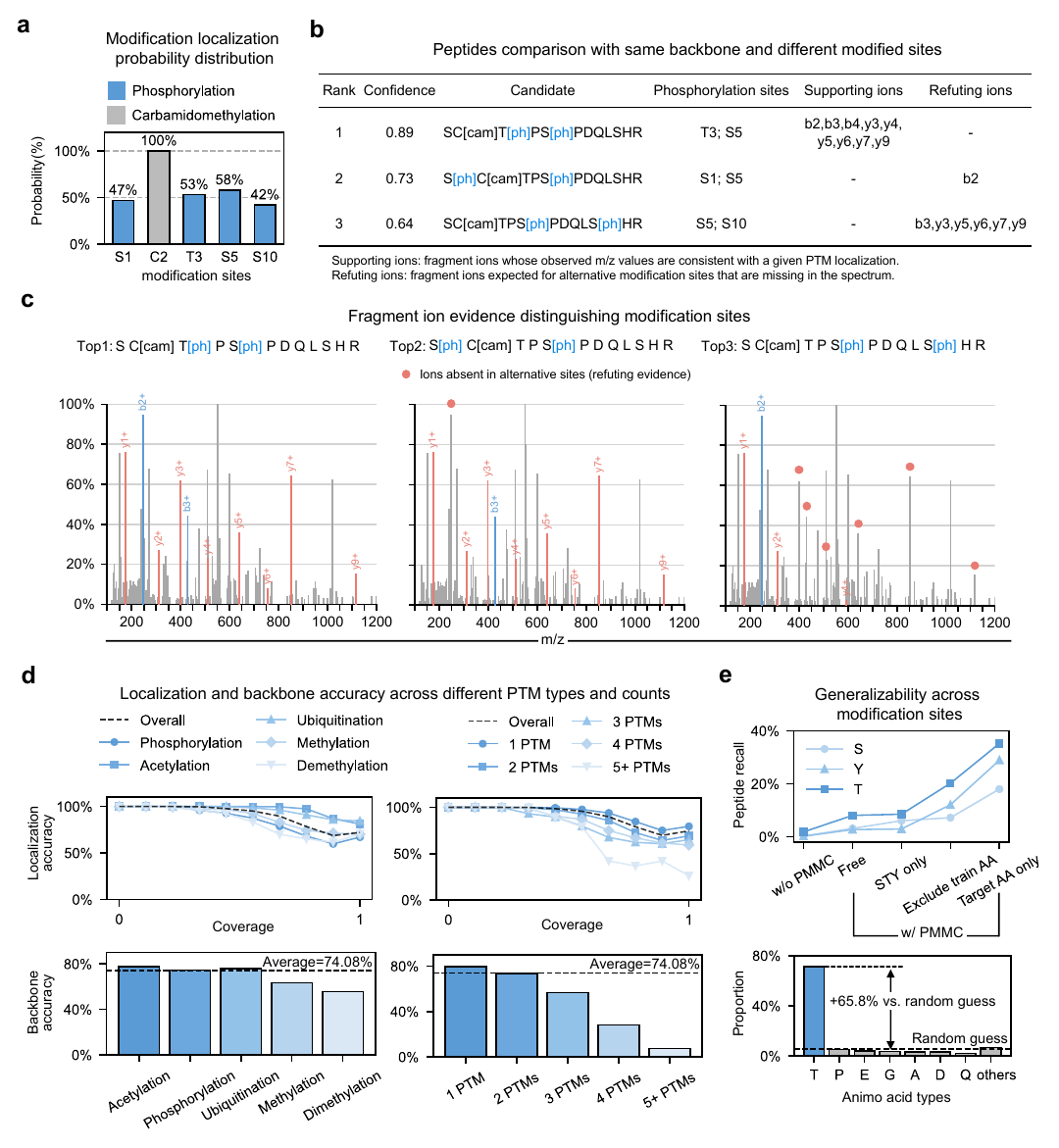}
    \caption{
    \textbf{Evaluation of PTM site localization and generalization ability.}
    \textbf{a}, Predicted probability distribution of phosphorylation (ph) and carbamidomethylation (cam) modifications across the amino acid sequence for a representative peptide.
    \textbf{b}, Top three candidate modified peptides for the spectrum in (a), ranked by model confidence, with their corresponding confidence scores, key supporting ions, and key refuting ions. 
    \textbf{c}, Annotated tandem mass spectra for the three candidate peptides from (b). Experimental spectra are aligned with theoretical spectra, highlighting matched b/y ions. Refuting ions listed in \textbf{b} are indicated, with missing peaks highlighting incorrect PTM localizations for lower-ranked candidates.
    \textbf{d}, Localization and backbone accuracy versus model confidence score quantiles (Qn). Accuracy is stratified by PTM type (left) and number of PTMs per peptide (right). Backbone accuracy is defined as the correctness of the amino acid sequence, irrespective of PTM placement.
    \textbf{e}, Model generalizability to unseen phosphorylation sites. \textbf{Top}, Peptide recall under various cross-residue training and testing conditions. 'S' indicates testing on Serine (S) phosphorylation after training on Tyrosine (Y) and Threonine (T) only; 'Y' and 'T' follow a similar logic. \textbf{Bottom}, Distribution of predicted phosphorylation sites for a model trained on S/Y-phosphorylated data and tested on T-phosphorylated data, compared to a random baseline. Predictions are constrained to amino acids other than S and Y.
}
    \label{fig:5}
\end{figure*}

\backmatter

\section*{Methods}

\subsection*{Data Curation and Benchmarking}
\textbf{Training Corpus Construction.} 

The training corpus, comprising 51.8 million PSMs, was constructed by curating and standardizing data from two primary sources: MassIVE-KB and 20 PTM-enriched datasets (detailed identifiers are provided in \textbf{Supplementary Table 1}). The MassIVE-KB (v1 and v2) datasets constitute approximately 75\% of the corpus and serve as the source for unmodified peptides; search results and spectra were retrieved directly from the repository as described previously\cite{wang2018assembling}. The PTM-enriched component was curated from projects in the PRIDE\cite{perez2025pride} and iProX\cite{ma2019iprox} repositories.

For the PTM-enriched datasets, we utilized the original database search results provided by the submitting authors. These searches were primarily performed using MaxQuant against human proteome databases; detailed metadata for each project, including organism, modification, instrument, MaxQuant versions, reference databases and other statistics, is provided in \textbf{Supplementary Data 1}. 

Raw mass spectrometry files were converted to MGF format using msconvert (ProteoWizard; Docker image \texttt{chambm/pwiz-skyline-i-agree-to-the-vendor-licenses:20.04}) with peak picking enabled. Spectra were mapped to the provided database search results (i.e., \texttt{msms.txt}) by matching raw filenames and scan identifiers. To standardize the training data, experimental neutral precursor masses were recalculated from the MGF precursor $m/z$ and charge state ($Mass = (m/z - 1.007276) \times \text{charge}$). PSMs exhibiting a deviation greater than 0.1 Da from the theoretical peptide mass were excluded to minimize incorrect mappings.

Data splitting for the PTM-enriched projects employed a stratified, peptide-level strategy to ensure strict separation between training and evaluation data. PSMs were categorized as unmodified, single-PTM, or multi-PTM. All unmodified peptides were assigned to the training set. For single- and multi-PTM categories, 30\% of unique peptide sequences (capped at 5,000 PSMs per project to prevent dataset bias) were allocated to the test set (PTMBench), with the remainder assigned to the training set. Importantly, this split was performed based on unique peptide sequences, ensuring that the set of modified peptides in the test set was strictly disjoint from that in the training set. The final OmniNovo training set was formed by merging the MassIVE-KB data with the training fraction of the PTM-enriched corpus.

\textbf{Evaluation Benchmarks.} 
We assessed OmniNovo across four distinct benchmarking scenarios to evaluate generalization across species and PTM handling capabilities:

(1) \textbf{PTMBench:} A dataset designed to evaluate the identification of five specific modifications: acetylation, dimethylation, methylation, ubiquitination, and phosphorylation. PTMBench is composed of two partitions.
The first partition originates from the held-out test subsets of the OmniNovo training corpus described above, containing peptides with single or multiple PTMs strictly disjoint from the training data.
The second partition comprises zero-shot datasets obtained from public repositories (identifiers listed in \textbf{Supplementary Table 1}). Detailed metadata for each project, including organism, modification, instrument, software versions, reference databases, and dataset statistics, is provided in \textbf{Supplementary Data 1}. These external datasets were subjected to the same quality control procedures as the training corpus, excluding unmodified peptides and removing PSMs with a precursor mass deviation greater than 0.1 Da.
Approximately 64\% of the benchmark data is derived from the held-out internal subsets, while the remaining 36\% originates from the external zero-shot datasets. In total, PTMBench comprises 257k PSMs and 46k unique peptides.

(2) \textbf{Nine-Species Benchmark:} 
The nine-species benchmark dataset was originally introduced in the DeepNovo\cite{tran2017novo} and has since become a standard reference for evaluating \textit{de novo} peptide sequencing models. This dataset comprises MS/MS spectra derived from nine phylogenetically diverse species—\textit{Apis mellifera}, \textit{Bacillus subtilis}, \textit{Candidatus endoloripes}, \textit{Homo sapiens}, \textit{Methanosarcina mazei}, \textit{Mus musculus}, \textit{Saccharomyces cerevisiae}, \textit{Solanum lycopersicum}, and \textit{Vigna mungo}. 

The dataset was directly downloaded as MGF files from the MassIVE repository (identifier: MSV000081382), originally curated and shared by DeepNovo\cite{tran2017novo}. Spectral identifications were performed using PEAKS DB software (version 8.0) under a FDR threshold of 1\%. Each species subset contains tens of thousands of annotated spectra with well-characterized peptides, offering a robust and balanced benchmark for both cross-species assessment and zero-shot evaluation of \textit{de novo} peptide sequencing models.

(3) \textbf{Revised Nine-Species Benchmark:} 
The revised nine-species benchmark dataset was constructed as an updated version of the original multi-species benchmark introduced in DeepNovo\cite{tran2017novo}, and formally described by Casanovo\cite{yilmaz2024sequence}. Raw mass spectrometry data were downloaded from the same nine PRIDE projects and converted into MGF files using ThermoRawFileParser (v1.3.4). Corresponding reference proteomes for the nine species were retrieved from UniProt to generate search indices. Each dataset was searched using the Tide search engine implemented in the Crux toolkit (version 4.1)\cite{mcilwain2014crux} with XCorr scoring and Tailor calibration. Static modifications were set to cysteine carbamidomethylation, while variable modifications included methionine oxidation, asparagine and glutamine deamidation, N-terminal acetylation, carbamylation, ammonia loss, and the combined N-terminal carbamylation plus ammonia loss.  

A 1\% FDR at the PSM level was applied using the Percolator algorithm, and PSMs corresponding to peptides shared across multiple species were removed to ensure species exclusivity. The final benchmark comprises approximately 2.8 million high-confidence PSMs derived from 343 RAW files. Compared with the original benchmark, the revised dataset incorporates higher-quality identifications, improved isotopic error handling, and explicitly controls for cross-species peptide redundancy. The finalized version of the revised nine-species benchmark dataset is publicly available in the MassIVE repository under the identifier MSV000090982, and we directly downloaded it from this source.

(4) \textbf{FGFR2:} The FGFR2 dataset was generated to assess model performance on complex biological phosphoproteomics data\cite{lauridsen2025instanovo}. It originates from experiments conducted on the T47D human breast cancer cell line, comparing wild-type (WT) cells with an FGFR2-knockout (KO) line created via CRISPR/Cas9. Following serum starvation, the cells were treated with the FGF7 ligand for either 40 minutes or 72 hours. After cell lysis and sequential enzymatic digestion with Lys-C and trypsin, phosphopeptides were specifically enriched using titanium dioxide (TiO$_2$) beads. The enriched samples were then analyzed on a Q Exactive HF mass spectrometer operating in a Data-Dependent Acquisition (DDA) mode with HCD fragmentation. To establish a ground truth for evaluation, the raw mass spectrometry data were processed using MaxQuant (v1.6.2.6), with peptide and protein identifications filtered to a 1\% FDR. The complete FGFR2 dataset comprises a total of 8,272,961 acquired spectra, of which 2,511,916 were matched to peptide sequences in the MaxQuant \texttt{msms.txt} output. For the phosphorylation zero-shot evaluation, we randomly sampled 10,000 spectra from each of the eight experimental conditions. Conversely, for the FDR analysis and benchmarking against MaxQuant, inference was performed on the full dataset of 8,272,961 spectra.

(5) \textbf{Phosphorylation Benchmarks:} To evaluate the capability of \textit{de novo} peptide sequencing models in identifying phosphorylated peptides, we compiled a benchmark collection comprising the FGFR2 dataset and six public datasets (detailed identifiers are provided in \textbf{Supplementary Table 1}). We employed a zero-shot evaluation setting; none of these datasets were included in the training data of any evaluated model (OmniNovo or baselines), allowing for an assessment of generalization to unseen phosphorylation events. Raw mass spectrometry files were converted to MGF format using ProteoWizard MSConvert\cite{adusumilli2017data} with default settings. PSMs were mapped to the provided search results (i.e., msms.txt) based on raw filenames and scan identifiers. PSMs with precursor mass deviations exceeding 0.1 Da from the theoretical peptide mass were excluded. Finally, non-phosphorylated PSMs were removed.

\textbf{Comparative Baselines.} 
We benchmarked OmniNovo against three state-of-the-art transformer-based models: Casanovo V2\cite{yilmaz2024sequence}, $\pi$-PrimeNovo\cite{zhang2025pi}, and InstaNovo-P\cite{lauridsen2025instanovo}. Casanovo V2 and $\pi$-PrimeNovo served as baselines for unmodified peptide sequencing, while InstaNovo-P was utilized as the specialized baseline for phosphorylation. 

To ensure reproducibility and eliminate confounding variables, we employed the official inference code, configuration files, and pre-trained model weights from the authors' public repositories for all standard evaluations. Independent Conda environments were constructed from scratch for each model, strictly adhering to their respective dependency requirements. All inference experiments were conducted on a uniform computing node equipped with 8 $\times$ NVIDIA H100 GPUs. To standardize performance metrics and avoid discrepancies arising from varying multi-GPU implementations across different codebases, execution was explicitly restricted to a single GPU (device 0) for all models. Batch sizes were maintained as specified in the original configurations. Detailed commands see \textbf{Supplementary Note 4}.

For the comprehensive PTM-aware evaluation involving modification types unsupported by InstaNovo-P, we established a custom baseline, $\pi$-PrimeNovo-PTM. To ensure equitable comparison, this model was initialized with the official pre-trained weights of $\pi$-PrimeNovo to leverage its existing fragmentation knowledge. It was subsequently fine-tuned on a PTM-enriched subset of our training corpus, from which all unmodified peptides were excluded to maximize the model's exposure to modification-specific features.

For the comprehensive PTM-aware evaluation involving modification types unsupported by InstaNovo-P, we established a custom baseline, $\pi$-PrimeNovo-PTM. To ensure equitable comparison, this model was initialized with the official pre-trained weights of $\pi$-PrimeNovo to leverage its existing fragmentation knowledge. It was subsequently fine-tuned on a PTM-enriched subset of our training corpus, from which all unmodified peptides were excluded to maximize the model's exposure to modification-specific features. This filtering process yielded a specialized dataset comprising 3 million PSMs, corresponding to 417k precursors and 350k unique peptides. The fine-tuning procedure utilized the same hardware configuration and hyperparameters as OmniNovo, as detailed in \textbf{Supplementary Table 2}.

\subsection*{OmniNovo Architecture}
\textbf{Model Overview.} 
OmniNovo employs a non-autoregressive Transformer architecture designed for simultaneous peptide sequencing and PTM localization. The Spectrum Encoder treats the mass spectrum as a sequence of peaks, mapping m/z and intensity values into a latent embedding space via sinusoidal encoding and self-attention layers. To balance spectral resolution with computational efficiency, input spectra are processed to retain the top 180 most intense peaks. The Peptide Decoder predicts the entire peptide sequence in parallel. The underlying network comprises a 12-layer Transformer with a hidden dimension ($d_{\text{model}}$) of 256, 16 attention heads, and a feed-forward dimension of 768. Regularization is applied via a dropout rate of 0.1, and the model utilizes FlashAttention-2 with bfloat16 mixed-precision to accelerate training.

Crucially, OmniNovo operates on a defined vocabulary of 31 tokens, comprising the 20 standard amino acid residues and 11 explicit PTM tokens. These PTM tokens represent specific mass shifts—including oxidation, phosphorylation, carbamidomethylation, and others (full list in \textbf{Supplementary Table 2})—allowing the model to learn PTM-specific fragmentation patterns and localization rules directly from spectral data.

\textbf{Training Objective.} 
To address the conditional independence assumption in non-autoregressive decoding, we utilized the Connectionist Temporal Classification (CTC) loss. The objective maximizes the log-likelihood of all valid alignment paths that collapse to the ground-truth peptide sequence $A$:
\begin{equation}
\mathcal{L}_{\text{ctc}} = - \log P(A|\mathcal{S}) = - \log \sum_{\pi \in \Gamma^{-1}(A)} \prod_{t=1}^{T} P(y_t = \pi_t | \mathcal{S})
\end{equation}
where $\Gamma$ is the CTC collapse function (merging repeats and removing blanks). Detailed derivations of the encoder embeddings and CTC dynamic programming are provided in \textbf{Supplementary Note 1}. The model was trained from scratch for 100 epochs using the AdamW optimizer (learning rate $3.5 \times 10^{-4}$, weight decay $3 \times 10^{-3}$) with a global batch size of 4,800. Detailed training configurations and hyperparameter settings are listed in \textbf{Supplementary Table 2}.

\textbf{Confidence Scoring.}
To quantify prediction reliability, OmniNovo calculates a confidence score representing the total probability of the predicted peptide sequence given the input spectrum. The non-autoregressive decoder outputs a probability matrix of fixed length ($T=40$) over the vocabulary, which includes amino acids, post-translational modifications, and the CTC blank token. After obtaining a candidate peptide via greedy decoding, the confidence score is computed by summing the probabilities of all possible latent alignment paths that reduce to this specific peptide sequence after merging consecutive identical tokens and removing blanks. This marginalization is performed using the forward algorithm dynamic programming, yielding a final score between 0 and 1 that accounts for all valid alignments corresponding to the prediction.

\subsection*{Precise Mass-and-Modification-Constrained (PMMC) Decoding}

Standard beam search decoding does not guarantee that the predicted peptide sequence matches the precursor mass measured by the mass spectrometer, nor does it inherently respect biochemical valency rules. To enforce these physical constraints, we developed the PMMC module, which reformulates the sequence generation as a mass-constrained knapsack optimization problem. The objective is to identify a sequence of tokens $\mathbf{y}$ (comprising amino acids and PTMs) that maximizes the cumulative log-probability while satisfying the precursor mass tolerance:

\begin{equation}
\text{maximize} \sum_{i=1}^{\mathbf{t}} \log P(\mathrm{y}_i | \mathcal{S})
\quad \text{subject to} \quad
m_{\text{pre}} - \delta \leq \sum_{j} w(a_j) \leq m_{\text{pre}} + \delta
\end{equation}

where $m_{\text{pre}}$ is the observed precursor mass, $\delta$ is the mass tolerance (accounting for instrument error), and $w(a_j)$ represents the mass of each residue or modification in the collapsed sequence.

We solve this optimization using a custom dynamic programming (DP) algorithm. We construct a discretized DP table where columns represent mass intervals of width $e$ (typically $10^{-4}$ Da) and rows represent decoding time steps $\tau$. Each cell $d_{\tau, l}$ stores the top-$B$ candidate partial sequences ending at time $\tau$ with a total mass falling within the interval $[e(l-1), el)$. The table is initialized at $\tau=1$ with valid starting amino acids or N-terminal modifications (e.g., acetylation) that fit the respective mass bins.

For subsequent steps, the state of each cell is updated based on three transition patterns derived from Connectionist Temporal Classification (CTC) logic:
\begin{enumerate}
    \item \textbf{Blank transition:} The sequence extends with a blank token ($\epsilon$), maintaining the current mass and sequence state.
    \item \textbf{Repeated token:} The sequence extends with a repeat of the previous non-blank token, which is merged during CTC collapsing, preserving the current mass.
    \item \textbf{New token addition:} A new amino acid or PTM token is appended. The algorithm calculates the new total mass and directs the candidate to the corresponding mass bin in the next time step.
\end{enumerate}

\textbf{Biochemical Constraint Enforcement.} Crucially, during the ``New token addition'' transition, we enforce hard biochemical constraints to prevent chemically impossible sequences. If the model proposes a PTM token, the algorithm verifies its compatibility with the preceding amino acid residue. Transitions are accepted only if they satisfy predefined rules, including: phosphorylation on Ser/Thr/Tyr; methylation, acetylation, and ubiquitination on Lys/Arg; and oxidation on Met. Invalid PTM-residue pairs are pruned immediately from the search space.

\textbf{Parallelized Implementation.} To ensure computational efficiency during inference, we implemented the PMMC algorithm as a parallelized CUDA kernel. The DP table is computed with mass bins processed in parallel threads. To handle the sequential dependency where step $\tau$ relies on $\tau-1$, we utilized a thread-level lock mechanism. This ensures that a cell at step $\tau$ waits for the completion of its dependency cells at $\tau-1$ before updating, preventing race conditions while maximizing GPU throughput.

\subsection*{False Discovery Rate Estimation via Database Calibration}

To estimate the error rate of \textit{de novo} sequencing results, we implemented a semi-supervised calibration method leveraging high-confidence identifications from standard database searches. 

For a given mass spectrometry project, we first performed a standard database search using MaxQuant, filtering results at 1\% FDR at both the PSM and peptide levels. Simultaneously, OmniNovo was applied to all MS/MS spectra in the dataset to generate \textit{de novo} peptide sequences, each associated with a model-derived confidence score, denoted as $S_{\text{conf}}$.

We then constructed a calibration set by intersecting the OmniNovo predictions with the high-confidence MaxQuant identifications. Within this subset, we evaluated the spectral quality of both the \textit{de novo} prediction and the database-identified peptide by computing their respective hyperscores against the experimental spectrum. Let $H_{\text{pred}}$ denote the hyperscore of the OmniNovo predicted peptide, and $H_{\text{db}}$ denote the hyperscore of the peptide identified by MaxQuant.

The calibration set was ranked in descending order based on the OmniNovo confidence score $S_{\text{conf}}$. We iterated through the ranked list to determine an empirical confidence threshold. For each PSM $i$ in the ranked list, we defined a binary indicator for a ``proxy false discovery'' ($I_{\text{FD}}$):
\begin{equation}
I_{\text{FD}}^{(i)} = 
\begin{cases} 
1 & \text{if } H_{\text{pred}}^{(i)} < H_{\text{db}}^{(i)} \\
0 & \text{if } H_{\text{pred}}^{(i)} \geq H_{\text{db}}^{(i)}
\end{cases}
\end{equation}
Here, a prediction is considered a ``true discovery'' if its spectral match is equal to or better than the high-confidence database assignment. Conversely, if the database search yields a significantly better spectral match, the \textit{de novo} prediction is flagged as a false discovery.

We calculated the cumulative FDR at rank $k$ as:
\begin{equation}
\text{FDR}_k = \frac{\sum_{i=1}^{k} I_{\text{FD}}^{(i)}}{k}
\end{equation}
The confidence score threshold, $\tau_{\text{1\%}}$, was determined as the score $S_{\text{conf}}^{(k)}$ of the PSM immediately preceding the first rank $k$ where $\text{FDR}_k > 0.01$. This empirically derived threshold $\tau_{\text{1\%}}$ was subsequently applied to filter the entire set of \textit{de novo} results, including spectra that were not identified by the database search. Specific confidence score thresholds determined for all datasets analyzed in Fig.~\ref{fig:4} are provided in \textbf{Supplementary Table 3}.

\subsection*{Assessment of FDR control via paired entrapment}
To rigorously evaluate the validity of False Discovery Rate (FDR) control and stress-test the specificity of the scoring algorithms, we conducted a paired entrapment-based False Discovery Proportion (FDP) estimation following the framework proposed by Wen et al.\cite{wen2025assessment}. 
This framework establishes a one-to-one pairing ($r=1$) between each original target and a unique entrapment sequence. While primarily demonstrated at the peptide level, the framework is equally applicable to PSM-level analyses provided the pairing assumptions hold.

We employed the \textbf{paired estimation method}, which provides a tighter upper bound on the FDP by evaluating the specific score relationship between each target and its paired mimic. The FDP is estimated as:
\begin{equation}
    \widehat{\text{FDP}}_{\mathcal{T} \cup \mathcal{E}_{\mathcal{T}}}^{*} = \frac{N_{\mathcal{E}} + N_{\mathcal{E} \geq s > \mathcal{T}} + 2N_{\mathcal{E} > \mathcal{T} \geq s}}{N_{\mathcal{T}} + N_{\mathcal{E}}},
\end{equation}
Here, the denominator represents the total number of reported discoveries at the threshold $s$, where $N_{\mathcal{T}}$ is the number of reported original targets and $N_{\mathcal{E}}$ is the number of reported entrapments. 
The numerator estimates the number of false discoveries by accounting for three specific scenarios:
(1) $N_{\mathcal{E}}$: the total count of reported entrapment sequences (score $\ge s$);
(2) $N_{\mathcal{E} \geq s > \mathcal{T}}$: the count of reported entrapments where the corresponding paired target was \textit{not} reported (target score $\tau < s$); and
(3) $2N_{\mathcal{E} > \mathcal{T} \geq s}$: twice the count of reported entrapments where the paired target was \textit{also} reported (target score $\tau \ge s$), but the entrapment achieved a higher score than the target (entrapment score $> \tau$).

To construct the entrapment database, we applied two distinct randomization strategies, each selected with a 50\% probability for any given target sequence, thereby ensuring diversity while preserving physicochemical properties.
In both strategies, the C-terminal amino acid was fixed to maintain enzymatic cleavage specificity, and PTMs were treated as integral parts of their corresponding amino acids.
The first strategy involved global shuffling of the peptide sequence, ensuring that the resulting sequence differed from the original. 
The second strategy employed a local permutation approach, where two adjacent amino acids were randomly swapped, provided they were not identical residues.

To challenge the discrimination power of the algorithms, we designed two experimental setups: a standard setting with a 1:1 ratio of entrapment-to-target sequences ($r=1$) where the paired estimator is applied, and a high-complexity stress test with a 100:1 ratio ($r=100$). 
In the $r=100$ setting, the search space was artificially expanded to create a rigorous stress test. 
Crucially, for a target peptide to be considered correctly identified in this scenario, it was required to be ranked as the top hit (Rank 1) against its 100 corresponding entrapment mimics. 
The search tools were applied to the concatenated target-plus-entrapment databases while blinded to the sequence labels. 
\textbf{Figure 4d} displays the FDP estimation results at the PSM level using the paired method, while the detailed peptide-level evaluation results are provided in \textbf{Supplementary Note 3}.

\subsection*{PTM Site Localization and Probability Estimation}

To rigorously quantify the confidence of PTM site assignments and resolve positional ambiguity, we implemented a post-inference localization strategy. Following the generation of the initial peptide sequence, the model extracts the peptide backbone by stripping all modification tokens while recording the types and counts of identified PTMs. We then perform a combinatorial expansion to generate the full set of theoretically plausible positional isomers. This process involves permuting the identified PTMs across all available residues on the backbone that satisfy user-defined biochemical constraints.

For each generated isomer $k$, the model computes a sequence-level confidence score, $S_k$, utilizing the path-marginalized scoring function described above. The site-specific localization probability for a given residue $i$ carrying a specific modification type $m$ is then derived via marginalization. Specifically, we calculate the ratio of the cumulative scores of all isomers where residue $i$ bears modification $m$ to the total probability mass of the entire isomer set:

\begin{equation}
P(\text{site}_{i,m}) = \frac{\sum_{k \in \mathcal{M}_{i,m}} S_k}{\sum_{j \in \mathcal{A}} S_j}
\end{equation}

where $\mathcal{A}$ represents the set of all generated positional isomers, and $\mathcal{M}_{i,m}$ is the subset of isomers where the modification $m$ is explicitly localized to residue $i$. Crucially, localization probabilities are computed independently for distinct PTM types; even if a single residue is capable of accepting multiple modification types, the numerator summation for a specific PTM $m$ includes only those isomers where residue $i$ is modified by $m$, thereby decoupling the confidence estimation for different modifications at the same site. This normalization yields a probability distribution across the peptide backbone, allowing for the precise distinguishing of high-confidence sites from ambiguous assignments based on spectral evidence.

\subsection*{Evaluation Metrics}

\textbf{Peptide Recall.} Defined as the percentage of ground-truth peptides correctly identified by the model. We require an exact string match between the predicted sequence and the ground truth. To account for isobaric ambiguity, Isoleucine (I) and Leucine (L) are treated as equivalent.

\textbf{Hyperscore.} Used as a quality metric for FDR calibration, the Hyperscore quantifies the match between the observed spectrum and the theoretical fragmentation. Formally, the hyperscore $H$ is defined as:
\begin{equation}
H = \log \left( n_b! \times n_y! \times \sum_{i=1}^{n_b} I_{b,i} \times \sum_{j=1}^{n_y} I_{y,j} \right),
\end{equation}
where $n_b$ and $n_y$ denote the numbers of matched $b$- and $y$-ions, respectively, and $I_{b,i}$ and $I_{y,j}$ represent the intensities of the matched fragment ions. All fragment ion intensities were normalized to the range of [1, 100] to ensure consistency and fairness across different spectra. This formulation follows the original definition described in MSFragger~\cite{kong2017msfragger}.

\textbf{Delta Hyperscore ($\Delta$Hyperscore).} To quantitatively assess cases where the de novo prediction disagrees with the database search label, we calculate the difference in spectral quality between the two assignments. For a given spectrum, let $H_{\text{pred}}$ be the hyperscore of the OmniNovo prediction and $H_{\text{db}}$ be the hyperscore of the database search label. The $\Delta$Hyperscore is defined as:
\begin{equation}
\Delta H = H_{\text{pred}} - H_{\text{db}}
\end{equation}
A positive $\Delta H$ indicates that the sequence predicted by OmniNovo provides a better statistical explanation of the experimental spectrum than the sequence assigned by the database search engine.

\textbf{Prediction Coverage.} Defined as the percentage of the total test spectra for which the model provides a prediction. Importantly, if the model fails to output a sequence for a given spectrum, it is not excluded from the calculation; instead, it is treated as an incorrect prediction with a confidence score of 0. Recall-coverage curves are generated by calculating peptide recall at varying confidence thresholds, illustrating the trade-off between the number of predictions made (coverage) and their accuracy.

\textbf{PTM Site Localization Precision.} We employed two distinct metrics to evaluate PTM localization performance:

\textit{1. Residue-level Precision (Fig. 2d):} Following the evaluation protocol established by Casanovo~\cite{yilmaz2022novo}, we assessed the precision of specific amino acids and modifications based on mass alignment. The predicted sequence and ground truth were tokenized, and the longest common prefix and suffix were identified based on cumulative mass matching (tolerance: 0.5 Da for cumulative mass, 0.1 Da for individual residues). A predicted residue or PTM is considered correct only if it aligns with the ground truth within the mass-matched segments. Precision is calculated as the number of correctly matched residues of a specific type divided by the total number of predicted residues of that type.

\textit{2. Peptide-level Localization Accuracy (Fig. 5d):} To evaluate the model's ability to correctly place PTMs on the correct backbone, we used a strict two-step verification. First, we calculate \textbf{Backbone Accuracy} by stripping all modifications and requiring an exact string match of the amino acid sequence (treating I and L as equivalent). Second, for the subset of peptides with correct backbones, we calculate \textbf{Modification Site Accuracy}. This requires an exact string match of the full sequence, including all modification positions. This metric strictly measures whether the PTMs are localized to the exact correct residues, conditional on the peptide backbone being correctly identified.

\newpage
\begin{appendices}

\setcounter{page}{1}
\renewcommand{\thefigure}{S\arabic{figure}}
\setcounter{figure}{0}  

\renewcommand{\thetable}{S\arabic{table}}
\setcounter{table}{0}  

\renewcommand{\theequation}{S\arabic{equation}}
\setcounter{equation}{0}

\section*{Supplementary Information}
For \textit{Accurate de novo sequencing of the modified proteome with OmniNovo} \\

The supplementary information includes:
\begin{itemize}
    \item \textbf{Supplementary Note 1:} Mathematical Details of OmniNovo
    \item \textbf{Supplementary Note 2:} PMMC Dynamic Programming
    \item \textbf{Supplementary Note 3:} Extended Performance Analysis
    \item \textbf{Supplementary Note 4:} Baseline Implementation Details
    \item \textbf{Supplementary Note 5:} Additional Spectral Alignment Examples for Conflicting Predictions
    \item \textbf{Supplementary Table 1:} Detailed Dataset Statistics
    \item \textbf{Supplementary Table 2:} Training Hyperparameters
    \item \textbf{Supplementary Table 3:} Confidence score thresholds
    \item Tables S1 to S3
    \item Figs. S1 to S30
\end{itemize}
\clearpage

\section*{Supplementary Note 1: Mathematical Details of OmniNovo}
\label{si:note1}

\textbf{Spectrum Encoder} 
The encoder maps the input spectrum $\mathcal{I}$ into a latent embedding $E$. 
We interpret the spectrum as a sequence, where each value $\text{mz}^{(i)}$ is projected through sinusoidal encodings: 
\begin{equation}
\text{e}^{0}_{i}(\text{mz}) =
\begin{cases}
\displaystyle
\sin\!\left(
  \dfrac{
    \text{mz}
  }{
    \dfrac{(\text{mz})_{\max}}{(\text{mz})_{\min}}
    \!\left(
      \dfrac{(\text{mz})_{\min}}{2\pi}
    \right)^{\tfrac{2i}{d}}
  }
\right),
& i \le \tfrac{d}{2}, \\[10pt]
\displaystyle
\cos\!\left(
  \dfrac{
    \text{mz}
  }{
    \dfrac{(\text{mz})_{\max}}{(\text{mz})_{\min}}
    \!\left(
      \dfrac{(\text{mz})_{\min}}{2\pi}
    \right)^{\tfrac{2i}{d}}
  }
\right),
& \text{otherwise.}
\end{cases}
\end{equation}
Here $d$ is the hidden dimension. Intensities $\text{p}^{(i)}$ are embedded with the same mapping and then summed with $\text{e}(\text{mz})$. 
The resulting sequence $E^{0} = (\text{e}^0_1, \ldots, \text{e}^0_k)$ is processed by $m$ Transformer encoder layers, where each layer refines the representation:
\begin{equation}
E^{j} = \text{SelfAttention}(\text{e}^{j-1}_0, \ldots, \text{e}^{j-1}_k).
\end{equation}
The final state $E^{(m)}$ serves as the compact spectral embedding used by the decoder. 

\textbf{Peptide Decoder} 
Our decoder follows a non-autoregressive design. Instead of predicting tokens sequentially, it outputs probabilities for all positions simultaneously. 
Each input token $\text{y}_i$ is embedded as $h^{0}_i = \text{EmbeddingLayer}(\text{y}_i)$ and then propagated through self-attention and cross-attention modules, where cross-attention incorporates the encoder features $E^{m}$. 
The last hidden states ${h}^{(L)}$ are mapped into token probabilities:
\begin{equation}
P_s(\cdot \mid \mathcal{I}) = \text{softmax}(W h_s^{(L)}).
\end{equation}
In contrast to earlier NAT designs that only relied on positional encodings, our decoder also uses partial sequence information (precursor m/z and charge) for glancing during training. 

\textbf{CTC Training} 
Direct cross-entropy training of parallel prediction models often produces multi-modal errors due to missing token dependencies. 
To overcome this, we adopt the Connectionist Temporal Classification (CTC) loss. 
CTC introduces a maximum decoding length $T$ and a reduction operator $\Gamma(\cdot)$ that merges consecutive identical tokens, removes the blank token $\epsilon$, and prevents merging across $\epsilon$. 
For example, the sequence AABC$\epsilon$C reduces to ABCC.  

The objective is to compute the probability of all alignment paths that collapse to the ground truth sequence $A$. 
Since enumerating all $(m+1)^{t}$ possible paths is infeasible, we compute this via dynamic programming. 
Let $\alpha(\tau, r)$ denote the probability of producing the prefix $A_{1:r}$ using only the first $\tau$ decoding steps:
\begin{equation}
\alpha(\tau, r) = \!\!\!\!\sum\nolimits_{y_{1:\tau} : \Gamma(y_{1:\tau}) = A_{1:r}} \prod_{i=1}^{\tau} P(\mathrm{y}_i|\mathcal{S}).
\end{equation}
Initialization is given by
\begin{equation}
\alpha(\tau, 0) = \prod_{i=1}^{\tau} P(\mathrm{y}_i = \epsilon|\mathcal{S}), \quad
\alpha(1,1) = P(\mathrm{y}_1 = a_1|\mathcal{S}), \quad
\alpha(1,r) = 0 \ \ (r>1).
\end{equation}

The recurrence splits by whether the current symbol equals the previous one:
\begin{equation}
\alpha(\tau, r) =
\begin{cases}
(\alpha(\tau-1, r|\mathrm{y}_{\tau-1}\neq\epsilon) + \alpha(\tau-1, r-1|\mathrm{y}_{\tau-1}=\epsilon)) P(\mathrm{y}_\tau = a_r|\mathcal{S}) \\ 
\quad + \alpha(\tau-1, r) P(\mathrm{y}_{\tau-1}=\epsilon|\mathcal{S}), & a_r = a_{r-1}, \\[6pt]
(\alpha(\tau-1, r|\mathrm{y}_\tau\neq\epsilon) + \alpha(\tau-1, r-1)) P(\mathrm{y}_{\tau-1}=a_r|\mathcal{S}) \\ 
\quad + \alpha(\tau-1, r) P(\mathrm{y}_\tau=\epsilon|\mathcal{S}), & a_r \neq a_{r-1}.
\end{cases}
\end{equation}

Finally, the training loss is defined as
\begin{equation}
\mathcal{L}_{\text{ctc}} = - \log P(A|\mathcal{S}) = - \log \alpha(t, |A|),
\end{equation}
which maximizes the probability of all valid alignment paths that yield the correct target sequence. 

\textbf{Confidence score}
To quantify the reliability of its predictions, OmniNovo employs a sophisticated confidence scoring mechanism derived directly from the output logits of its non-autoregressive decoder. The core of this mechanism is analogous to the forward algorithm used in CTC training. For each input spectrum, the decoder generates a probability matrix of size $T \times V$, where $T=40$ is the predetermined sequence length and $V$ is the vocabulary size, which includes all amino acids, PTMs, and a special CTC blank token ($\epsilon$).

After the model produces a final peptide prediction, $Y_{pred}$, by applying a greedy decoding strategy to this probability matrix, we proceed to calculate its confidence score. This score is defined as the total probability of all possible raw output sequences of length $T=40$ that can be reduced to the final prediction $Y_{pred}$ according to the CTC collapsing rules (i.e., merging consecutive identical non-blank tokens and removing all blank tokens). This summation over all valid alignment paths is computed efficiently using a dynamic programming approach, mirroring the calculation of the total probability $P(A|\mathcal{S})$ as defined in our CTC loss function. The resulting value, representing the marginalized probability of observing the predicted peptide given the spectrum, naturally falls within the range and serves as our final confidence score. This approach provides a more robust confidence measure than single-path-based methods, as it holistically considers the entire landscape of latent alignments that correspond to the same final peptide sequence.

\textbf{Training configuration}
The OmniNovo model, which comprises approximately 35 million parameters, was trained from scratch for 60 epochs. For model validation and hyperparameter tuning, a dedicated validation set was created by randomly sampling 100,000 PSMs from the complete training dataset, ensuring a balanced composition of 50,000 unmodified and 50,000 post-translationally modified peptides. All experiments were conducted with a fixed random seed of 42 to ensure reproducibility.

We employed the AdamW optimizer for model training with a learning rate of 3.5e-4 and a weight decay of 3e-3. Additional AdamW hyperparameters were set to standard values of $\beta_1 = 0.9$, $\beta_2 = 0.999$, and $\epsilon = 1 \times 10^{-8}$. A linear learning rate warm-up schedule was applied during the first epoch to facilitate a stable start to the training process. The optimization objective was the CTC loss, as previously described. To prevent exploding gradients and ensure training stability, we applied gradient norm clipping with a maximum norm of 2.5.

The training was executed on a Linux server equipped with 8x NVIDIA H100 GPUs. We utilized a distributed data parallel strategy, orchestrated by the PyTorch Lightning framework, with a global training batch size of 4,800. To maximize computational throughput and reduce memory footprint, training was performed using bfloat16 (bf16) mixed-precision, and the self-attention mechanism was accelerated using FlashAttention-2. The entire training process for 60 epochs completed in approximately five days.

\clearpage
\section*{Supplementary Note 2: PMMC Dynamic Programming}
\label{si:note2}

To ensure that the generated \textit{de novo} peptide sequence aligns with the molecular mass measured by the mass spectrometer as well as PTM formation under biochemical constraints, we impose a precise mass-and-modification constraint during decoding. The true peptide mass $\mathrm{m}_{tr}$ must fall within the range $[m - \sigma, m + \sigma]$, where $m$ denotes the precursor mass reported by the mass spectrometer and $\sigma$ represents its measurement error (typically around $10^{-3}$), and each PTM token must correspond to a chemically compatible amino acid tokens—for example, phosphorylation is typically limited to serine (S), threonine (T), and tyrosine (Y) residues. 

Because neural sequence generation models are often opaque and difficult to steer, it is nontrivial to guarantee that the predicted peptide strictly satisfies both mass and PTM rules. To achieve this, we reinterpret the decoding process as a \textbf{knapsack-like optimization problem}.
In this setting, each amino acid as well as PTM token is treated as an “item” with an associated mass and predicted log-probability. The objective is to select a combination of such items so that their total mass lies within the target range while maximizing the accumulated probability:

\begin{equation}
\text{maximize} \sum_{i=1}^{\mathbf{t}} \log P(\mathrm{y}_i | \mathcal{S})
\quad \text{subject to} \quad
\mathcal{L} \leq \sum\nolimits_{\forall a_j \in \Gamma(\mathbf{y})} w(a_j) \leq \mathcal{U},
\end{equation}

where $\mathcal{L} = m - tol$ and $\mathcal{U} = m + tol$ define the allowed mass interval, with $tol$ accounting for both decoding tolerance and instrumental uncertainty.

Inspired by the dynamic programming principle of $\pi$-PrimeNovo, we solve this constrained optimization efficiently through a structured decoding table. Let $e$ denote the mass precision step. The DP table consists of $\lceil \mathcal{U} / e \rceil$ columns, each corresponding to a narrow mass interval of width $e$. The cell $\mathbf{d}^{\tau, l}$ holds the most probable peptide fragment of length $\tau$ whose total mass lies within $[e(l-1), e l)$.

At the first decoding step ($\tau = 1$), the dynamic programming (DP) table is initialized by assigning each cell according to its corresponding mass interval. The first cell ($l = 0$) is initialized with the empty sequence $\epsilon$, representing zero mass. 
For other cells, if there exist amino acids (or PTM) whose masses fall within the interval $[e(l-1), e \cdot l)$, all such residues are stored in that cell as potential starting points. By default, only acetylation is allowed as the initial PTM, but this setting can be easily modified in the configuration file according to user requirements.
Cells that do not contain any residues within their mass range remain empty. This initialization ensures that every feasible amino acid or PTM that can serve as the first element of a valid peptide is properly considered.

For each subsequent decoding step $\tau$, we update $\mathbf{d}^{\tau, l}$ using three transition patterns:

\begin{enumerate}
    \item \textbf{Blank transition:}  
    When $y_{\tau} = \epsilon$, the sequence and its cumulative mass remain unchanged:
    \begin{equation}
        \mathcal{H}_{\tau,l}^{(1)} = \{\mathbf{y} \oplus \epsilon \ | \ \forall \mathbf{y} \in \mathbf{d}^{\tau-1,l}\}.
    \end{equation}

    \item \textbf{Repeated token:}  
    If the new symbol repeats the previous non-$\epsilon$ token, CTC collapsing keeps the sequence identical, yielding:
    \begin{equation}
        \mathcal{H}_{\tau,l}^{(2)} = \{\mathbf{y} \oplus y_{\tau-1} \ | \ \forall \mathbf{y} \in \mathbf{d}^{\tau-1,l}, \ y_{\tau-1} \neq \epsilon \}.
    \end{equation}

    \item \textbf{New residue or PTM addition:}  
    When a new amino acid or PTM is appended, the updated total mass must place the sequence into the current interval:
    \begin{align}
        \mathcal{H}_{\tau,l}^{(3)} = \Big\{
        \mathbf{y} \oplus y_{\tau} \ \Big| \
        &\forall \, 1 \leq l_0 < l, \ \mathbf{y} \in \mathbf{d}^{\tau-1,l_0}, \ y_{\tau} \neq \epsilon, \\[-2pt]
        &e(l-1) \leq \sum\nolimits_{\forall a_j \in \Gamma(\mathbf{y} \oplus y_{\tau})} w(a_j) < e l
        \Big\}.
    \end{align}

    During this update, a strict PTM control rule is enforced: if $y_\tau$ corresponds to a PTM, its preceding token $y_{\tau-1}$ is examined, and $y_\tau$ is accepted only when the combination of $y_{\tau-1}$ and $y_\tau$ constitutes a biochemically valid amino acid–PTM pair. Consequently, only chemically feasible PTMs—those predicted by the model and validated against established biochemical constraints—are allowed to contribute to $\mathcal{H}_{\tau,l}^{(3)}$.

   By default, we pre-define a set of  rules to control valid PTM generation during PMMC decoding: 1) phosphorylation is permitted on serine (S), threonine (T), and tyrosine (Y); 2) methylation on lysine (K), arginine (R), and serine (S); 3) dimethylation on lysine (K) and arginine (R); 4) trimethylation on lysine (K); 5) acetylation on lysine (K) and arginine (R); 6) oxidation on methionine (M); ubiquitination on lysine (K); 7) carbamidomethylation on cysteine (C); 8) and deamidation on asparagine (N) and glutamine (Q). These default settings can be easily adjusted in the configuration file according to specific experimental settings or user-defined modification schemes.
\end{enumerate}

After collecting all candidates, the $l$-th cell at step $\tau$ is updated by keeping the top-$B$ highest-scoring sequences:
\begin{equation}
\mathbf{d}^{\tau, l} = 
\operatorname*{top_B}_{\mathbf{y} \in 
\mathcal{H}_{\tau,l}^{(1)} \cup \mathcal{H}_{\tau,l}^{(2)} \cup \mathcal{H}_{\tau,l}^{(3)}
} \Big(\sum\nolimits_{y_j \in \mathbf{y}} \log P(y_j | \mathcal{S}) \Big).
\end{equation}

Once decoding reaches the final time step $\mathbf{t}$, the peptide sequence from the cell corresponding to the mass closest to the target precursor mass is chosen as the final prediction.

We also have CUDA algorithm designed for PMMC decode, which begins by constructing a two-dimensional dynamic programming table. In this table, we store the most probable partial sequence at time step $t$, ensuring that it satisfies the mass constraint for each cell. Each cell independently carries out its calculations. Since the computation of each cell relies on results from the previous time step, we employ a lock mechanism to prevent computational race conditions.
Starting from the first time step, we select the amino acid or permissible initial PTM (e.g., acetylation) whose mass aligns with the mass constraint for each mass bin, filling in the respective cell. After completing its computation, the cell releases its dependency lock, allowing the subsequent cell to utilize its results for further computation. Throughout the process, we perform recursive updates in accordance with the PMMC transition rules, extending sequences with either amino acids or PTMs. Crucially, during these updates, we enforce biochemical validity: if a PTM token is proposed, we verify it against the preceding amino acid (e.g., ensuring phosphorylation only follows S, T, or Y) before accepting the transition.
Once all cells in the table have completed their calculations, we can retrieve the value in the last cell, which represents the final sequence we seek.

\begin{algorithm}[H]
\caption{CUDA PMMC Algorithm.}
\begin{algorithmic}[1]
\Procedure{$PMMC\_inference$}{$lock, token\_mass, grid\_size, rule$}
\State $w \gets blockIdx.x$ \Comment{Mass index}
\State $dim \gets blockDim.x$
\State $h \gets threadIdx.x$ \Comment{Time step index (The [w,h] grid)}
\If{$w = 0 \lor h = 0$}
    \State \Return
\EndIf \Comment{The zero position and zero mass do not need calculation.}

\If{$w > maxW$}
    \State $lock[w \times dim + h] \gets 1$
    \State \Return
\EndIf 

\If{$h = 1$}
    \State $Calculate\_h1\_WithInitPTM(w,h,rule)$ 
    \Comment{Initialize with valid AAs or allowed initial PTMs fitting mass [w].}
    \State $lock[w \times dim + h] \gets 1$
    \State \Return
\EndIf

\While{$lock[w \times dim + h - 1] \neq 1$} \EndWhile \Comment{Wait for dependency lock(previous time step) to release.}

\State $PutIn\epsilon OrRepeat()$
\Comment{Handle Blank transition ($\epsilon$) and Repeated Token (CTC collapse).}

\For{$i \gets 1$ \textbf{to} $Vocab\_Size$} \Comment{Iterate over all AAs and PTMs}
    \State $minw \gets \lfloor {w \times grid\_size - token\_mass[i]}/{grid\_size} \rfloor$ 
    \For{\textbf{each} $src\_w \in \{minw, minw+1\}$}
        \While{$lock[src\_w \times dim + h - 1] \neq 1$} \EndWhile
        \Comment{Wait for source cell dependency.}
        
        \If{$IsPTM(i)$}
            \State $prev\_token \gets GetLastToken(src\_w, h-1)$
            \If{$CheckBioConstraint(prev\_token, i,rule)$}
                \State $PutInToken(i)$ 
                \Comment{Add PTM only if biochemically valid.}
            \EndIf
        \Else
            \State $PutInToken(i)$
            \Comment{Add Amino Acid.}
        \EndIf
    \EndFor
\EndFor

\State $lock[w \times dim + h] \gets 1$ \Comment{Release lock for current cell.}
\State \Return
\EndProcedure
\end{algorithmic}
\end{algorithm}

\clearpage
\section*{Supplementary Note 3: Extended Performance Analysis}
\label{si:note3}

This note details performance metrics across individual species and modification types, supplementing the aggregated results in the main text.

\subsection*{Species-Specific Performance}
We analyzed peptide recall rates for the nine species comprising the original and revised benchmarks (Fig.~\ref{fig:supp_granular}a,b). OmniNovo consistently outperformed baseline models (Casanovo-V2, InstaNovo-P, and $\pi$-PrimeNovo) across diverse organisms. The analysis covers mammalian species (\textit{H. sapiens}, \textit{M. musculus}) as well as phylogenetically distinct organisms, including \textit{A. mellifera}, \textit{B. subtilis}, \textit{C. endoloripes}, \textit{M. mazei}, \textit{S. cerevisiae}, \textit{S. lycopersicum}, and \textit{V. mungo}. Furthermore, the ablation study (OmniNovo w/o PMMC) confirms the substantial contribution of the PMMC module to recall improvements across all tested species.

\subsection*{PTM Site Localization Precision}
We further evaluated the precision of modification site localization on the PTMBench dataset (Fig.~\ref{fig:supp_granular}c). OmniNovo demonstrated superior precision compared to $\pi$-PrimeNovo-PTM across four major modification types: acetylation, phosphorylation, methylation, and ubiquitinization.

\subsection*{Zero-Shot Phosphorylation Site Localization}
\begin{figure}[htbp]
    \centering
    \begin{subfigure}[b]{0.49\textwidth}
        \centering
        \includegraphics[width=\textwidth]{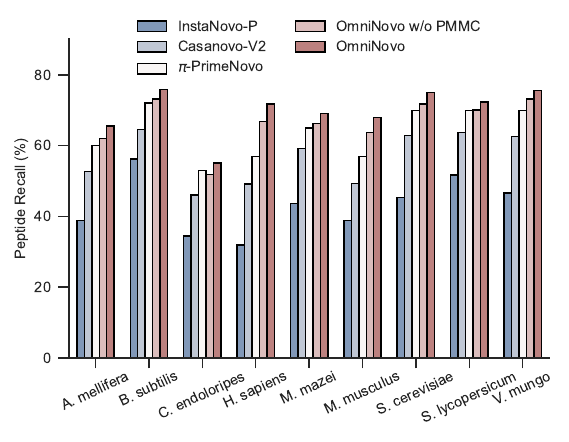}
        \caption{Original nine-species breakdown}
        \label{fig:supp_granular_a}
    \end{subfigure}
    \hfill
    \begin{subfigure}[b]{0.49\textwidth}
        \centering
        \includegraphics[width=\textwidth]{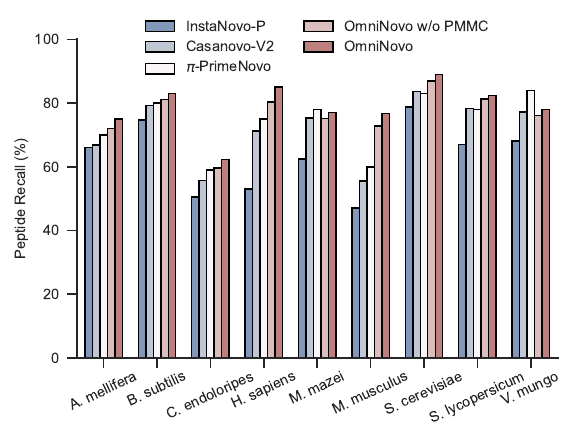}
        \caption{Revised nine-species breakdown}
        \label{fig:supp_granular_b}
    \end{subfigure}
    
    \begin{subfigure}[b]{0.39\textwidth}
        \centering
        \includegraphics[width=\textwidth]{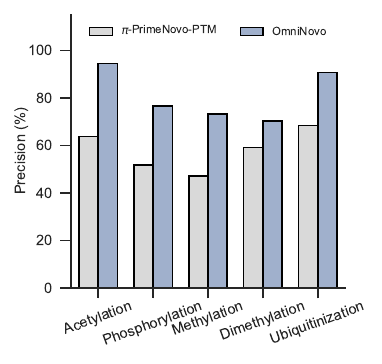}
        \caption{PTM site identification precision}
        \label{fig:supp_granular_c}
    \end{subfigure}
    \hfill
    \begin{subfigure}[b]{0.6\textwidth}
        \centering
        \includegraphics[width=\textwidth]{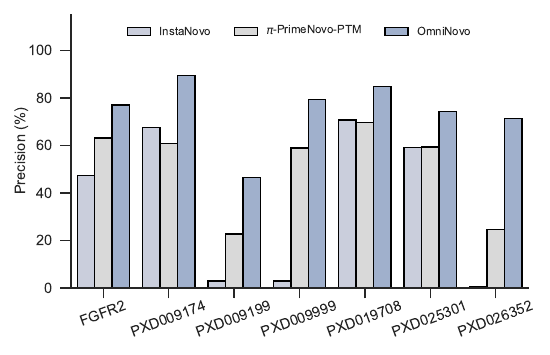}
        \caption{Zero-shot phosphorylation datasets}
        \label{fig:supp_granular_d}
    \end{subfigure}
    
    \caption{\textbf{Detailed performance metrics.} 
    \textbf{a, b,} Peptide recall for each of the nine species in the original (\textbf{a}) and revised (\textbf{b}) benchmarks. Comparisons include OmniNovo, OmniNovo (w/o PMMC), and baseline models.
    \textbf{c,} Modification site precision on PTMBench.
    \textbf{d,} Phosphorylation site precision on zero-shot phosphorylation datasets.}
    \label{fig:supp_granular}
\end{figure}

To assess zero-shot generalization, we analyzed phosphorylation site precision on external datasets representing experimental conditions not seen during training (Fig.~\ref{fig:supp_granular}d). OmniNovo maintained high precision across these independent datasets (e.g., PXD009174, PXD019708), significantly outperforming InstaNovo-P and $\pi$-PrimeNovo-PTM, thereby validating its robustness in handling unseen experimental environments.

\subsection*{Peptide-level FDP estimation}
Complementing the PSM-level analysis (Fig. 4d), we evaluated peptide-level FDR control using the paired entrapment framework described in the Methods. Supplementary Figure X compares the estimated False Discovery Proportion (FDP) against the nominal FDR thresholds reported by OmniNovo and MaxQuant.

In the standard setting ($r=1$), OmniNovo demonstrates rigorous control with an estimated FDP near zero across all thresholds. In contrast, MaxQuant exhibits systematic FDR inflation, with a constant estimated FDP of approximately 19.85\%, independent of the nominal threshold.

Under the high-complexity stress test ($r=100$), OmniNovo maintains effective discrimination at strict thresholds (e.g., 1\% FDR), with FDP increasing monotonically as thresholds relax. Conversely, MaxQuant fails to control FDR in this expanded search space, yielding an estimated FDP of 113.29\%, implying that top-ranking entrapment sequences outnumbered target identifications. These results confirm that OmniNovo offers superior specificity and robust error rate estimation at the peptide level.


\begin{figure*}[htbp]
    \centering
    \includegraphics[width=0.4\textwidth]{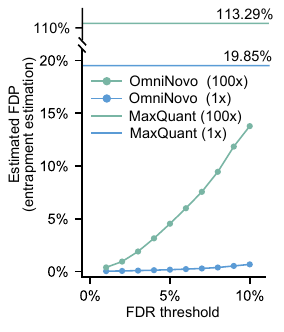} 
    \caption{Comparison of peptide-level estimated false discovery proportions on the FGFR2 dataset as a function of the FDR threshold. FDPs were estimated for OmniNovo and MaxQuant using either one (1×) or one hundred (100×) entrapment sequences per PSM.}
    \label{fig:supp_fdp}
\end{figure*}
\section*{Supplementary Note 4: Baseline Implementation Details}
\label{si:note4}

To ensure fair comparisons, all baseline models were executed using their official repositories with standardized configurations.

\subsection*{InstaNovo-P.} 
We utilized the official implementation of InstaNovo provided in the \href{https://github.com/instadeepai/InstaNovo}{authors' GitHub repository}. A dedicated Conda environment was established to satisfy all \href{https://instadeepai.github.io/InstaNovo/tutorials/getting_started/#installation}{dependency requirements}. All inference tasks were executed on a computing node equipped with 8$\times$ NVIDIA H100 GPU. But we only use GPU 0 for inference.

We employed the pre-trained checkpoint \texttt{instanovo-phospho-v1.0.0.ckpt} with the corresponding \texttt{instanovo-p} configuration. The prediction was performed with the \texttt{--no-refinement} flag, as the refinement module is not supported for the instanovo-phospho model variant. The exact command line argument used for inference is as follows:

\begin{verbatim}
CUDA_VISIBLE_DEVICES=0 instanovo predict
--data-path test.mgf
--config-name instanovo-p 
--output-path test.csv
--instanovo-model instanovo-phospho-v1.0.0.ckpt
--no-refinement
\end{verbatim}

\subsection*{$\pi$-PrimeNovo}
We utilized the official implementation of $\pi$-PrimeNovo provided in the \href{https://github.com/PHOENIXcenter/pi-PrimeNovo}{authors' GitHub repository}. A dedicated Conda environment was established to satisfy all dependency requirements. All inference tasks were executed on a computing node equipped with a single NVIDIA H100 GPU. All inference tasks were executed on a computing node equipped with 8$\times$ NVIDIA H100 GPU. But we only use GPU 0 for inference.

We employed the pre-trained checkpoint \texttt{model\_massive.ckpt} with the default configuration. The exact command line argument used for inference is as follows:

\begin{verbatim}
CUDA_VISIBLE_DEVICES=0 python -m PrimeNovo.PrimeNovo 
--mode=eval
--peak_path=./test.mgf 
--model=./model_massive.ckpt
\end{verbatim}
\section*{Supplementary Note 5: Additional Spectral Alignment Examples for Conflicting Predictions}

To provide a comprehensive assessment of cases where OmniNovo's high-confidence predictions (confidence score $> 0.8$) conflict with labels from database search engines (MaxQuant, MSFragger) or the baseline model ($\pi$-PrimeNovo-PTM), we present an extended catalog of spectral evidence in Supplementary Figures S3 to S30.

These figures comprise a total of 112 representative case studies (4 cases per figure across 28 figures), categorized by PTM type. For each case, the experimental spectrum is visualized against the peptide sequences identified by four different methods. The subplots are arranged vertically in the following order:
\begin{enumerate}
    \item OmniNovo
    \item MaxQuant
    \item MSFragger
    \item $\pi$-PrimeNovo-PTM
\end{enumerate}

Spectral annotation was performed using \texttt{spectrum-utils} (v0.4.2). Peak matching was restricted to b- and y-ions with a mass tolerance of 20 ppm. In the resulting plots, y-ions are annotated in red, and b-ions are annotated in blue. The identified peptide sequence is displayed in the upper-right corner of each subplot.

PTMs within the sequences are indicated by specific abbreviations. The correspondence between these abbreviations and the full modification names is listed below:
\begin{itemize}
    \item \textbf{cam}: Carbamidomethylation
    \item \textbf{ox}: Oxidation
    \item \textbf{ac}: Acetylation
    \item \textbf{deam}: Deamidation
    \item \textbf{car}: Carbamylation
    \item \textbf{ph}: Phosphorylation
    \item \textbf{me}: Methylation
    \item \textbf{di}: Dimethylation
    \item \textbf{tr}: Trimethylation
    \item \textbf{ub}: Ubiquitination
\end{itemize}

Collectively, these visualizations permit a detailed inspection of peak coverage and intensity matching across different methods for spectra where traditional database search results diverge from OmniNovo's predictions.

\begin{table}[htbp]
\centering
\caption{\textbf{Summary of dataset identifiers used in this study.}}
\label{tab:compact_datasets}
\renewcommand{\arraystretch}{1.5} 
\begin{tabularx}{\textwidth}{@{}l>{\RaggedRight\arraybackslash}X@{}} 
\toprule
\textbf{Dataset Category} & \textbf{Identifiers} \\
\midrule
\textbf{Training set} & MassIVE-KB v1, MassIVE-KB v2, PXD037285, PXD039441, IPX001804000, PXD020271, PXD052627, PXD002425, PXD012007, PXD000559, PXD005536, PXD021188, PXD046571, PXD003700, PXD011926, PXD003281, PXD014719, PXD027949, PXD009994, PXD002800, PXD009070, PXD005252 \\
\midrule
\textbf{PTMBench} & PXD037285, PXD039441, IPX001804000, PXD020271, PXD052627, PXD002425, PXD012007, PXD000559, PXD005536, PXD021188, PXD046571, PXD003700, PXD011926, PXD003281, PXD014719, PXD027949, PXD009994, PXD002800, PXD009070, PXD005252, PXD015782, PXD005903, PXD023468, PXD024309, PXD014799 \\
\midrule
\textbf{Phosphorylation datasets} & PXD009174, PXD009999, PXD019708, PXD025301, PXD026352, PXD009199, PXD062859(FGFR2) \\
\bottomrule
\end{tabularx}
\end{table}
\begin{table}[htbp]
    \centering
    \caption{\textbf{Supplementary Table 2 | Hyperparameters and configuration details for the model training and inference.} }
    \label{tab:hyperparams}
    \renewcommand{\arraystretch}{1.3} 
    \begin{tabularx}{\textwidth}{@{}l>{\RaggedRight\arraybackslash}X@{}}
        \toprule
        \textbf{Parameter} & \textbf{Value} \\
        \midrule
        \multicolumn{2}{@{}l}{\textit{\textbf{Model Architecture}}} \\
        Embedding dimension ($d_{\text{model}}$) & 256 \\
        Feedforward dimension ($d_{\text{ff}}$) & 768 \\
        Number of layers ($N_{\text{layers}}$) & 12 \\
        Number of attention heads ($h$) & 16 \\
        Number of key-value heads & 8 \\
        Head dimension & 48 \\
        Dropout rate & 0.1 \\
        Activation function & SiLU \\ 
        Normalization layer & RMSNorm \\ 
        Flash Attention & True \\
        Precision & bfloat16 (bf16) \\
        
        \midrule
        \multicolumn{2}{@{}l}{\textit{\textbf{Spectrum Processing}}} \\
        Maximum peaks per spectrum & 180 \\
        Precursor mass tolerance & 50 ppm \\
        Fragment mass range & 1 -- 6500 Da \\
        Maximum precursor charge & 10+ \\
        Remove precursor peak tolerance & 1.0 Da \\
        Isotope error range & [0, 1] \\
        
        \midrule
        \multicolumn{2}{@{}l}{\textit{\textbf{Training Optimization}}} \\
        Optimizer & AdamW \\ 
        Learning rate & $3.5 \times 10^{-4}$ \\
        Weight decay & $3.0 \times 10^{-3}$ \\
        Gradient clipping & 2.5 (Norm) \\
        Batch size (Training) & 4800 spectra \\
        Maximum epochs & 60 \\
        Warm-up epochs & 1 \\
        Learning rate scheduler & Cosine Decay \\ 
        Loss function & torch.nn.CTCLoss \\ 
        Label smoothing & False \\ 
        
        \midrule
        \multicolumn{2}{@{}l}{\textit{\textbf{Inference \& Decoding}}} \\
        Inference batch size & 4096 \\
        Beam search size & 0 (Greedy search) \\
        PMMC mass control tolerance & 0.1 Da \\
        
        \midrule
        \multicolumn{2}{@{}l}{\textit{\textbf{Hardware \& Environment}}} \\
        GPU Model & NVIDIA H100 (80GB) \\ 
        Number of GPUs & 8 \\ 
        Training time (approx.) & \~144 hours \\ 
        PyTorch version & 2.3.1 \\ 
        Random seed & 42 \\

        \midrule
        \textbf{Vocabulary \& Masses} & 
        \footnotesize 
        \textbf{Amino Acids:} 
        G: 57.0215, A: 71.0371, S: 87.0320, P: 97.0528, V: 99.0684, T: 101.0477, C: 103.0092, L/I: 113.0841, N: 114.0429, D: 115.0269, Q: 128.0586, K: 128.0950, E: 129.0426, M: 131.0405, H: 137.0589, F: 147.0684, R: 156.1011, Y: 163.0633, W: 186.0793. \newline 
        \textbf{Modifications:} 
        Carbamylation (car): 43.0058, Deamidation (deam): 0.9840, Oxidation (ox): 15.9949, Carbamidomethylation (cam): 57.0215, NH$_3$ loss (-nh3): -17.0265, Acetylation (ac): 42.0106, Di-methylation (di): 28.0313, Methylation (me): 14.0157, Trimethylation (tr): 42.0470, Ubiquitination (ub): 114.0429, Phosphorylation (ph): 79.9663. \\
        \bottomrule
    \end{tabularx}
\end{table}
\begin{table}[htbp]
    \centering
    \caption{\textbf{Supplementary Table 3 | Confidence score thresholds for different datasets.} }
    \label{tab:confidence_thresholds}
    
    \renewcommand{\arraystretch}{1.3} 
    
    \begin{tabular}{lcc}
        \toprule
        \multirow{2.5}{*}{\textbf{Dataset}} & \multicolumn{2}{c}{\textbf{Confidence score threshold}} \\
        
        \cmidrule(lr){2-3} 
        
         & \textbf{1\% FDR} & \textbf{5\% FDR} \\
        \midrule
        
        PXD015782 & 0.9214 & 0.7124 \\
        PXD005903 & 0.8167 & 0.3154 \\
        PXD023468 & 0.8014 & 0.4057 \\
        PXD024309 & 0.9101 & 0.3686 \\
        PXD014799 & 0.7753 & 0.1816 \\
        PXD062859 (FGFR2) & 0.9401 & 0.6538 \\
        \bottomrule
    \end{tabular}
\end{table}

\begin{figure}[htbp]
    \centering
    \begin{subfigure}[b]{0.48\linewidth}
        \centering
        \includegraphics[width=\linewidth]{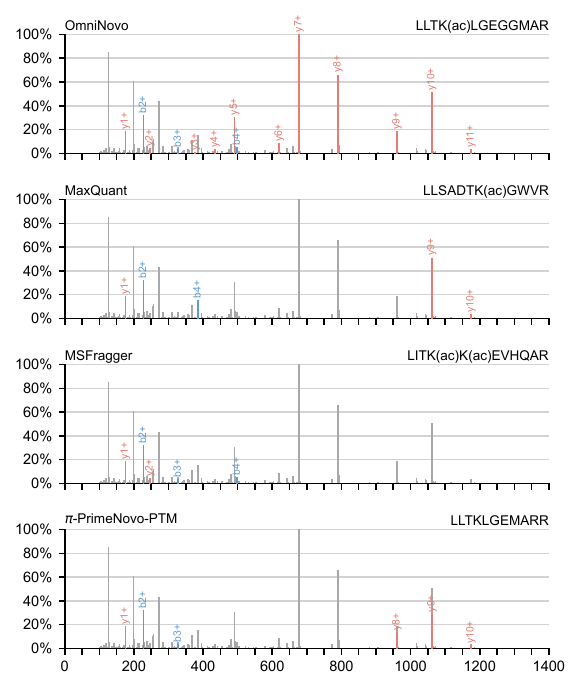}
        \caption{Case 1}
        \label{fig:sup_ac_case_1}
    \end{subfigure}\hfill
    \begin{subfigure}[b]{0.48\linewidth}
        \centering
        \includegraphics[width=\linewidth]{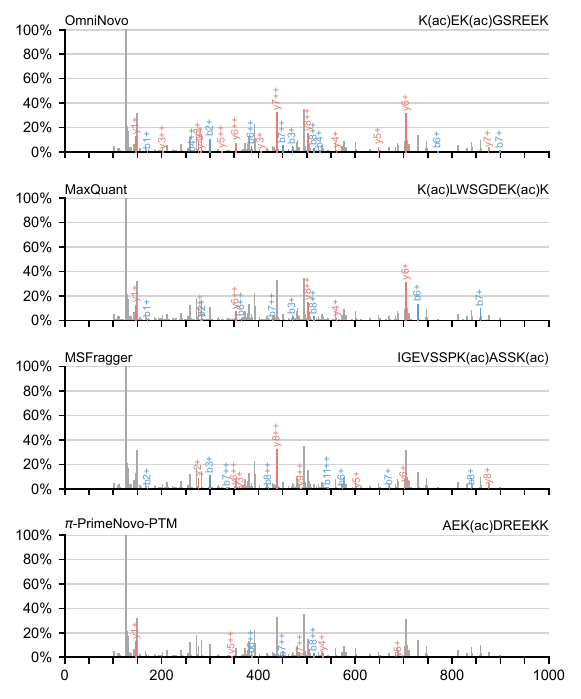}
        \caption{Case 2}
        \label{fig:sup_ac_case_2}
    \end{subfigure}
    \vspace{1em}

    \begin{subfigure}[b]{0.48\linewidth}
        \centering
        \includegraphics[width=\linewidth]{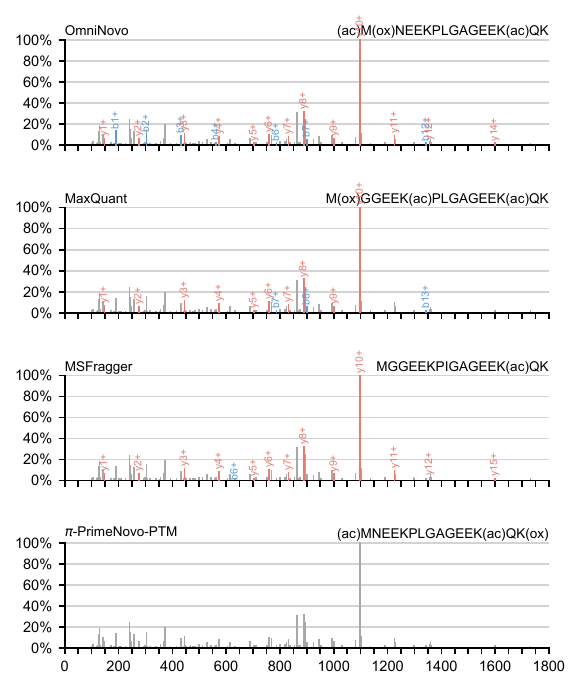}
        \caption{Case 3}
        \label{fig:sup_ac_case_3}
    \end{subfigure}\hfill
    \begin{subfigure}[b]{0.48\linewidth}
        \centering
        \includegraphics[width=\linewidth]{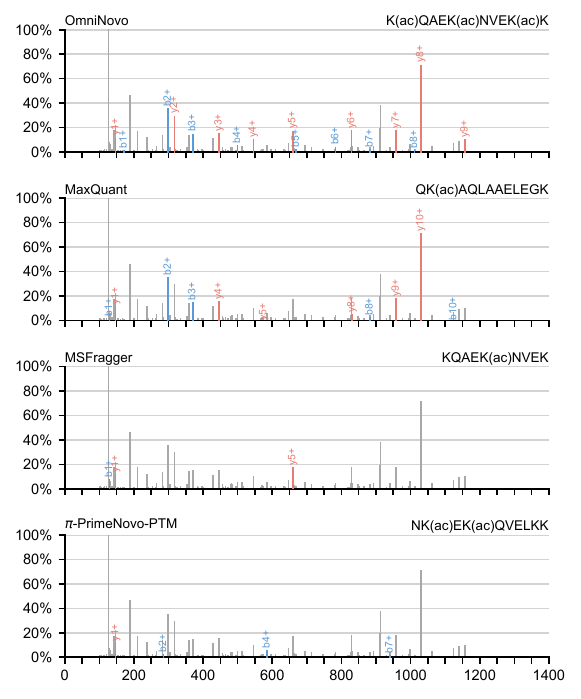}
        \caption{Case 4}
        \label{fig:sup_ac_case_4}
    \end{subfigure}
    \caption{PSM comparison for Acetylation (Cases 1--4).}
\end{figure}
\clearpage
\begin{figure}[htbp]
    \centering
    \begin{subfigure}[b]{0.48\linewidth}
        \centering
        \includegraphics[width=\linewidth]{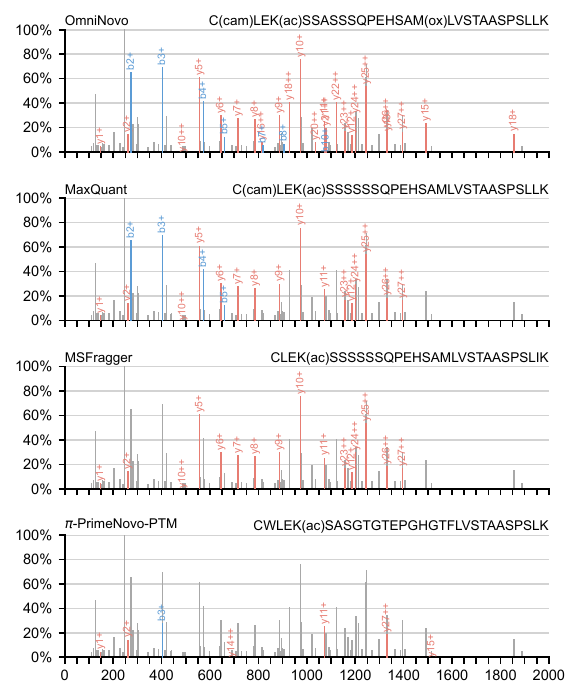}
        \caption{Case 5}
        \label{fig:sup_ac_case_5}
    \end{subfigure}\hfill
    \begin{subfigure}[b]{0.48\linewidth}
        \centering
        \includegraphics[width=\linewidth]{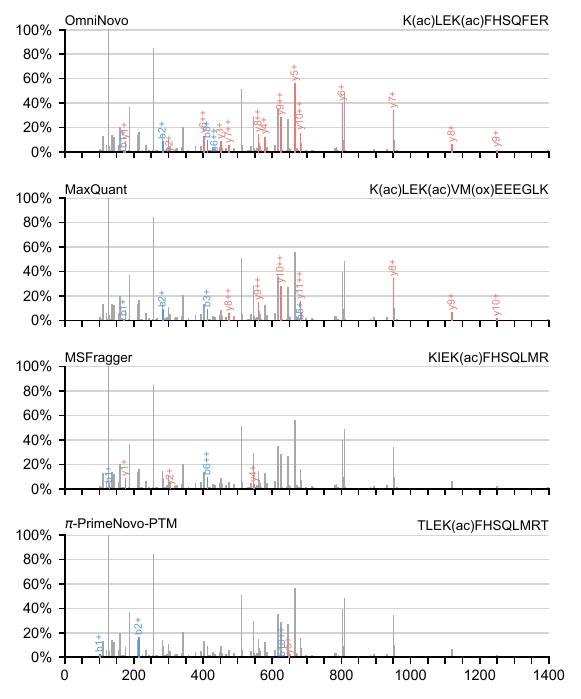}
        \caption{Case 6}
        \label{fig:sup_ac_case_6}
    \end{subfigure}
    \vspace{1em}

    \begin{subfigure}[b]{0.48\linewidth}
        \centering
        \includegraphics[width=\linewidth]{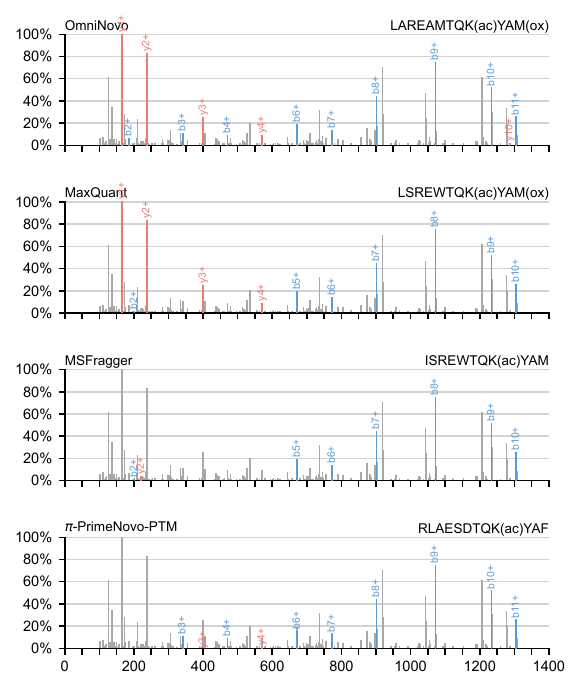}
        \caption{Case 7}
        \label{fig:sup_ac_case_7}
    \end{subfigure}\hfill
    \begin{subfigure}[b]{0.48\linewidth}
        \centering
        \includegraphics[width=\linewidth]{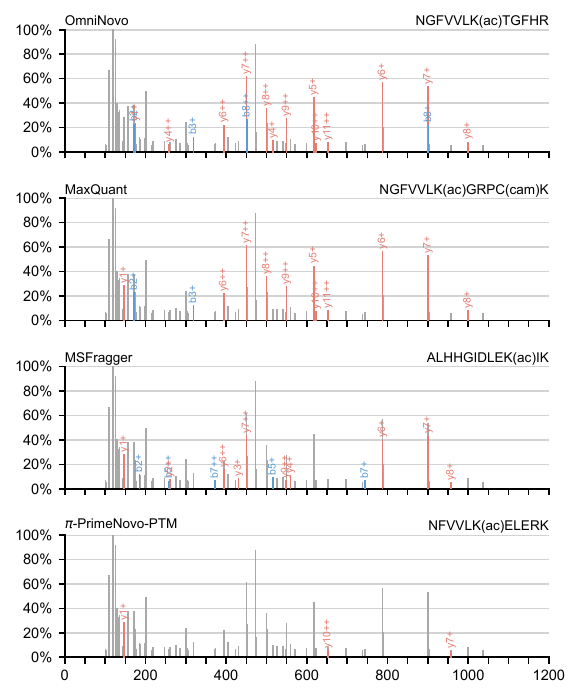}
        \caption{Case 8}
        \label{fig:sup_ac_case_8}
    \end{subfigure}
    \caption{PSM comparison for Acetylation (Cases 5--8).}
\end{figure}
\clearpage
\begin{figure}[htbp]
    \centering
    \begin{subfigure}[b]{0.48\linewidth}
        \centering
        \includegraphics[width=\linewidth]{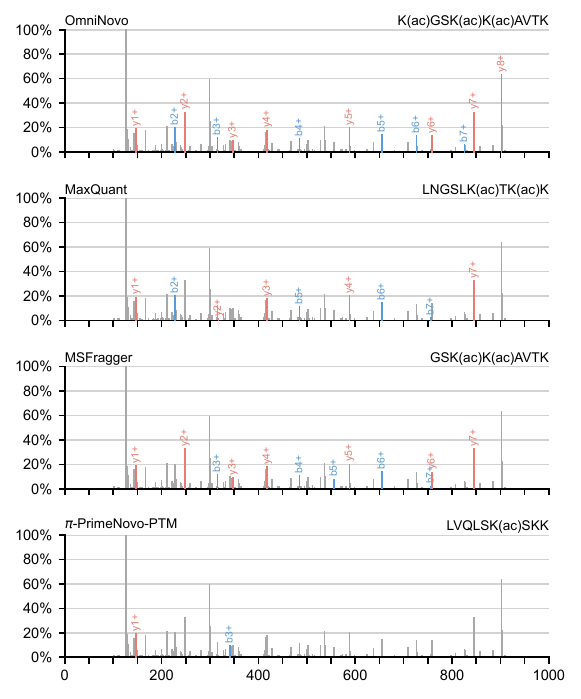}
        \caption{Case 9}
        \label{fig:sup_ac_case_9}
    \end{subfigure}\hfill
    \begin{subfigure}[b]{0.48\linewidth}
        \centering
        \includegraphics[width=\linewidth]{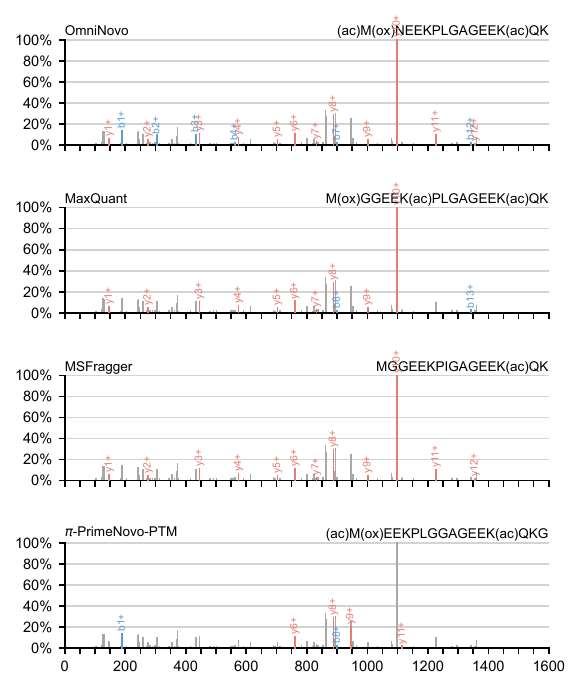}
        \caption{Case 10}
        \label{fig:sup_ac_case_10}
    \end{subfigure}
    \vspace{1em}

    \begin{subfigure}[b]{0.48\linewidth}
        \centering
        \includegraphics[width=\linewidth]{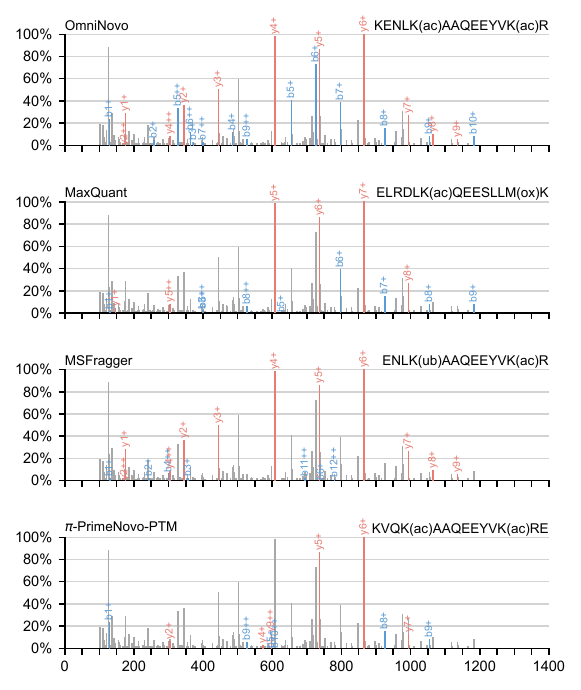}
        \caption{Case 11}
        \label{fig:sup_ac_case_11}
    \end{subfigure}\hfill
    \begin{subfigure}[b]{0.48\linewidth}
        \centering
        \includegraphics[width=\linewidth]{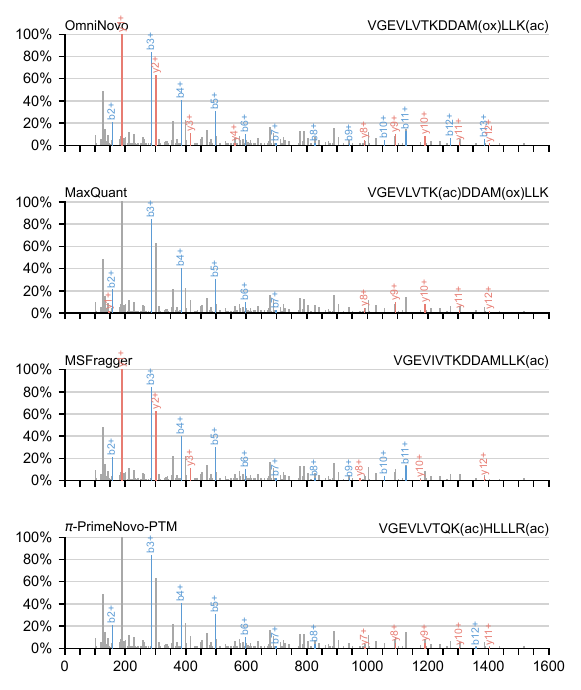}
        \caption{Case 12}
        \label{fig:sup_ac_case_12}
    \end{subfigure}
    \caption{PSM comparison for Acetylation (Cases 9--12).}
\end{figure}
\clearpage
\begin{figure}[htbp]
    \centering
    \begin{subfigure}[b]{0.48\linewidth}
        \centering
        \includegraphics[width=\linewidth]{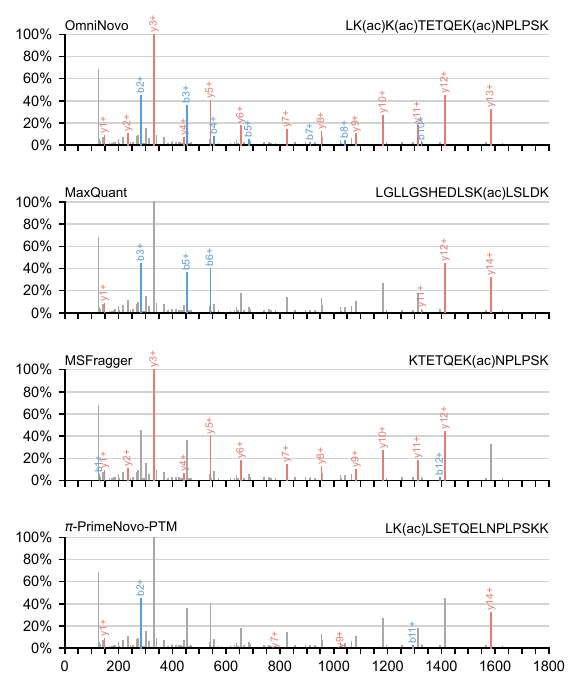}
        \caption{Case 13}
        \label{fig:sup_ac_case_13}
    \end{subfigure}\hfill
    \begin{subfigure}[b]{0.48\linewidth}
        \centering
        \includegraphics[width=\linewidth]{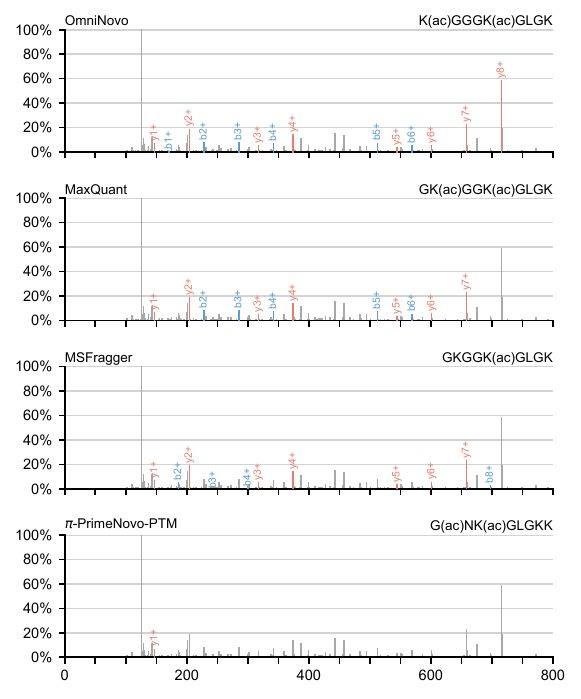}
        \caption{Case 14}
        \label{fig:sup_ac_case_14}
    \end{subfigure}
    \vspace{1em}

    \begin{subfigure}[b]{0.48\linewidth}
        \centering
        \includegraphics[width=\linewidth]{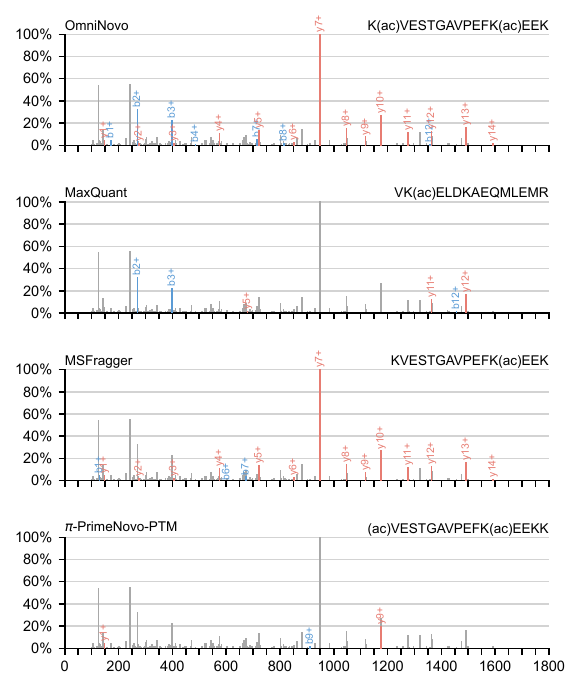}
        \caption{Case 15}
        \label{fig:sup_ac_case_15}
    \end{subfigure}\hfill
    \begin{subfigure}[b]{0.48\linewidth}
        \centering
        \includegraphics[width=\linewidth]{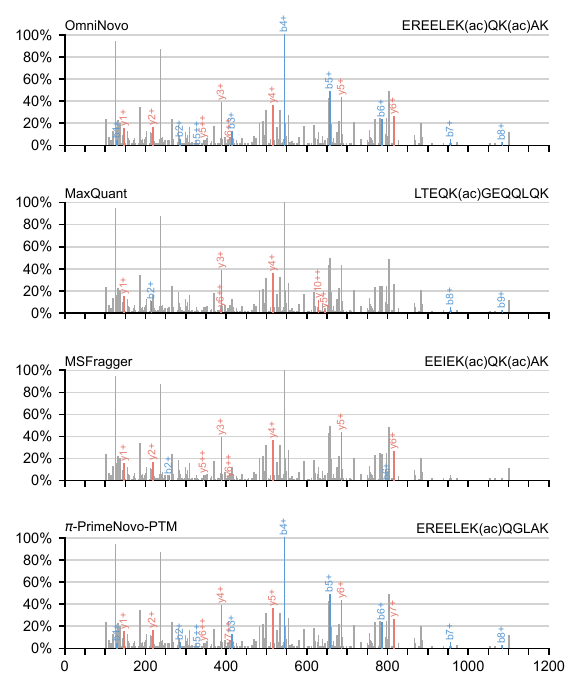}
        \caption{Case 16}
        \label{fig:sup_ac_case_16}
    \end{subfigure}
    \caption{PSM comparison for Acetylation (Cases 13--16).}
\end{figure}
\clearpage

\clearpage
\begin{figure}[htbp]
    \centering
    \begin{subfigure}[b]{0.48\linewidth}
        \centering
        \includegraphics[width=\linewidth]{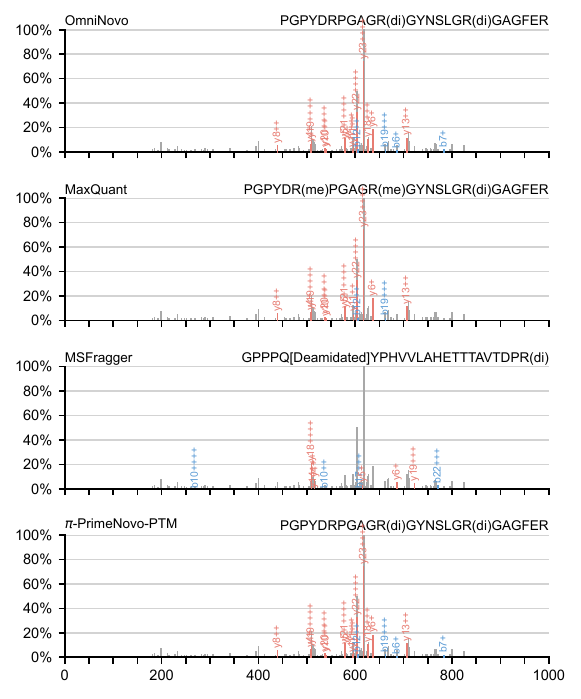}
        \caption{Case 1}
        \label{fig:sup_di_case_1}
    \end{subfigure}\hfill
    \begin{subfigure}[b]{0.48\linewidth}
        \centering
        \includegraphics[width=\linewidth]{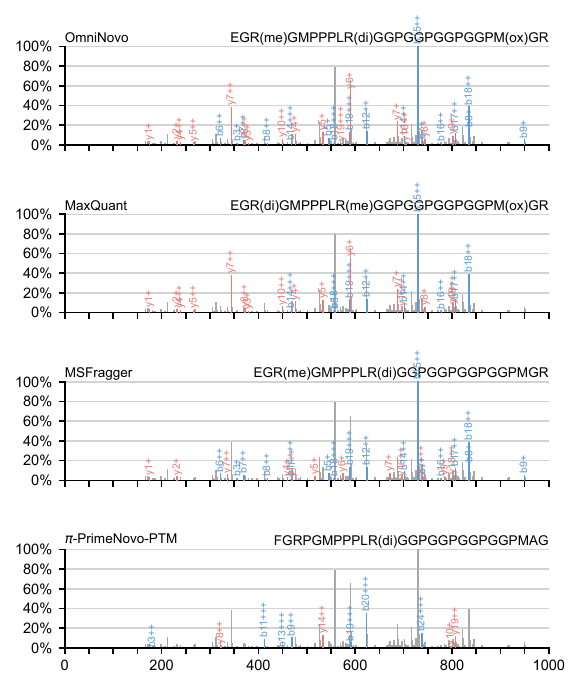}
        \caption{Case 2}
        \label{fig:sup_di_case_2}
    \end{subfigure}
    \vspace{1em}

    \begin{subfigure}[b]{0.48\linewidth}
        \centering
        \includegraphics[width=\linewidth]{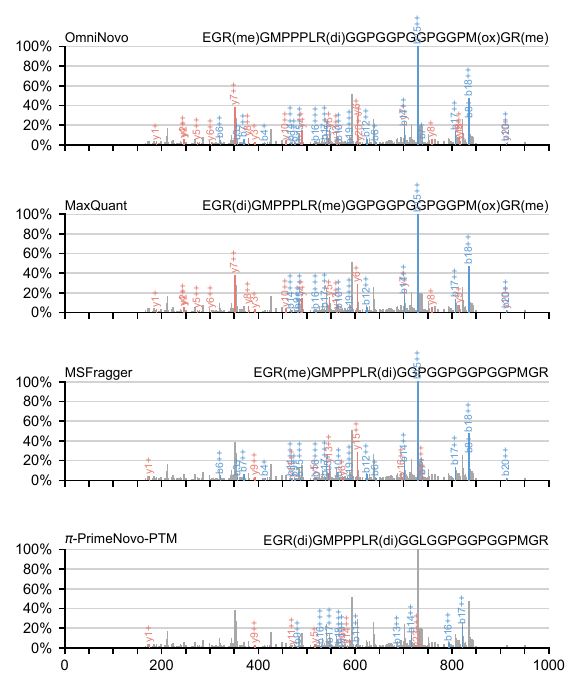}
        \caption{Case 3}
        \label{fig:sup_di_case_3}
    \end{subfigure}\hfill
    \begin{subfigure}[b]{0.48\linewidth}
        \centering
        \includegraphics[width=\linewidth]{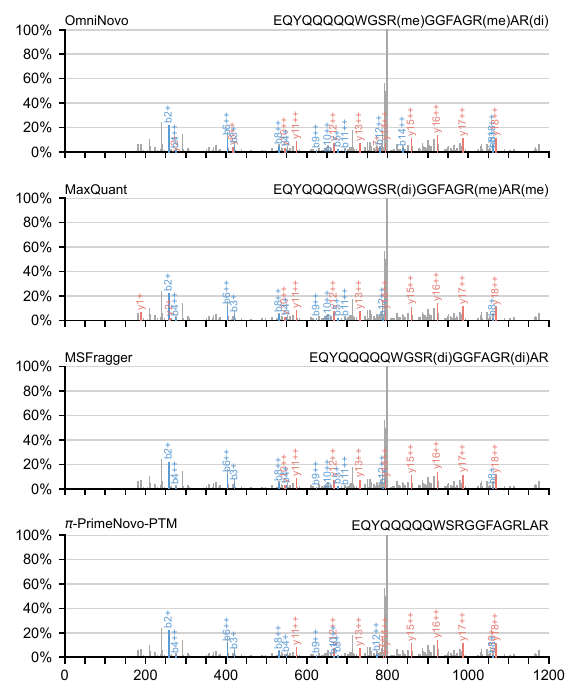}
        \caption{Case 4}
        \label{fig:sup_di_case_4}
    \end{subfigure}
    \caption{PSM comparison for Dimethylation (Cases 1--4).}
\end{figure}
\clearpage
\begin{figure}[htbp]
    \centering
    \begin{subfigure}[b]{0.48\linewidth}
        \centering
        \includegraphics[width=\linewidth]{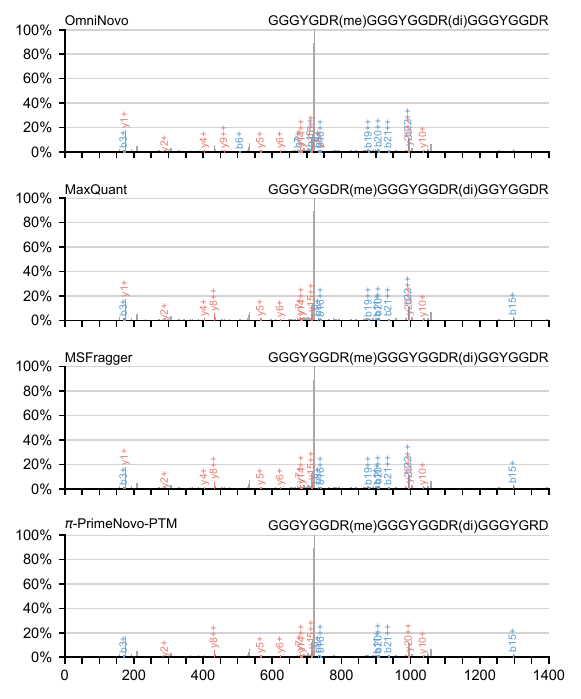}
        \caption{Case 5}
        \label{fig:sup_di_case_5}
    \end{subfigure}\hfill
    \begin{subfigure}[b]{0.48\linewidth}
        \centering
        \includegraphics[width=\linewidth]{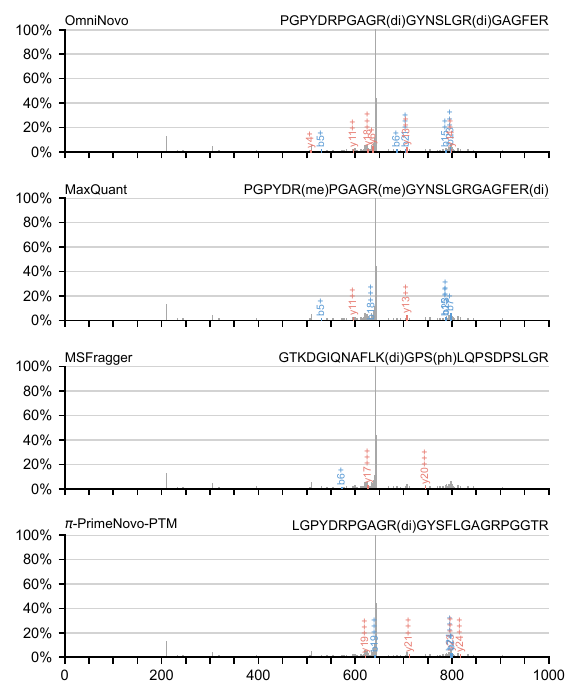}
        \caption{Case 6}
        \label{fig:sup_di_case_6}
    \end{subfigure}
    \vspace{1em}

    \begin{subfigure}[b]{0.48\linewidth}
        \centering
        \includegraphics[width=\linewidth]{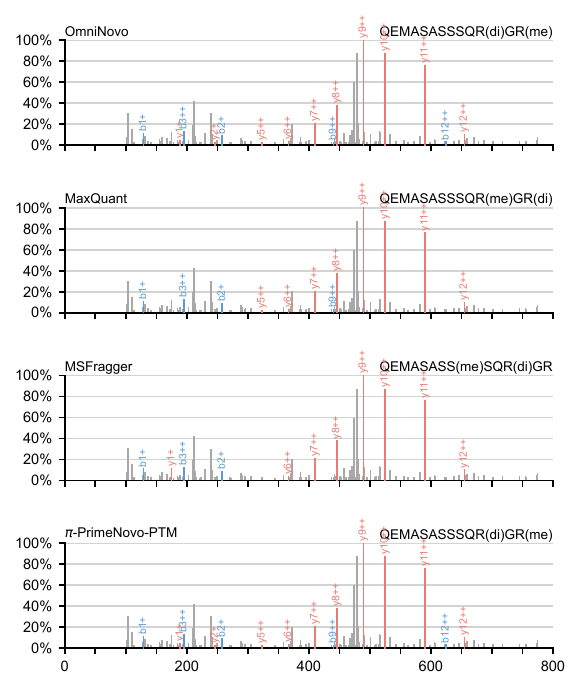}
        \caption{Case 7}
        \label{fig:sup_di_case_7}
    \end{subfigure}\hfill
    \begin{subfigure}[b]{0.48\linewidth}
        \centering
        \includegraphics[width=\linewidth]{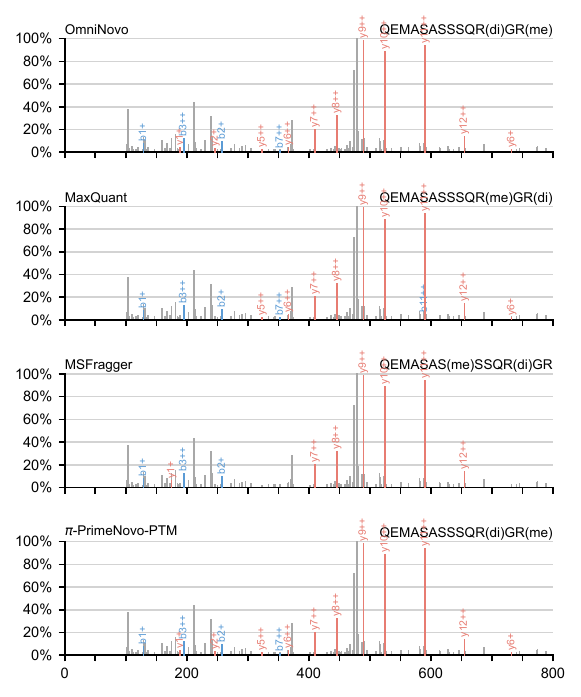}
        \caption{Case 8}
        \label{fig:sup_di_case_8}
    \end{subfigure}
    \caption{PSM comparison for Dimethylation (Cases 5--8).}
\end{figure}
\clearpage
\begin{figure}[htbp]
    \centering
    \begin{subfigure}[b]{0.48\linewidth}
        \centering
        \includegraphics[width=\linewidth]{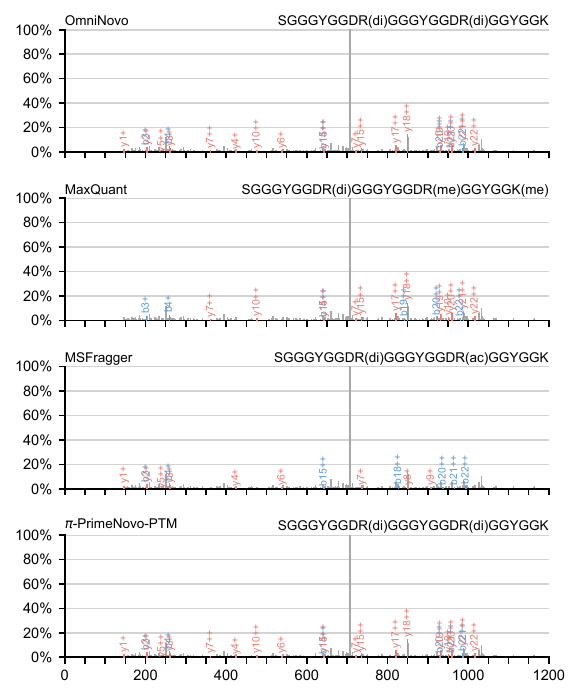}
        \caption{Case 9}
        \label{fig:sup_di_case_9}
    \end{subfigure}\hfill
    \begin{subfigure}[b]{0.48\linewidth}
        \centering
        \includegraphics[width=\linewidth]{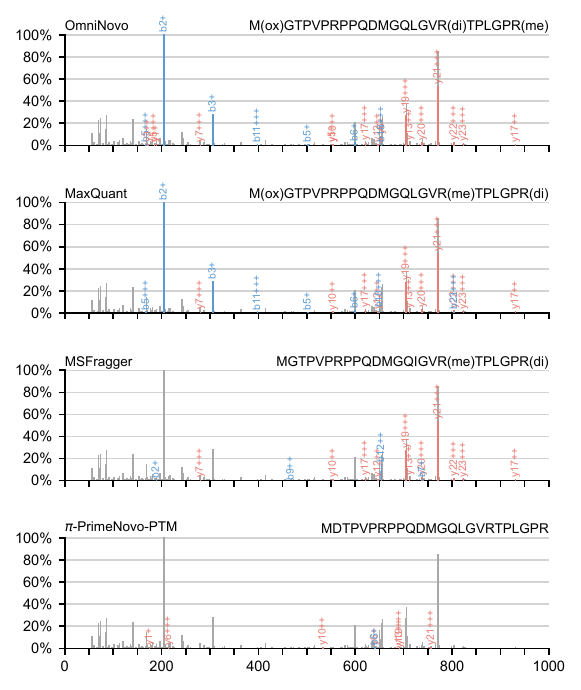}
        \caption{Case 10}
        \label{fig:sup_di_case_10}
    \end{subfigure}
    \vspace{1em}

    \begin{subfigure}[b]{0.48\linewidth}
        \centering
        \includegraphics[width=\linewidth]{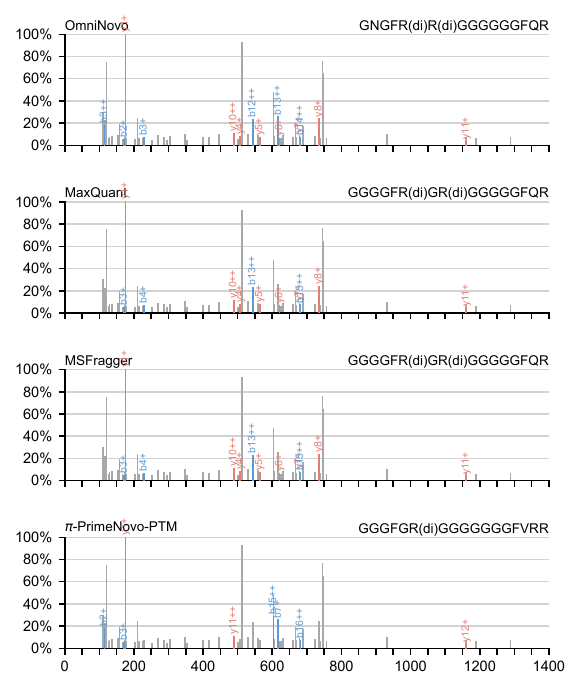}
        \caption{Case 11}
        \label{fig:sup_di_case_11}
    \end{subfigure}\hfill
    \begin{subfigure}[b]{0.48\linewidth}
        \centering
        \includegraphics[width=\linewidth]{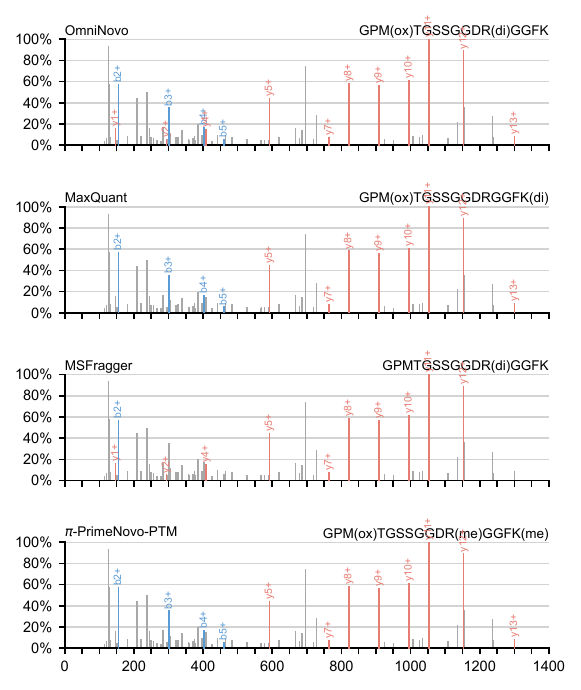}
        \caption{Case 12}
        \label{fig:sup_di_case_12}
    \end{subfigure}
    \caption{PSM comparison for Dimethylation (Cases 9--12).}
\end{figure}
\clearpage
\begin{figure}[htbp]
    \centering
    \begin{subfigure}[b]{0.48\linewidth}
        \centering
        \includegraphics[width=\linewidth]{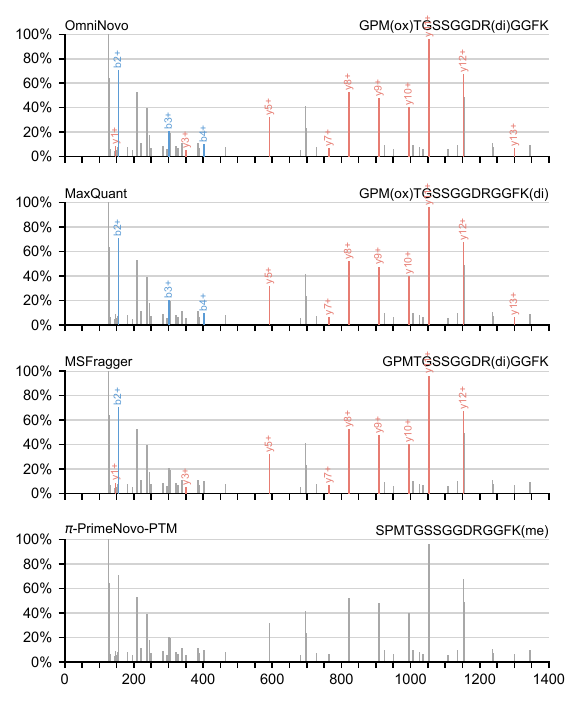}
        \caption{Case 13}
        \label{fig:sup_di_case_13}
    \end{subfigure}\hfill
    \begin{subfigure}[b]{0.48\linewidth}
        \centering
        \includegraphics[width=\linewidth]{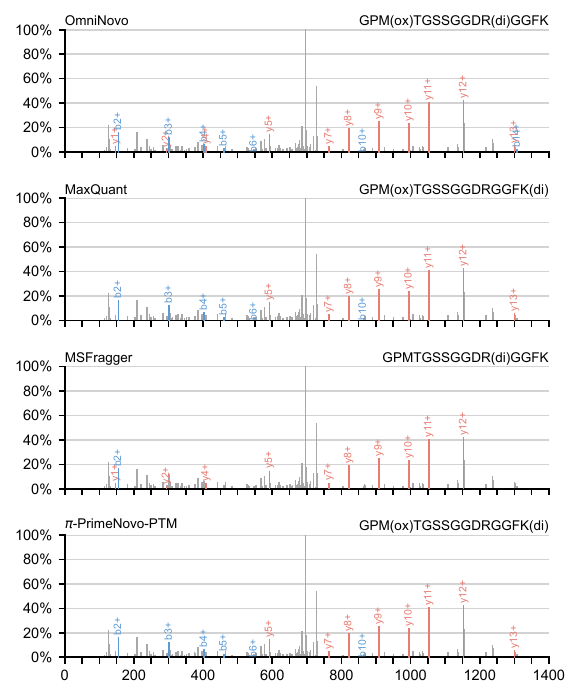}
        \caption{Case 14}
        \label{fig:sup_di_case_14}
    \end{subfigure}
    \vspace{1em}

    \begin{subfigure}[b]{0.48\linewidth}
        \centering
        \includegraphics[width=\linewidth]{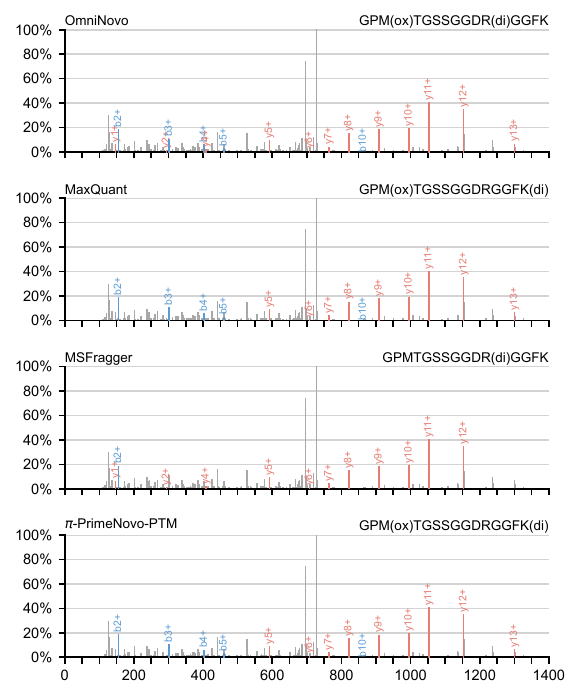}
        \caption{Case 15}
        \label{fig:sup_di_case_15}
    \end{subfigure}\hfill
    \begin{subfigure}[b]{0.48\linewidth}
        \centering
        \includegraphics[width=\linewidth]{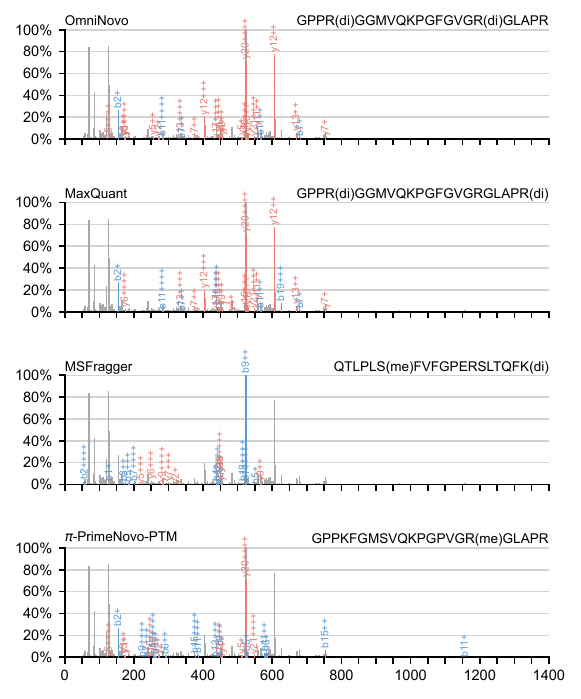}
        \caption{Case 16}
        \label{fig:sup_di_case_16}
    \end{subfigure}
    \caption{PSM comparison for Dimethylation (Cases 13--16).}
\end{figure}
\clearpage
\begin{figure}[htbp]
    \centering
    \begin{subfigure}[b]{0.48\linewidth}
        \centering
        \includegraphics[width=\linewidth]{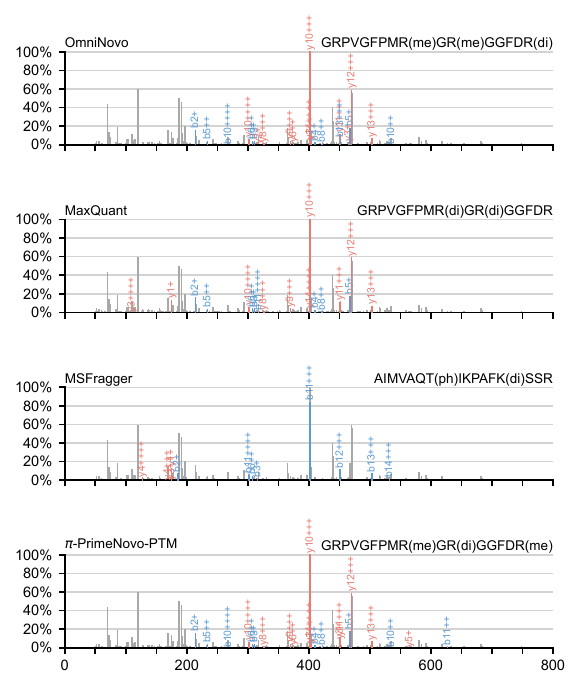}
        \caption{Case 17}
        \label{fig:sup_di_case_17}
    \end{subfigure}\hfill
    \begin{subfigure}[b]{0.48\linewidth}
        \centering
        \includegraphics[width=\linewidth]{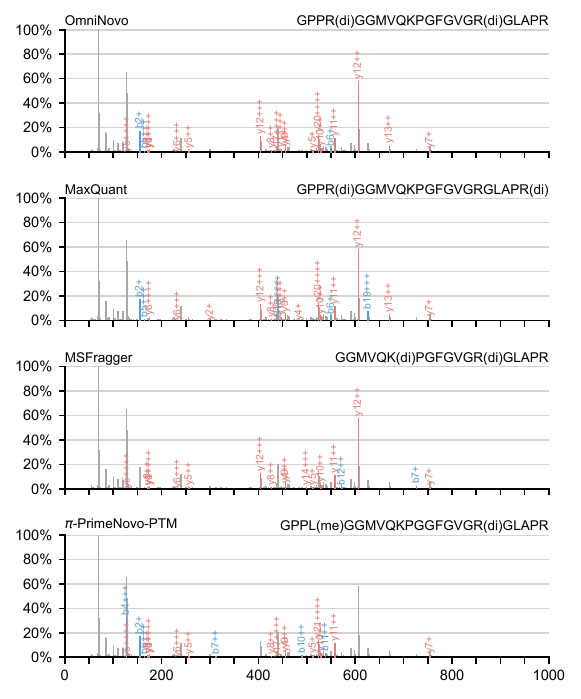}
        \caption{Case 18}
        \label{fig:sup_di_case_18}
    \end{subfigure}
    \vspace{1em}

    \begin{subfigure}[b]{0.48\linewidth}
        \centering
        \includegraphics[width=\linewidth]{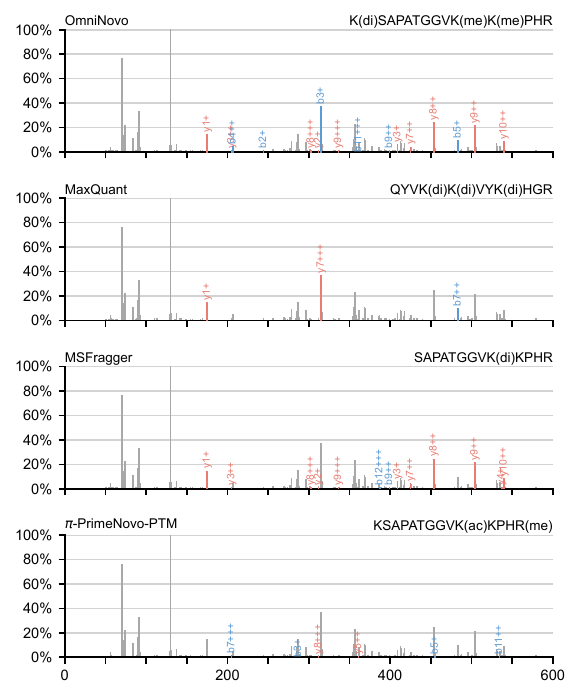}
        \caption{Case 19}
        \label{fig:sup_di_case_19}
    \end{subfigure}\hfill
    \begin{subfigure}[b]{0.48\linewidth}
        \centering
        \includegraphics[width=\linewidth]{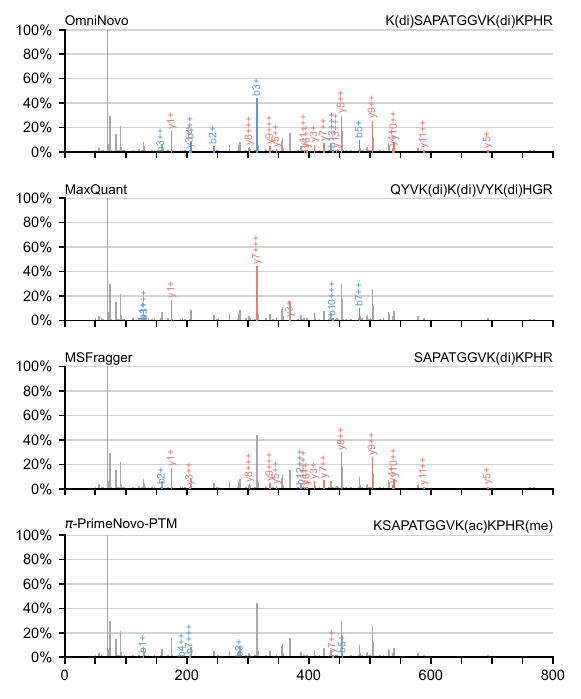}
        \caption{Case 20}
        \label{fig:sup_di_case_20}
    \end{subfigure}
    \caption{PSM comparison for Dimethylation (Cases 17--20).}
\end{figure}
\clearpage
\begin{figure}[htbp]
    \centering
    \begin{subfigure}[b]{0.48\linewidth}
        \centering
        \includegraphics[width=\linewidth]{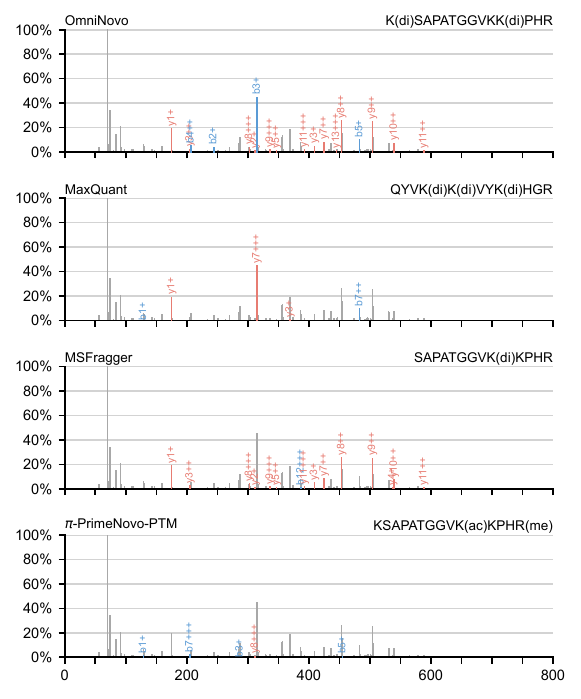}
        \caption{Case 21}
        \label{fig:sup_di_case_21}
    \end{subfigure}\hfill
    \begin{subfigure}[b]{0.48\linewidth}
        \centering
        \includegraphics[width=\linewidth]{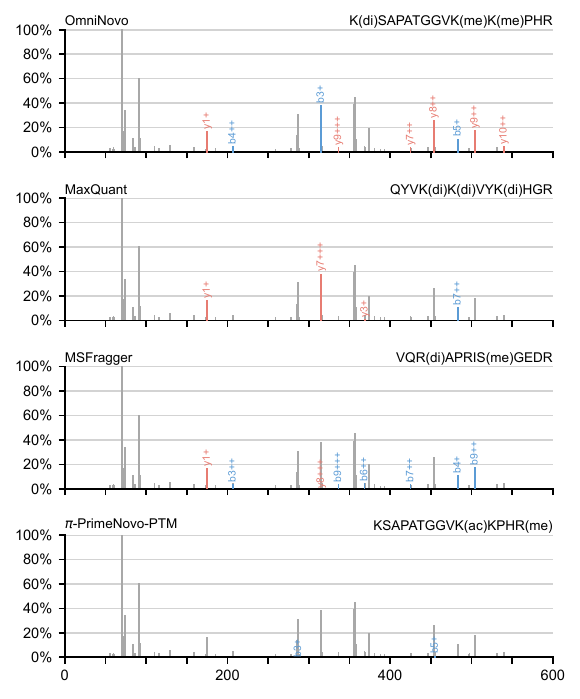}
        \caption{Case 22}
        \label{fig:sup_di_case_22}
    \end{subfigure}
    \vspace{1em}

    \begin{subfigure}[b]{0.48\linewidth}
        \centering
        \includegraphics[width=\linewidth]{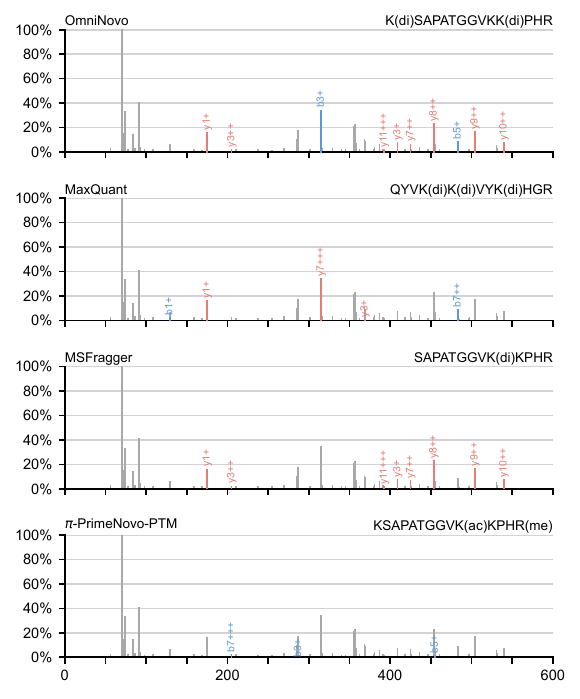}
        \caption{Case 23}
        \label{fig:sup_di_case_23}
    \end{subfigure}\hfill
    \begin{subfigure}[b]{0.48\linewidth}
        \centering
        \includegraphics[width=\linewidth]{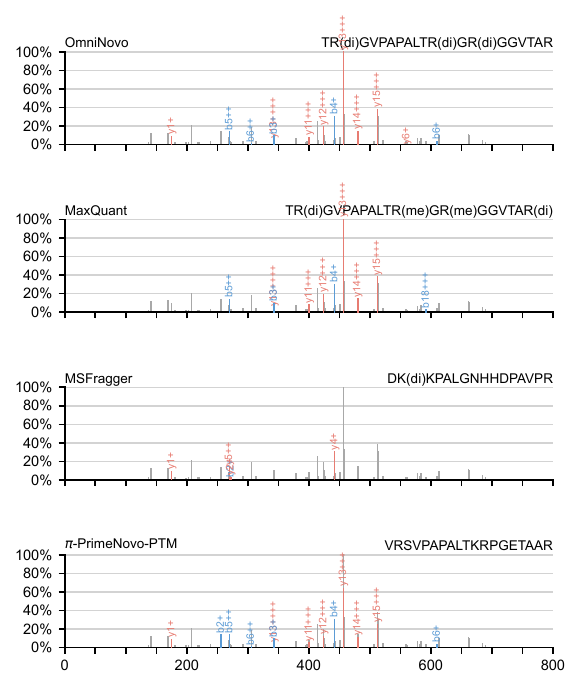}
        \caption{Case 24}
        \label{fig:sup_di_case_24}
    \end{subfigure}
    \caption{PSM comparison for Dimethylation (Cases 21--24).}
\end{figure}
\clearpage
\begin{figure}[htbp]
    \centering
    \begin{subfigure}[b]{0.48\linewidth}
        \centering
        \includegraphics[width=\linewidth]{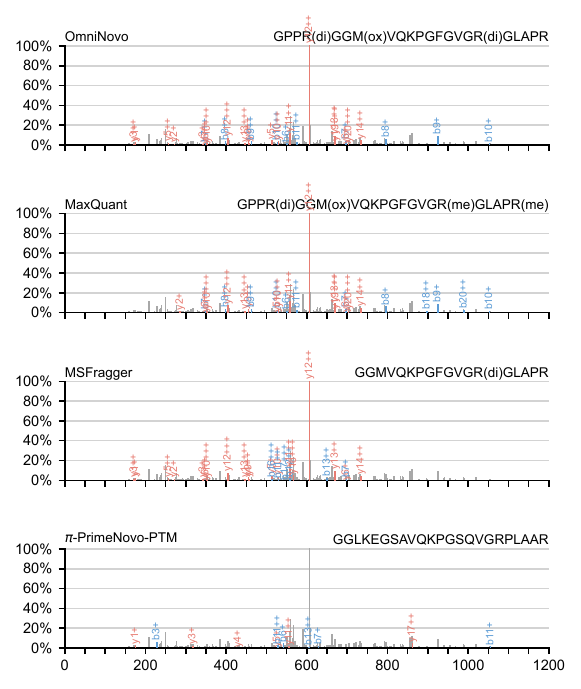}
        \caption{Case 25}
        \label{fig:sup_di_case_25}
    \end{subfigure}\hfill
    \begin{subfigure}[b]{0.48\linewidth}
        \centering
        \includegraphics[width=\linewidth]{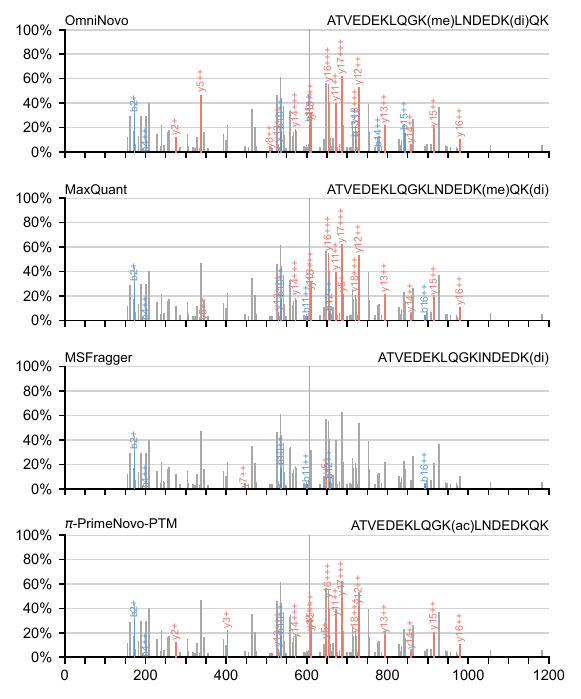}
        \caption{Case 26}
        \label{fig:sup_di_case_26}
    \end{subfigure}
    \vspace{1em}

    \begin{subfigure}[b]{0.48\linewidth}
        \centering
        \includegraphics[width=\linewidth]{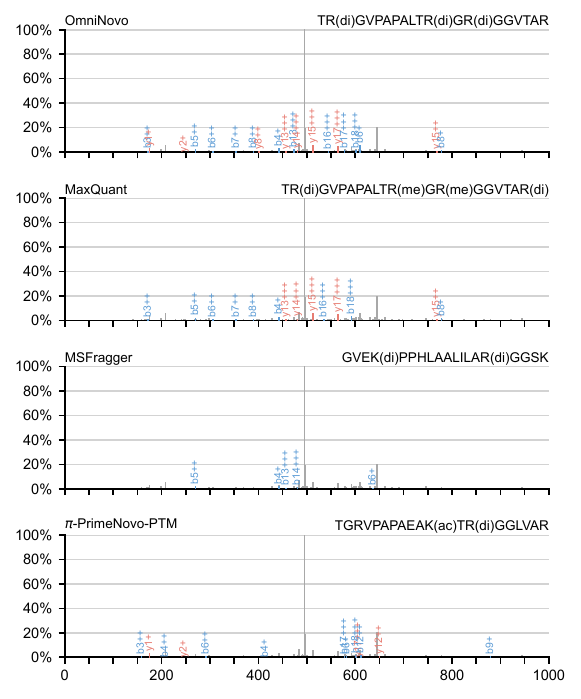}
        \caption{Case 27}
        \label{fig:sup_di_case_27}
    \end{subfigure}\hfill
    \begin{subfigure}[b]{0.48\linewidth}
        \centering
        \includegraphics[width=\linewidth]{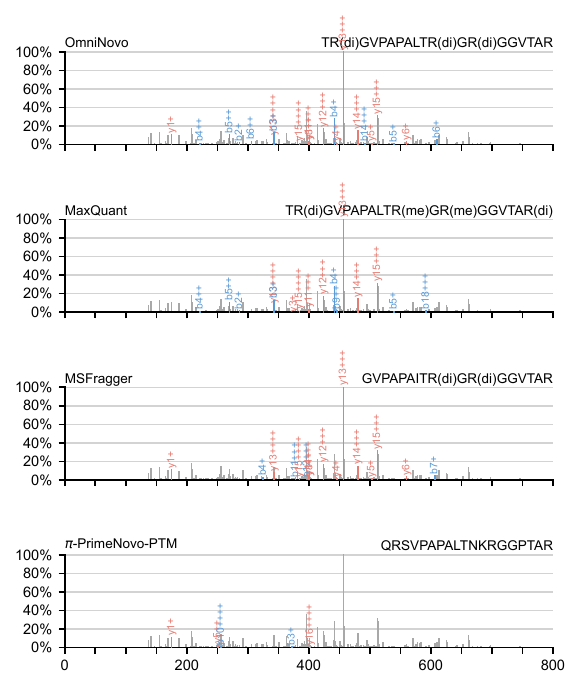}
        \caption{Case 28}
        \label{fig:sup_di_case_28}
    \end{subfigure}
    \caption{PSM comparison for Dimethylation (Cases 25--28).}
\end{figure}
\clearpage
\begin{figure}[htbp]
    \centering
    \begin{subfigure}[b]{0.48\linewidth}
        \centering
        \includegraphics[width=\linewidth]{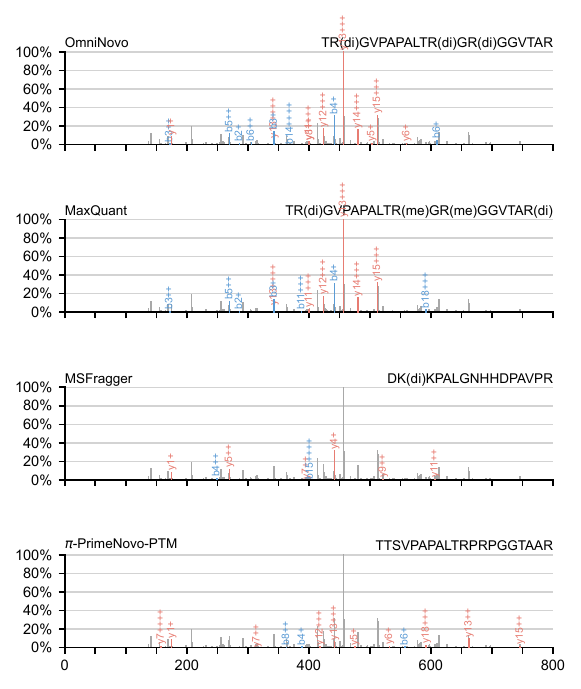}
        \caption{Case 29}
        \label{fig:sup_di_case_29}
    \end{subfigure}\hfill
    \begin{subfigure}[b]{0.48\linewidth}
        \centering
        \includegraphics[width=\linewidth]{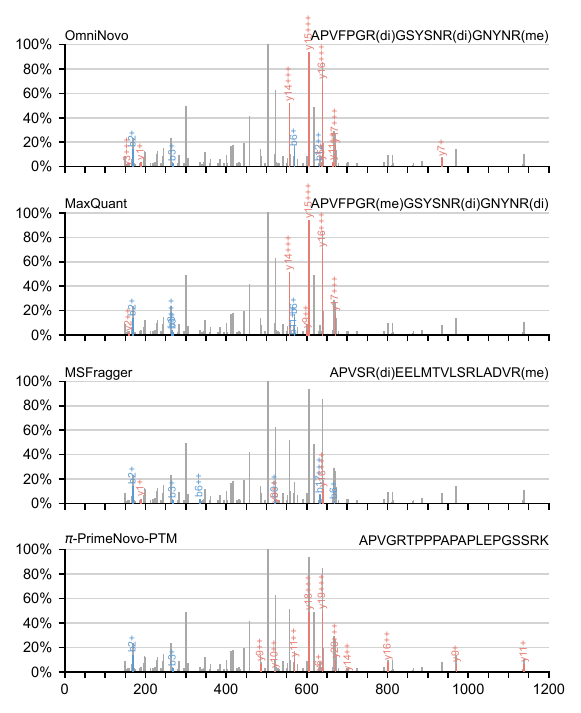}
        \caption{Case 30}
        \label{fig:sup_di_case_30}
    \end{subfigure}
    \vspace{1em}

    \begin{subfigure}[b]{0.48\linewidth}
        \centering
        \includegraphics[width=\linewidth]{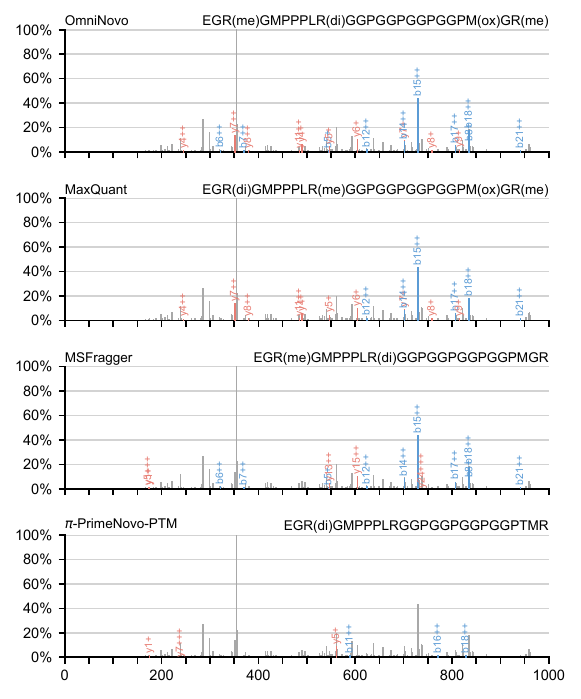}
        \caption{Case 31}
        \label{fig:sup_di_case_31}
    \end{subfigure}\hfill
    \begin{subfigure}[b]{0.48\linewidth}
        \centering
        \includegraphics[width=\linewidth]{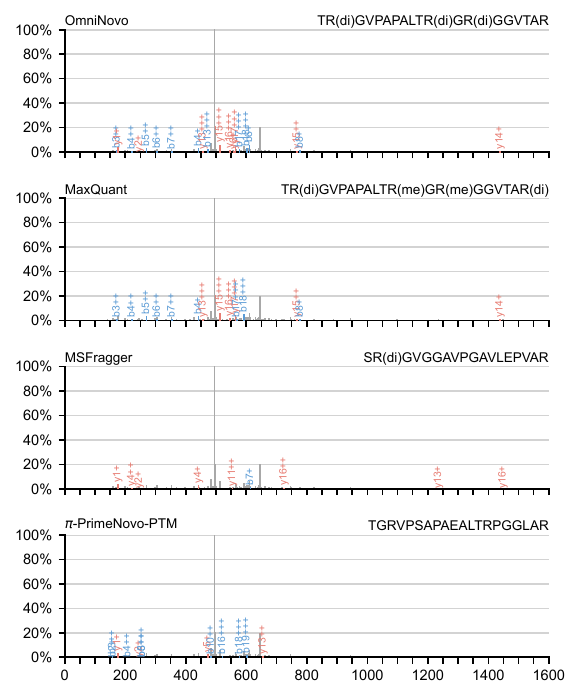}
        \caption{Case 32}
        \label{fig:sup_di_case_32}
    \end{subfigure}
    \caption{PSM comparison for Dimethylation (Cases 29--32).}
\end{figure}
\clearpage
\begin{figure}[htbp]
    \centering
    \begin{subfigure}[b]{0.48\linewidth}
        \centering
        \includegraphics[width=\linewidth]{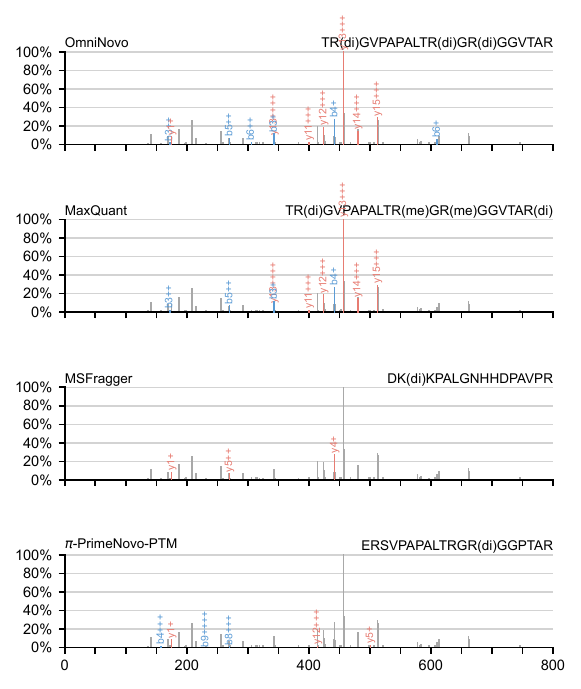}
        \caption{Case 33}
        \label{fig:sup_di_case_33}
    \end{subfigure}\hfill
    \begin{subfigure}[b]{0.48\linewidth}
        \centering
        \includegraphics[width=\linewidth]{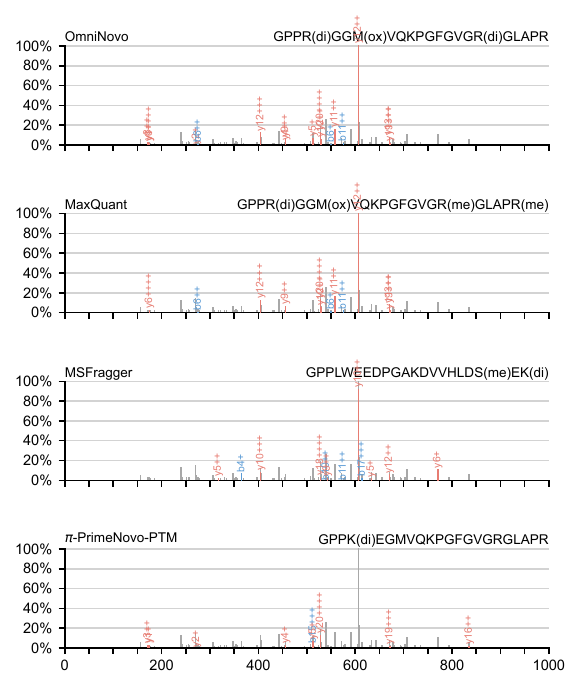}
        \caption{Case 34}
        \label{fig:sup_di_case_34}
    \end{subfigure}
    \vspace{1em}

    \begin{subfigure}[b]{0.48\linewidth}
        \centering
        \includegraphics[width=\linewidth]{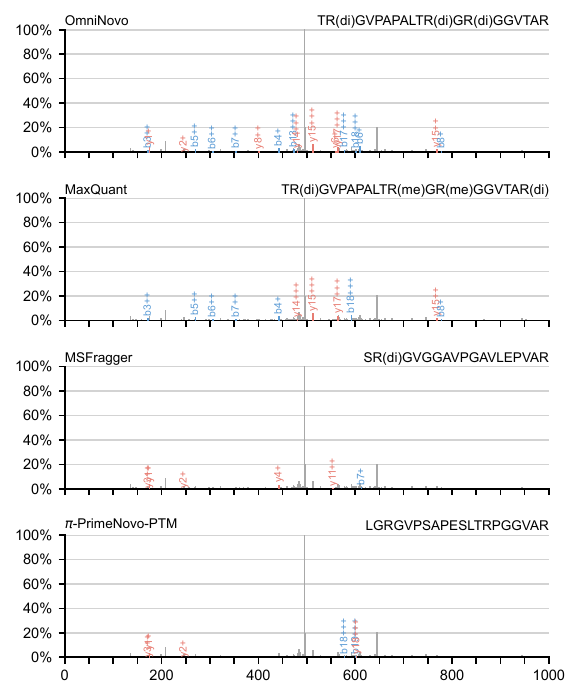}
        \caption{Case 35}
        \label{fig:sup_di_case_35}
    \end{subfigure}\hfill
    \begin{subfigure}[b]{0.48\linewidth}
        \centering
        \includegraphics[width=\linewidth]{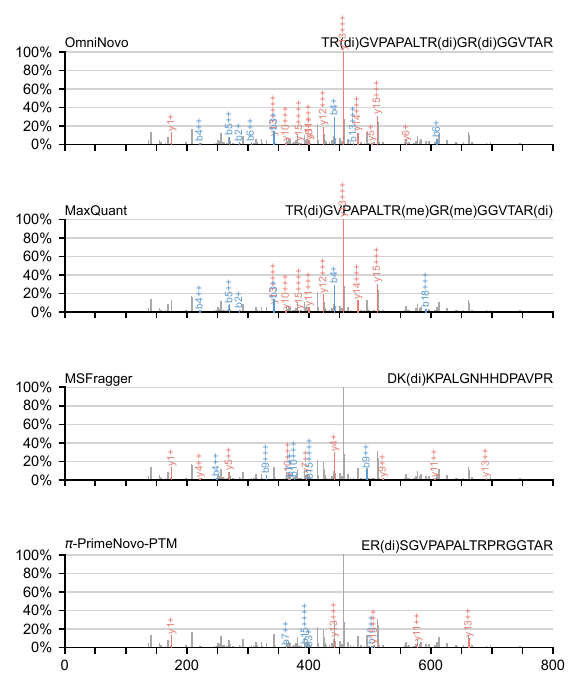}
        \caption{Case 36}
        \label{fig:sup_di_case_36}
    \end{subfigure}
    \caption{PSM comparison for Dimethylation (Cases 33--36).}
\end{figure}
\clearpage
\begin{figure}[htbp]
    \centering
    \begin{subfigure}[b]{0.48\linewidth}
        \centering
        \includegraphics[width=\linewidth]{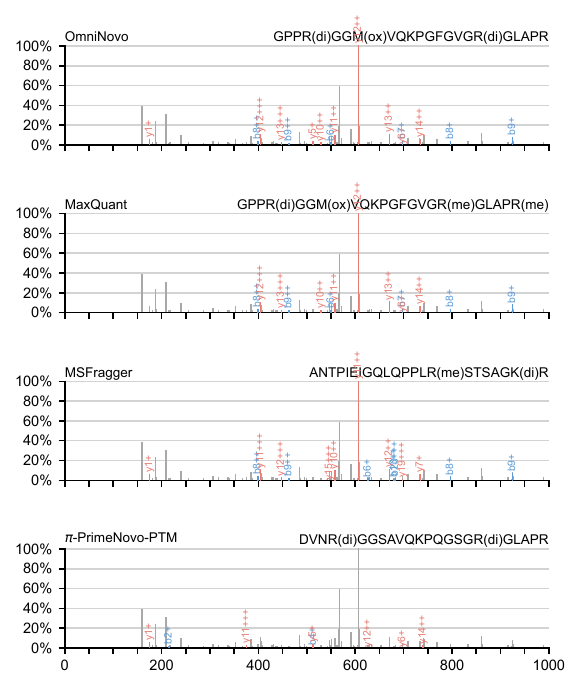}
        \caption{Case 37}
        \label{fig:sup_di_case_37}
    \end{subfigure}\hfill
    \begin{subfigure}[b]{0.48\linewidth}
        \centering
        \includegraphics[width=\linewidth]{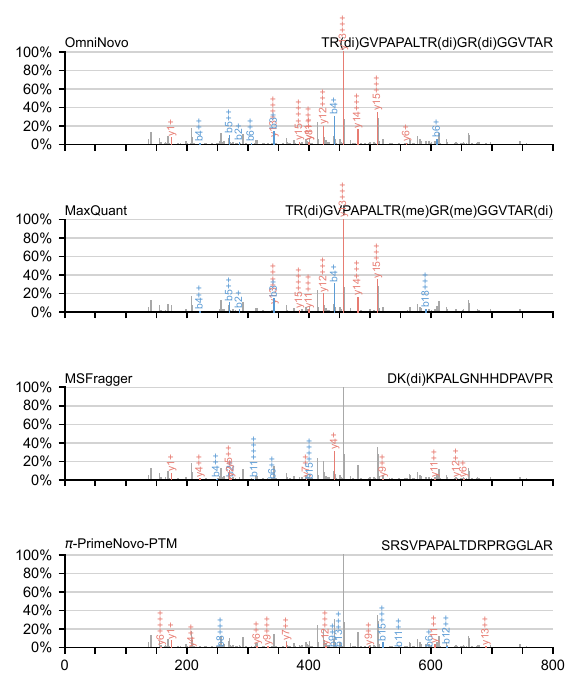}
        \caption{Case 38}
        \label{fig:sup_di_case_38}
    \end{subfigure}
    \vspace{1em}

    \begin{subfigure}[b]{0.48\linewidth}
        \centering
        \includegraphics[width=\linewidth]{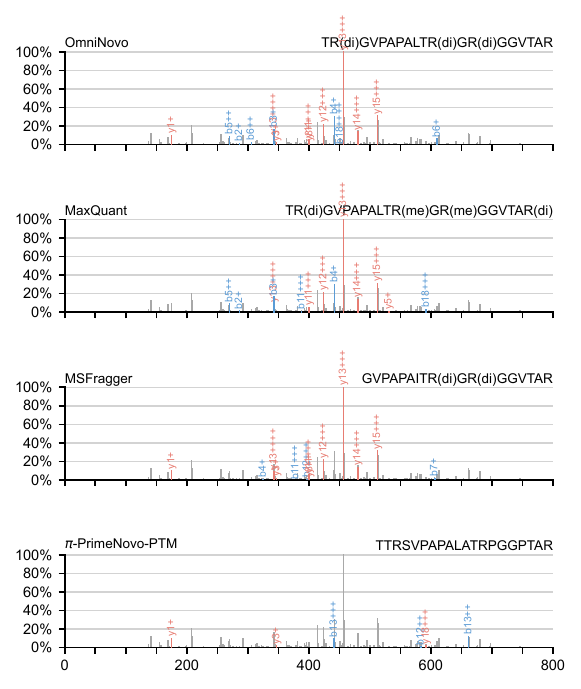}
        \caption{Case 39}
        \label{fig:sup_di_case_39}
    \end{subfigure}\hfill
    \begin{subfigure}[b]{0.48\linewidth}
        \centering
        \includegraphics[width=\linewidth]{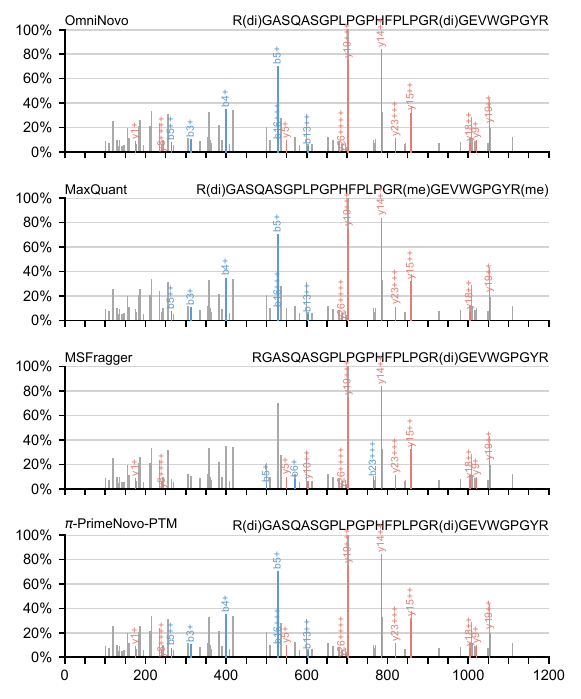}
        \caption{Case 40}
        \label{fig:sup_di_case_40}
    \end{subfigure}
    \caption{PSM comparison for Dimethylation (Cases 37--40).}
\end{figure}
\clearpage
\begin{figure}[htbp]
    \centering
    \begin{subfigure}[b]{0.48\linewidth}
        \centering
        \includegraphics[width=\linewidth]{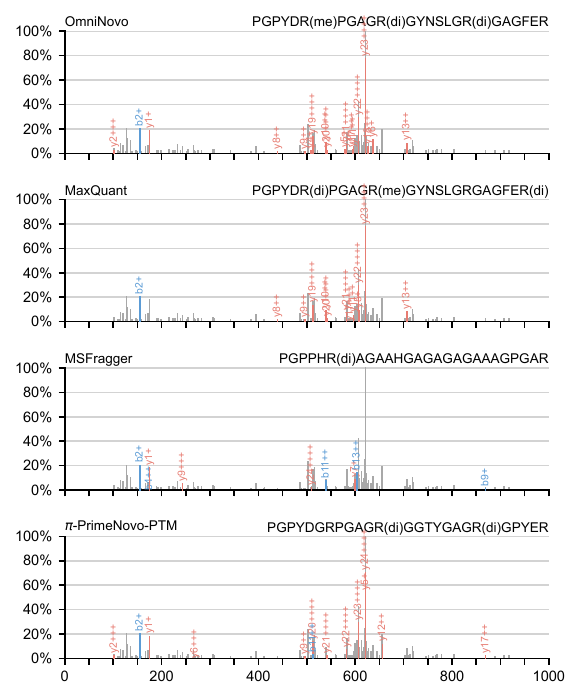}
        \caption{Case 41}
        \label{fig:sup_di_case_41}
    \end{subfigure}\hfill
    \begin{subfigure}[b]{0.48\linewidth}
        \centering
        \includegraphics[width=\linewidth]{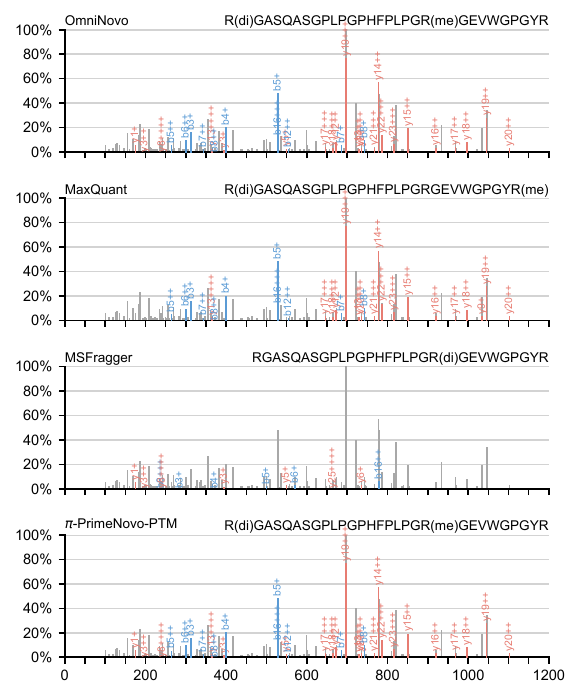}
        \caption{Case 42}
        \label{fig:sup_di_case_42}
    \end{subfigure}
    \vspace{1em}

    \begin{subfigure}[b]{0.48\linewidth}
        \centering
        \includegraphics[width=\linewidth]{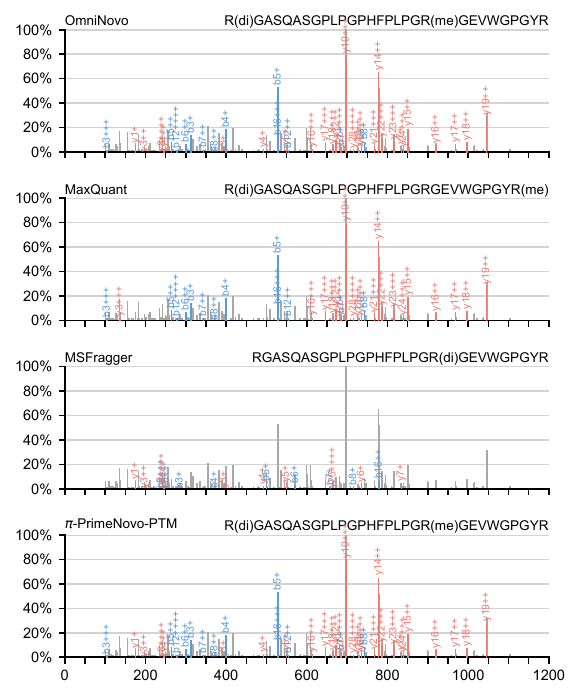}
        \caption{Case 43}
        \label{fig:sup_di_case_43}
    \end{subfigure}\hfill
    \begin{subfigure}[b]{0.48\linewidth}
        \centering
        \includegraphics[width=\linewidth]{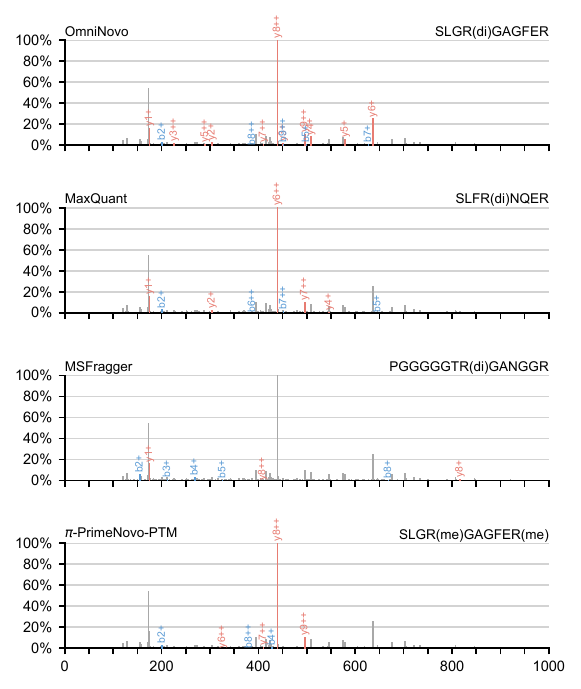}
        \caption{Case 44}
        \label{fig:sup_di_case_44}
    \end{subfigure}
    \caption{PSM comparison for Dimethylation (Cases 41--44).}
\end{figure}
\clearpage
\begin{figure}[htbp]
    \centering
    \begin{subfigure}[b]{0.48\linewidth}
        \centering
        \includegraphics[width=\linewidth]{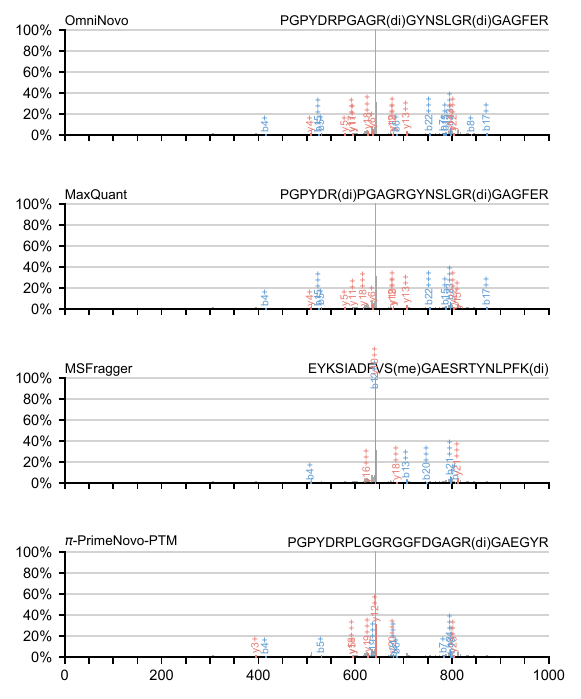}
        \caption{Case 45}
        \label{fig:sup_di_case_45}
    \end{subfigure}\hfill
    \begin{subfigure}[b]{0.48\linewidth}
        \centering
        \includegraphics[width=\linewidth]{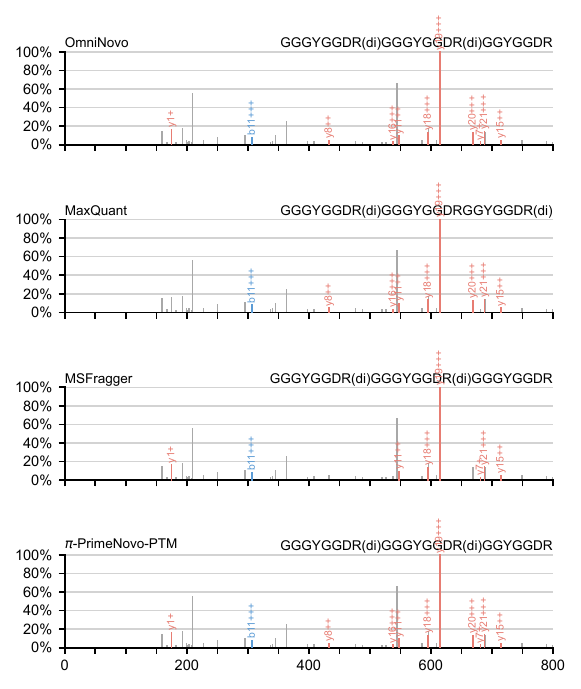}
        \caption{Case 46}
        \label{fig:sup_di_case_46}
    \end{subfigure}
    \vspace{1em}

    \begin{subfigure}[b]{0.48\linewidth}
        \centering
        \includegraphics[width=\linewidth]{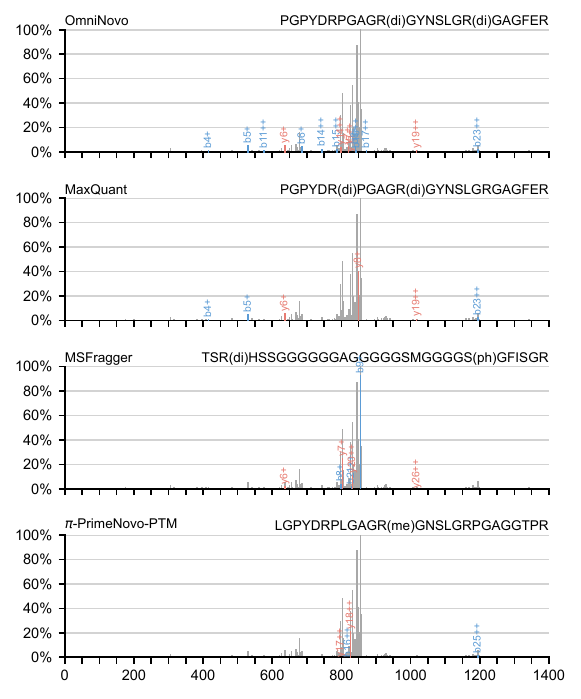}
        \caption{Case 47}
        \label{fig:sup_di_case_47}
    \end{subfigure}\hfill
    \begin{subfigure}[b]{0.48\linewidth}
        \centering
        \includegraphics[width=\linewidth]{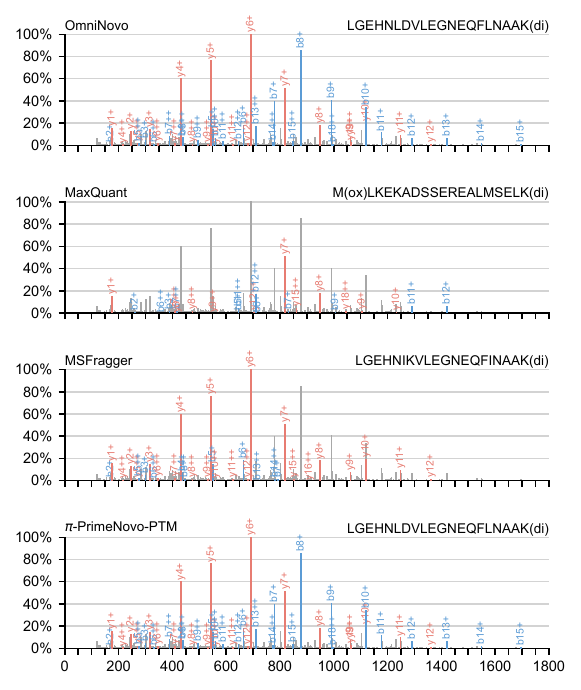}
        \caption{Case 48}
        \label{fig:sup_di_case_48}
    \end{subfigure}
    \caption{PSM comparison for Dimethylation (Cases 45--48).}
\end{figure}
\clearpage
\begin{figure}[htbp]
    \centering
    \begin{subfigure}[b]{0.48\linewidth}
        \centering
        \includegraphics[width=\linewidth]{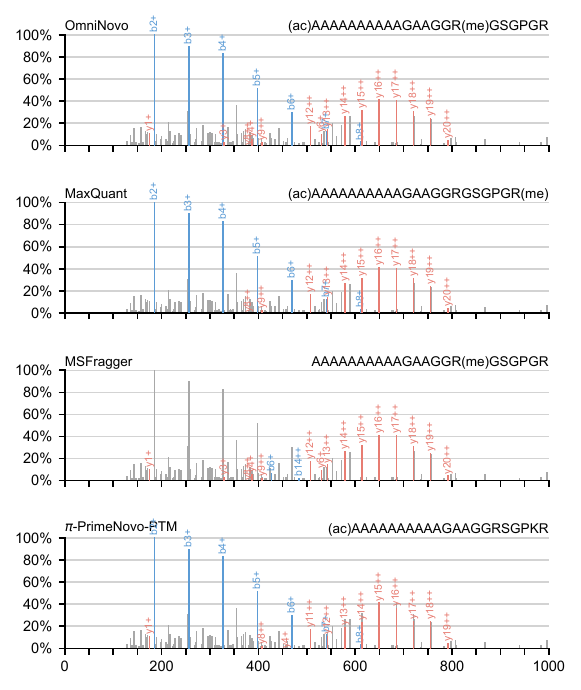}
        \caption{Case 1}
        \label{fig:sup_me_case_1}
    \end{subfigure}\hfill
    \begin{subfigure}[b]{0.48\linewidth}
        \centering
        \includegraphics[width=\linewidth]{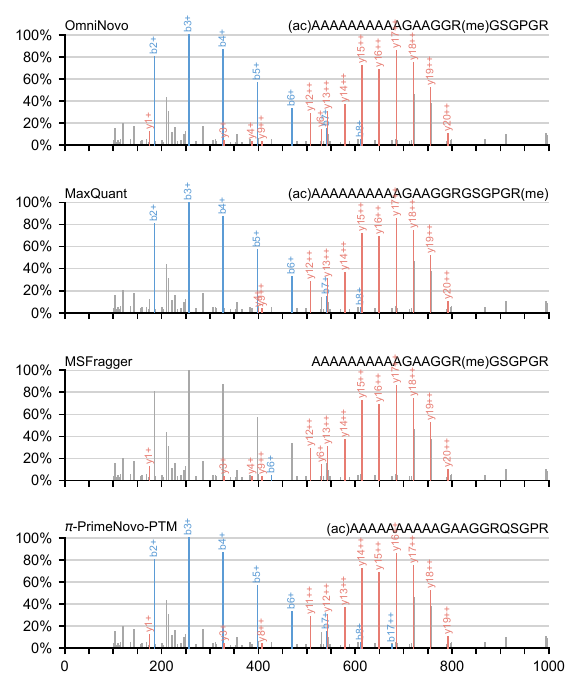}
        \caption{Case 2}
        \label{fig:sup_me_case_2}
    \end{subfigure}
    \vspace{1em}

    \begin{subfigure}[b]{0.48\linewidth}
        \centering
        \includegraphics[width=\linewidth]{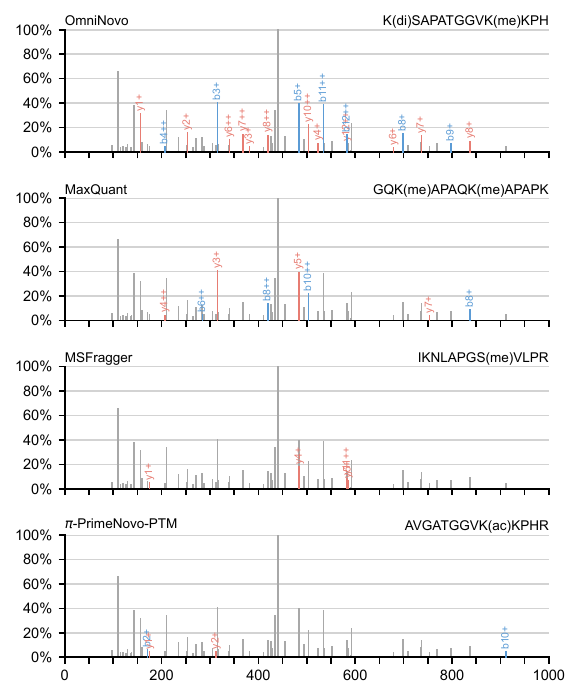}
        \caption{Case 3}
        \label{fig:sup_me_case_3}
    \end{subfigure}\hfill
    \begin{subfigure}[b]{0.48\linewidth}
        \centering
        \includegraphics[width=\linewidth]{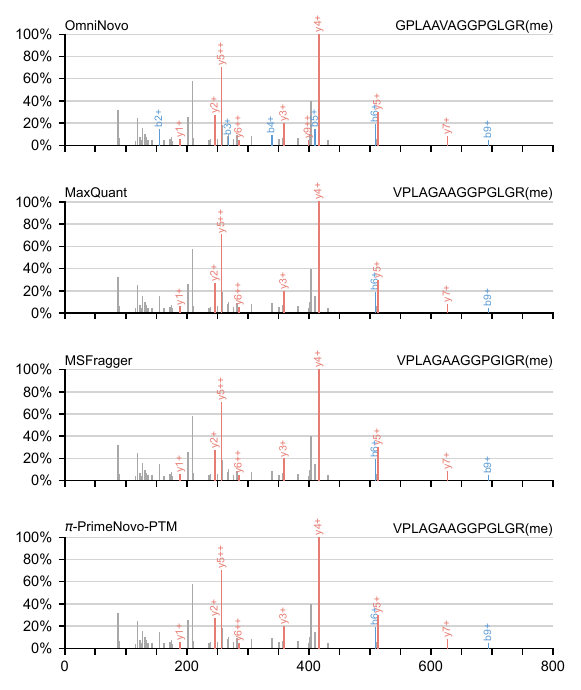}
        \caption{Case 4}
        \label{fig:sup_me_case_4}
    \end{subfigure}
    \caption{PSM comparison for Methylation (Cases 1--4).}
\end{figure}
\clearpage
\begin{figure}[htbp]
    \centering
    \begin{subfigure}[b]{0.48\linewidth}
        \centering
        \includegraphics[width=\linewidth]{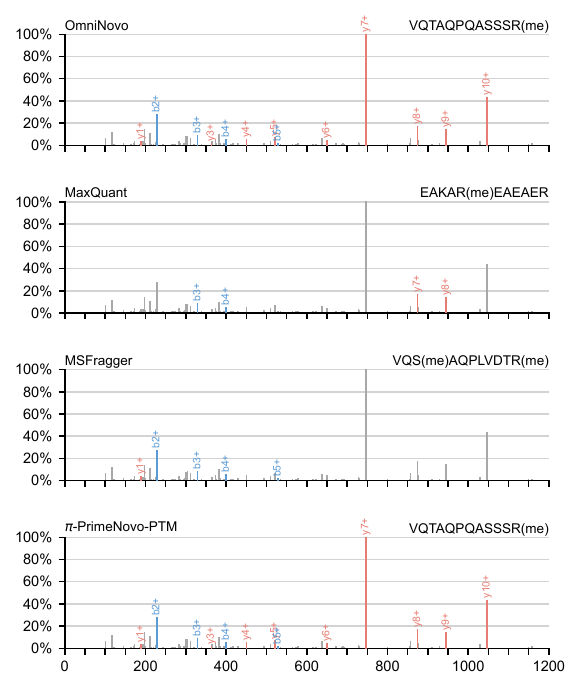}
        \caption{Case 5}
        \label{fig:sup_me_case_5}
    \end{subfigure}\hfill
    \begin{subfigure}[b]{0.48\linewidth}
        \centering
        \includegraphics[width=\linewidth]{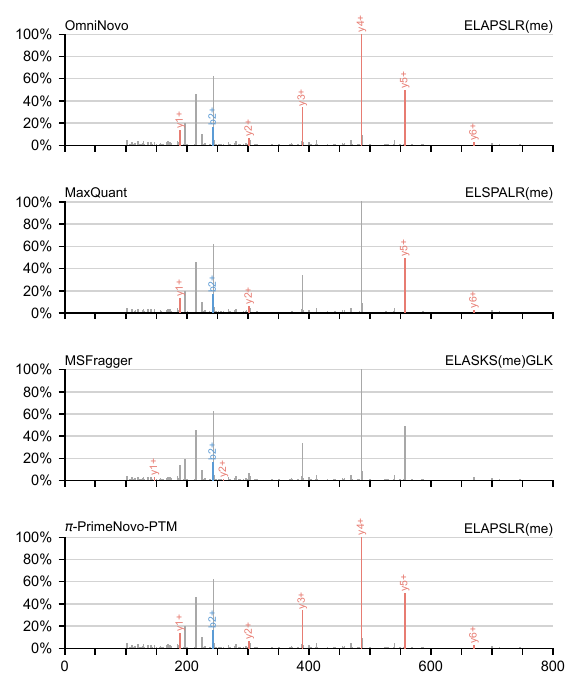}
        \caption{Case 6}
        \label{fig:sup_me_case_6}
    \end{subfigure}
    \vspace{1em}

    \begin{subfigure}[b]{0.48\linewidth}
        \centering
        \includegraphics[width=\linewidth]{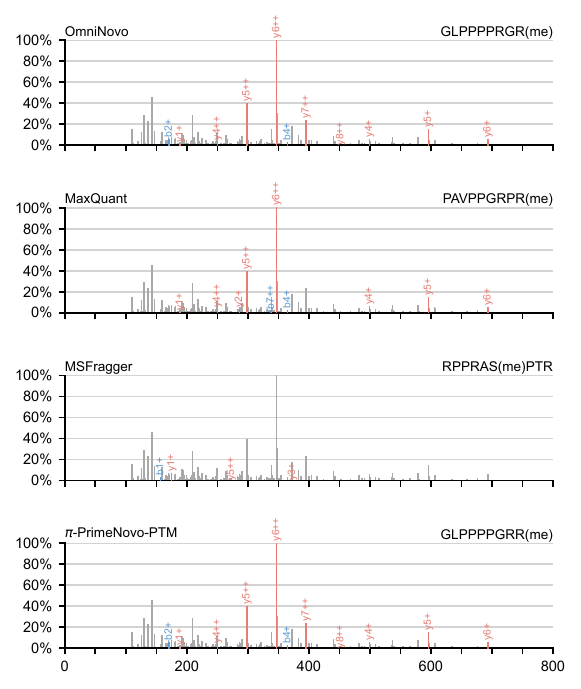}
        \caption{Case 7}
        \label{fig:sup_me_case_7}
    \end{subfigure}\hfill
    \begin{subfigure}[b]{0.48\linewidth}
        \centering
        \includegraphics[width=\linewidth]{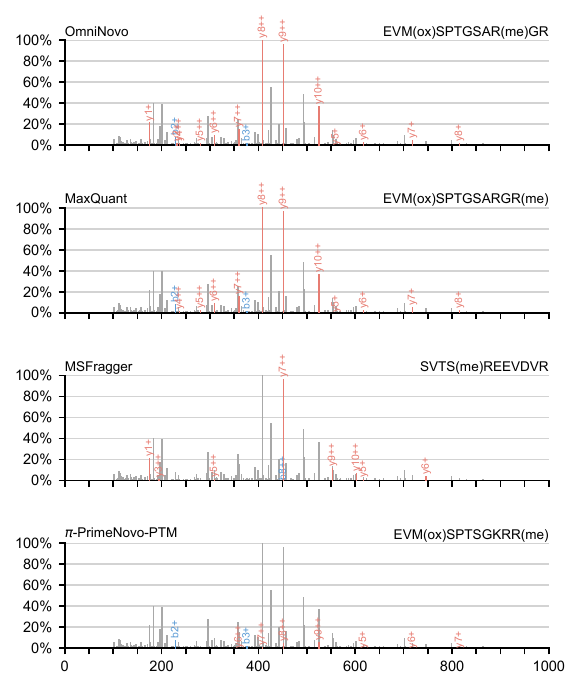}
        \caption{Case 8}
        \label{fig:sup_me_case_8}
    \end{subfigure}
    \caption{PSM comparison for Methylation (Cases 5--8).}
\end{figure}
\clearpage
\begin{figure}[htbp]
    \centering
    \begin{subfigure}[b]{0.48\linewidth}
        \centering
        \includegraphics[width=\linewidth]{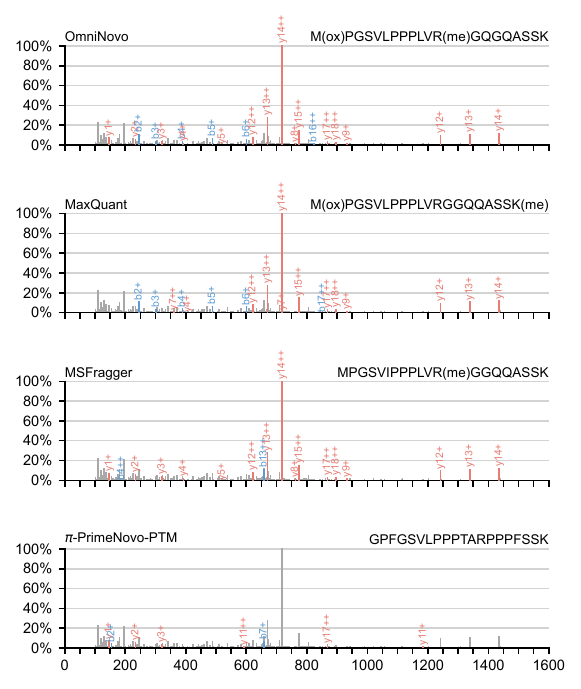}
        \caption{Case 9}
        \label{fig:sup_me_case_9}
    \end{subfigure}\hfill
    \begin{subfigure}[b]{0.48\linewidth}
        \centering
        \includegraphics[width=\linewidth]{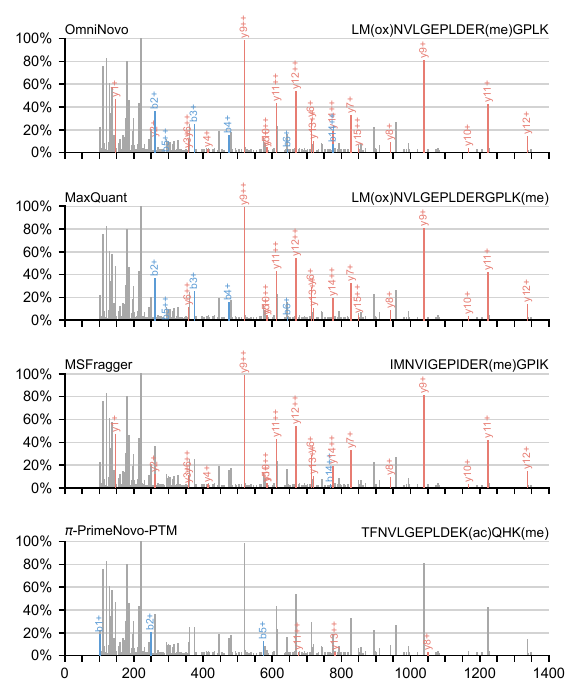}
        \caption{Case 10}
        \label{fig:sup_me_case_10}
    \end{subfigure}
    \vspace{1em}

    \begin{subfigure}[b]{0.48\linewidth}
        \centering
        \includegraphics[width=\linewidth]{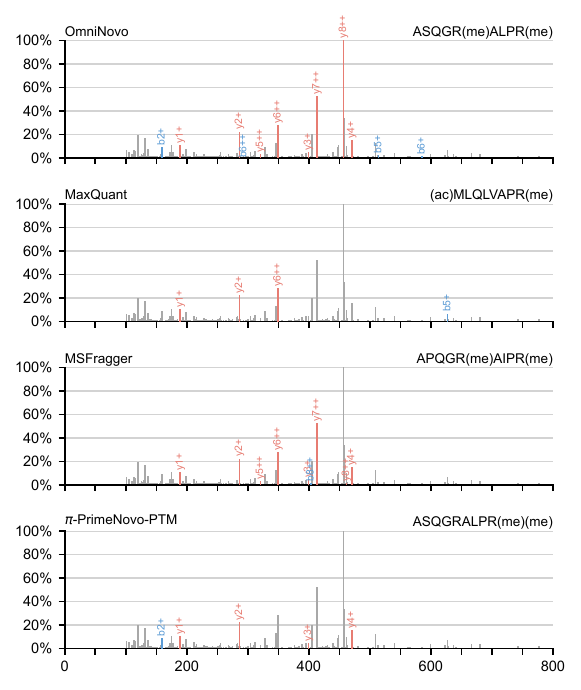}
        \caption{Case 11}
        \label{fig:sup_me_case_11}
    \end{subfigure}\hfill
    \begin{subfigure}[b]{0.48\linewidth}
        \centering
        \includegraphics[width=\linewidth]{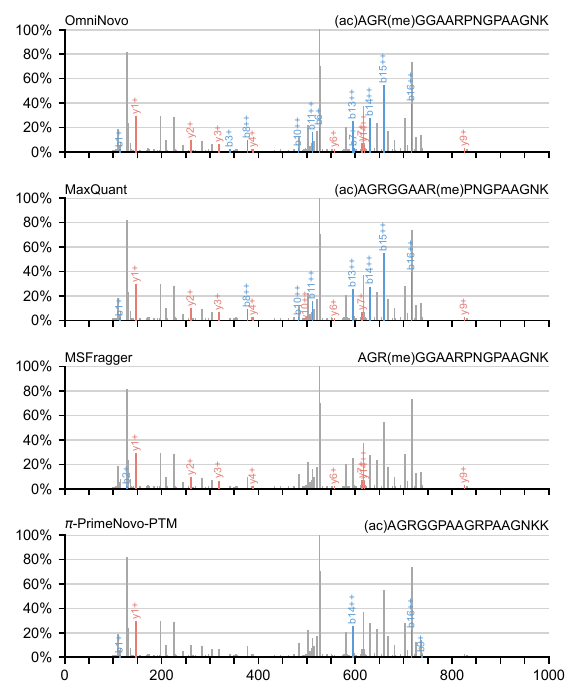}
        \caption{Case 12}
        \label{fig:sup_me_case_12}
    \end{subfigure}
    \caption{PSM comparison for Methylation (Cases 9--12).}
\end{figure}
\clearpage
\begin{figure}[htbp]
    \centering
    \begin{subfigure}[b]{0.48\linewidth}
        \centering
        \includegraphics[width=\linewidth]{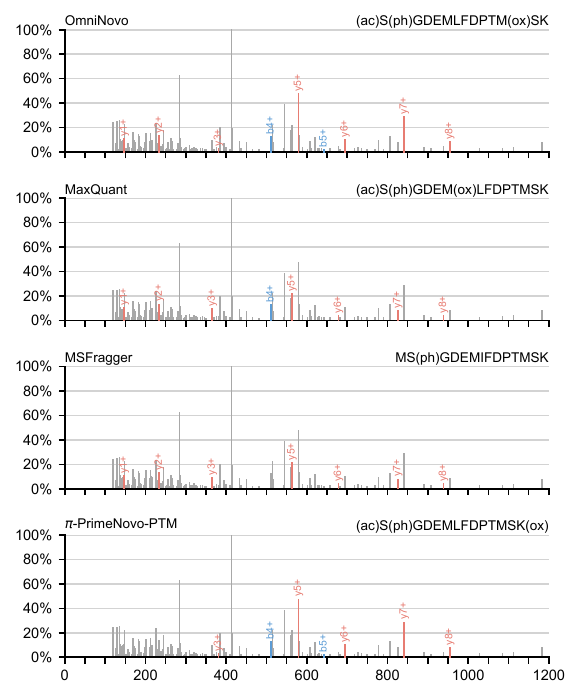}
        \caption{Case 1}
        \label{fig:sup_ph_case_1}
    \end{subfigure}\hfill
    \begin{subfigure}[b]{0.48\linewidth}
        \centering
        \includegraphics[width=\linewidth]{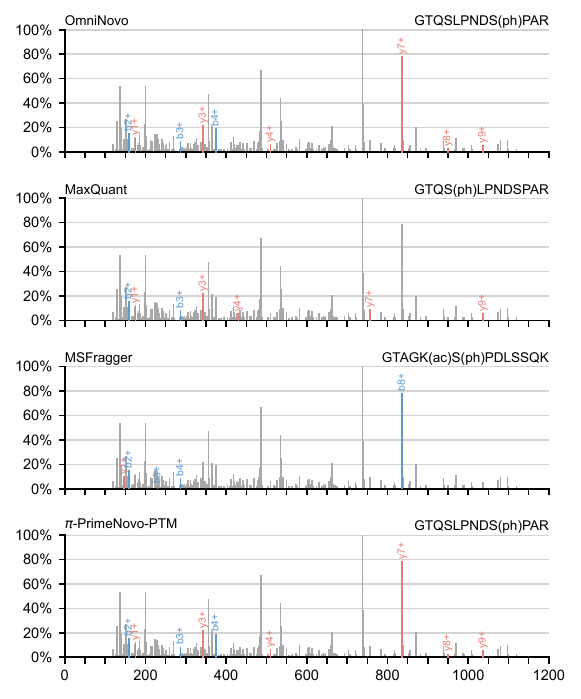}
        \caption{Case 2}
        \label{fig:sup_ph_case_2}
    \end{subfigure}
    \vspace{1em}

    \begin{subfigure}[b]{0.48\linewidth}
        \centering
        \includegraphics[width=\linewidth]{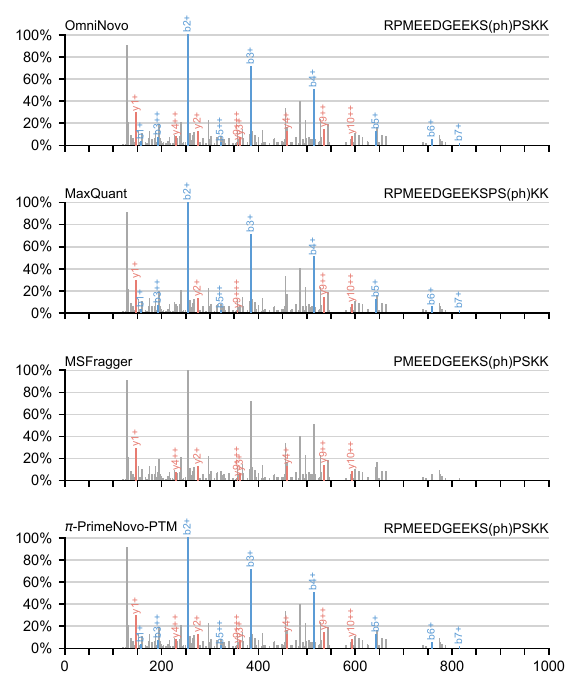}
        \caption{Case 3}
        \label{fig:sup_ph_case_3}
    \end{subfigure}\hfill
    \begin{subfigure}[b]{0.48\linewidth}
        \centering
        \includegraphics[width=\linewidth]{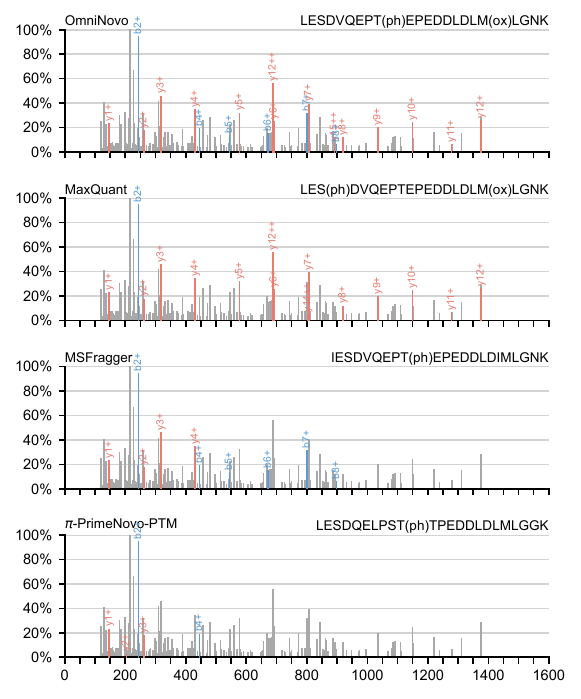}
        \caption{Case 4}
        \label{fig:sup_ph_case_4}
    \end{subfigure}
    \caption{PSM comparison for Phosphorylation (Cases 1--4).}
\end{figure}
\clearpage
\begin{figure}[htbp]
    \centering
    \begin{subfigure}[b]{0.48\linewidth}
        \centering
        \includegraphics[width=\linewidth]{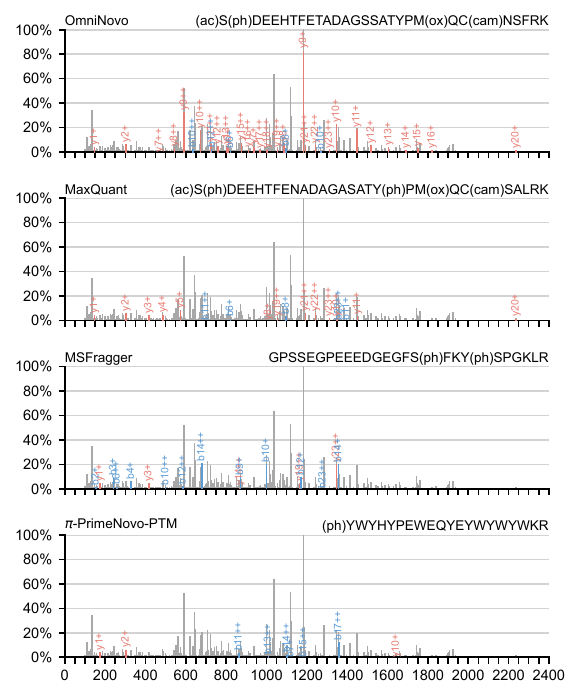}
        \caption{Case 5}
        \label{fig:sup_ph_case_5}
    \end{subfigure}\hfill
    \begin{subfigure}[b]{0.48\linewidth}
        \centering
        \includegraphics[width=\linewidth]{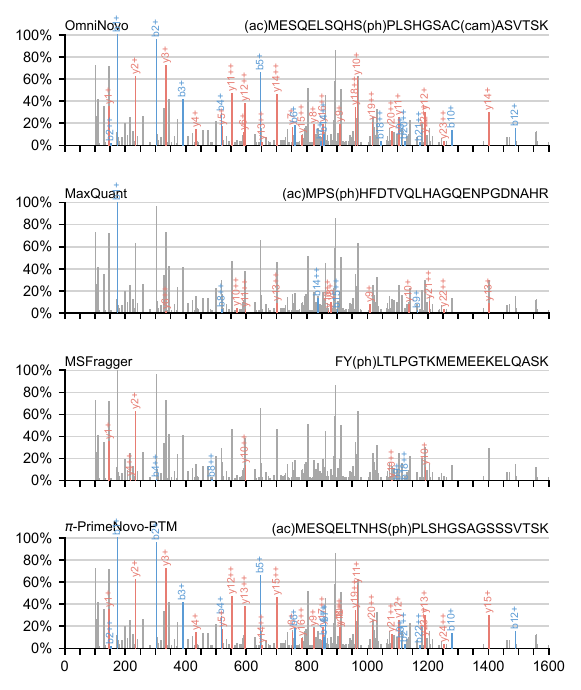}
        \caption{Case 6}
        \label{fig:sup_ph_case_6}
    \end{subfigure}
    \vspace{1em}

    \begin{subfigure}[b]{0.48\linewidth}
        \centering
        \includegraphics[width=\linewidth]{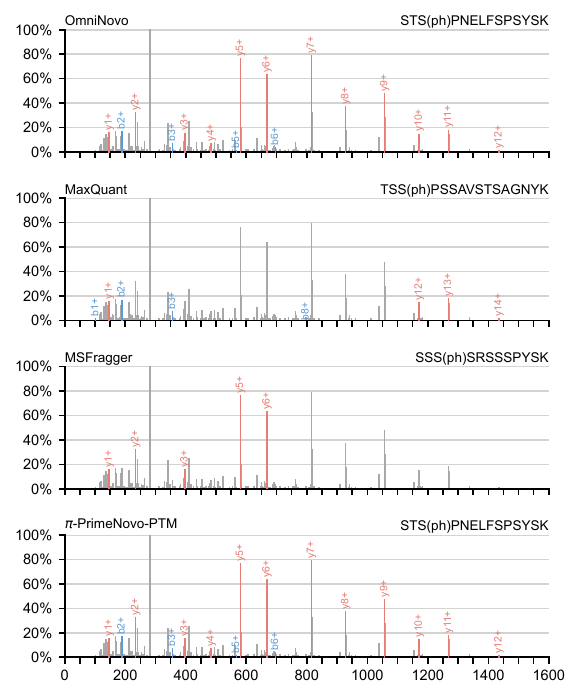}
        \caption{Case 7}
        \label{fig:sup_ph_case_7}
    \end{subfigure}\hfill
    \begin{subfigure}[b]{0.48\linewidth}
        \centering
        \includegraphics[width=\linewidth]{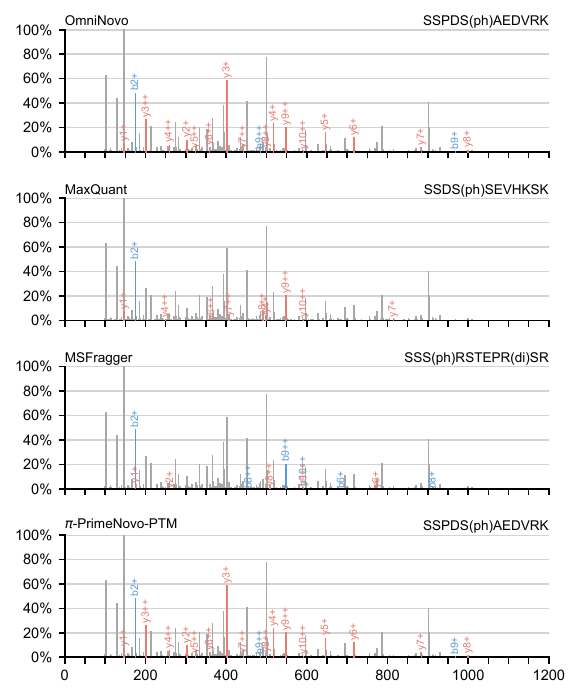}
        \caption{Case 8}
        \label{fig:sup_ph_case_8}
    \end{subfigure}
    \caption{PSM comparison for Phosphorylation (Cases 5--8).}
\end{figure}
\clearpage
\begin{figure}[htbp]
    \centering
    \begin{subfigure}[b]{0.48\linewidth}
        \centering
        \includegraphics[width=\linewidth]{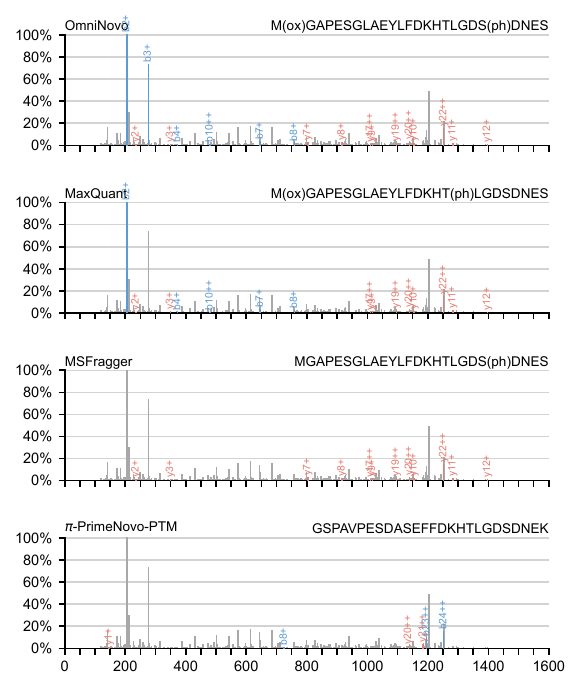}
        \caption{Case 9}
        \label{fig:sup_ph_case_9}
    \end{subfigure}\hfill
    \begin{subfigure}[b]{0.48\linewidth}
        \centering
        \includegraphics[width=\linewidth]{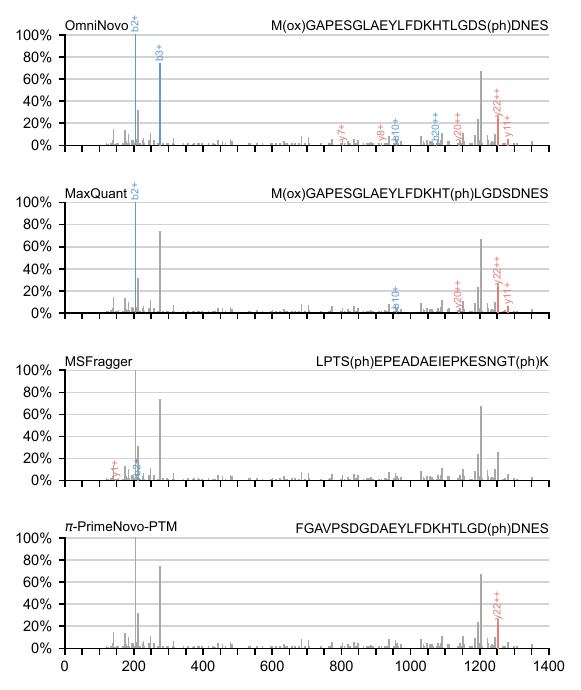}
        \caption{Case 10}
        \label{fig:sup_ph_case_10}
    \end{subfigure}
    \vspace{1em}

    \begin{subfigure}[b]{0.48\linewidth}
        \centering
        \includegraphics[width=\linewidth]{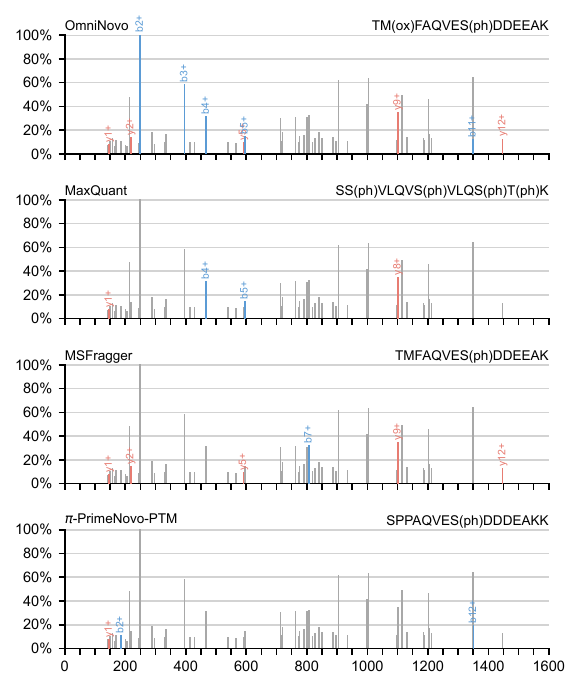}
        \caption{Case 11}
        \label{fig:sup_ph_case_11}
    \end{subfigure}\hfill
    \begin{subfigure}[b]{0.48\linewidth}
        \centering
        \includegraphics[width=\linewidth]{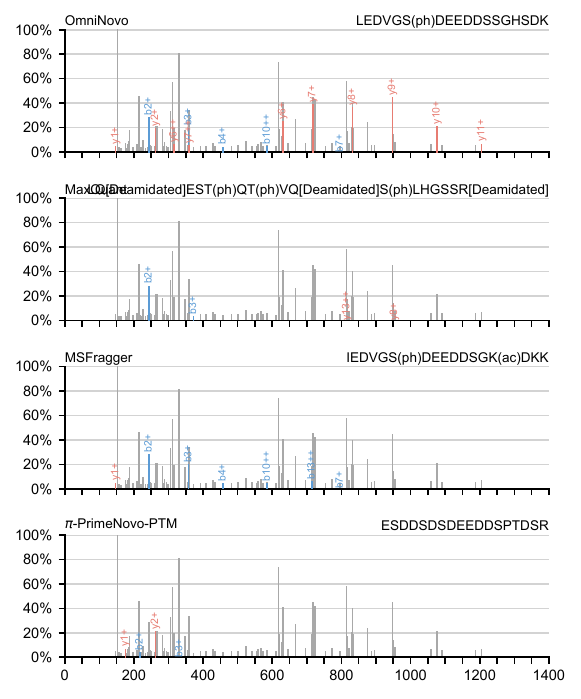}
        \caption{Case 12}
        \label{fig:sup_ph_case_12}
    \end{subfigure}
    \caption{PSM comparison for Phosphorylation (Cases 9--12).}
\end{figure}
\clearpage
\begin{figure}[htbp]
    \centering
    \begin{subfigure}[b]{0.48\linewidth}
        \centering
        \includegraphics[width=\linewidth]{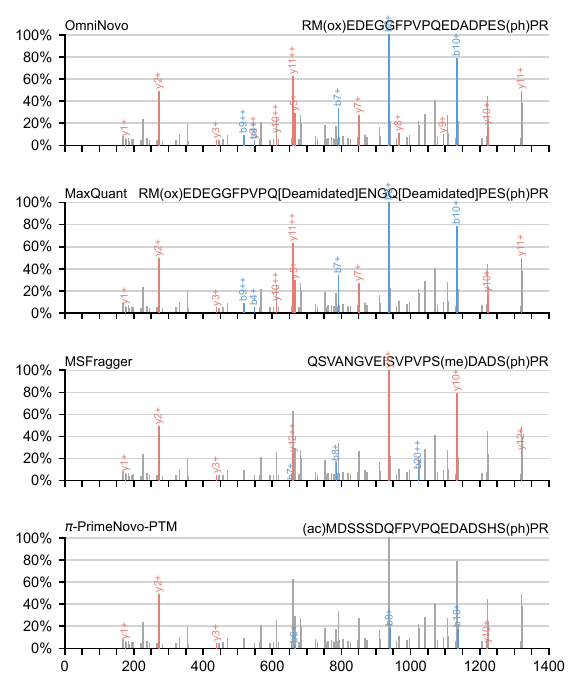}
        \caption{Case 13}
        \label{fig:sup_ph_case_13}
    \end{subfigure}\hfill
    \begin{subfigure}[b]{0.48\linewidth}
        \centering
        \includegraphics[width=\linewidth]{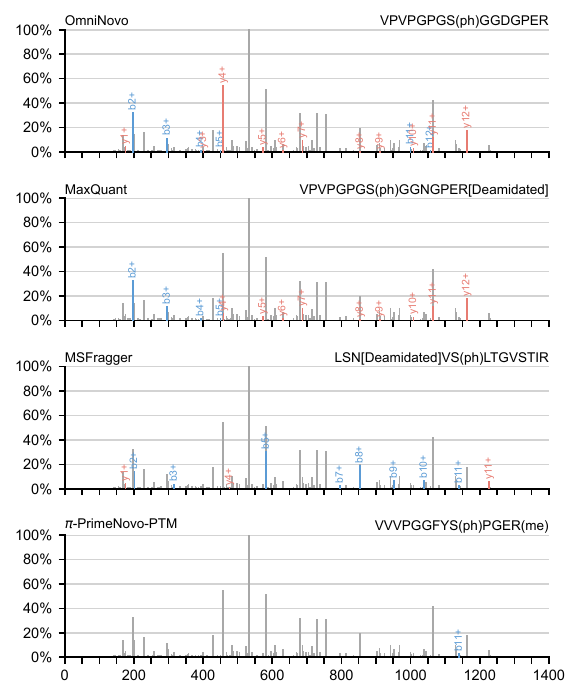}
        \caption{Case 14}
        \label{fig:sup_ph_case_14}
    \end{subfigure}
    \vspace{1em}

    \caption{PSM comparison for Phosphorylation (Cases 13--14).}
\end{figure}
\clearpage
\begin{figure}[htbp]
    \centering
    \begin{subfigure}[b]{0.48\linewidth}
        \centering
        \includegraphics[width=\linewidth]{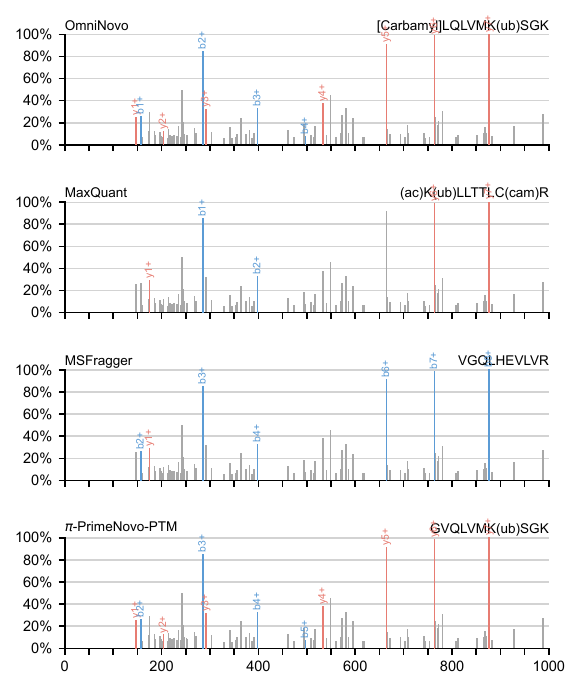}
        \caption{Case 1}
        \label{fig:sup_ub_case_1}
    \end{subfigure}\hfill
    \begin{subfigure}[b]{0.48\linewidth}
        \centering
        \includegraphics[width=\linewidth]{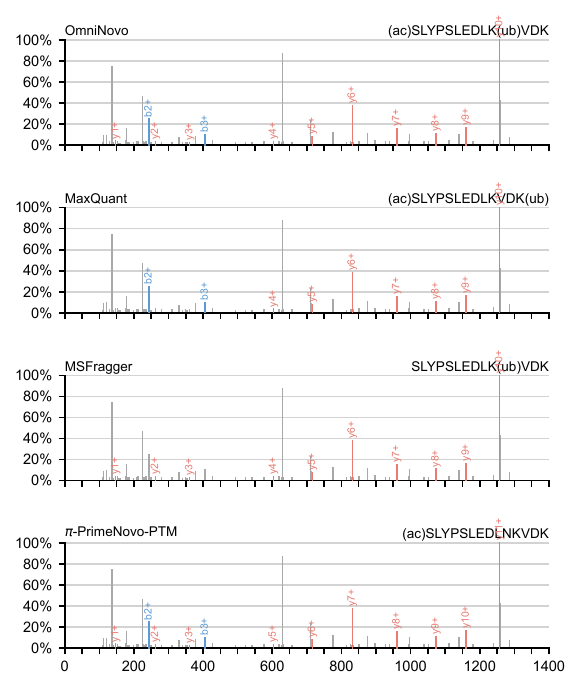}
        \caption{Case 2}
        \label{fig:sup_ub_case_2}
    \end{subfigure}
    \vspace{1em}

    \begin{subfigure}[b]{0.48\linewidth}
        \centering
        \includegraphics[width=\linewidth]{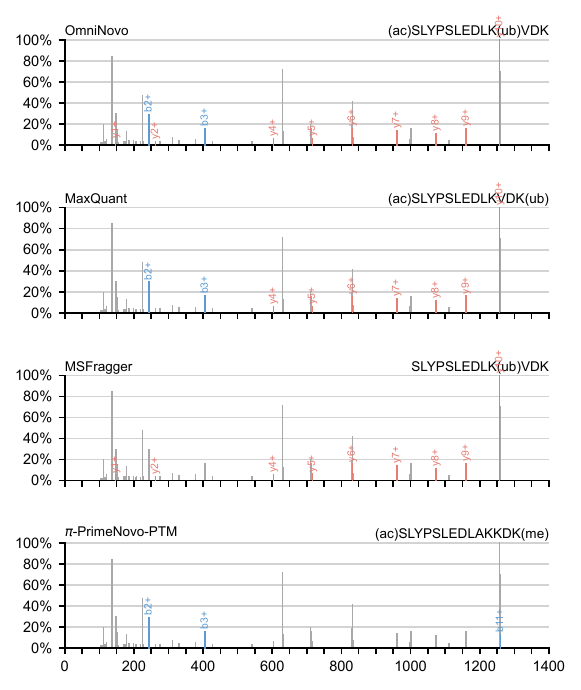}
        \caption{Case 3}
        \label{fig:sup_ub_case_3}
    \end{subfigure}\hfill
    \begin{subfigure}[b]{0.48\linewidth}
        \centering
        \includegraphics[width=\linewidth]{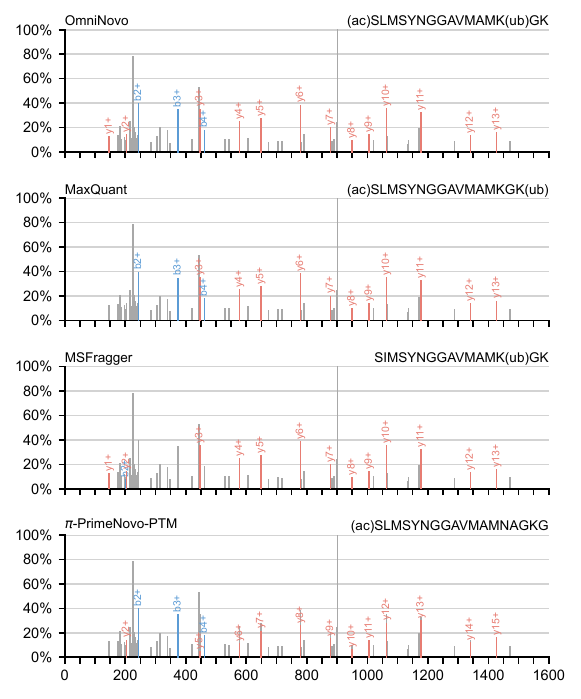}
        \caption{Case 4}
        \label{fig:sup_ub_case_4}
    \end{subfigure}
    \caption{PSM comparison for Ubiquitination (Cases 1--4).}
\end{figure}
\clearpage
\begin{figure}[htbp]
    \centering
    \begin{subfigure}[b]{0.48\linewidth}
        \centering
        \includegraphics[width=\linewidth]{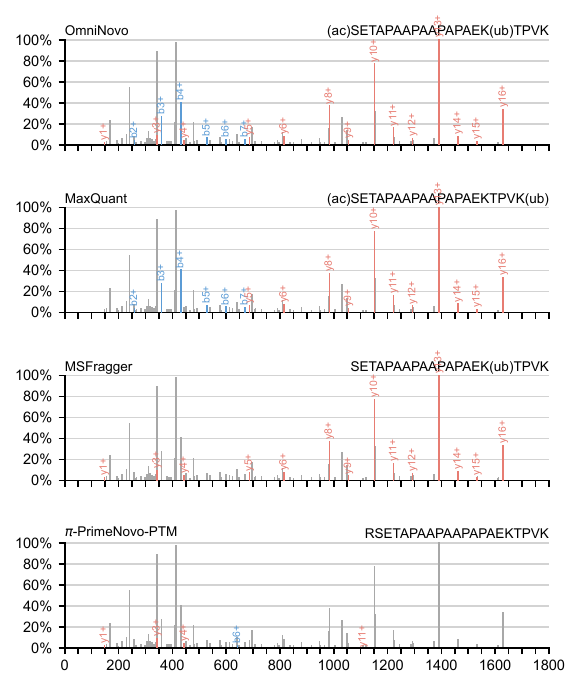}
        \caption{Case 5}
        \label{fig:sup_ub_case_5}
    \end{subfigure}\hfill
    \begin{subfigure}[b]{0.48\linewidth}
        \centering
        \includegraphics[width=\linewidth]{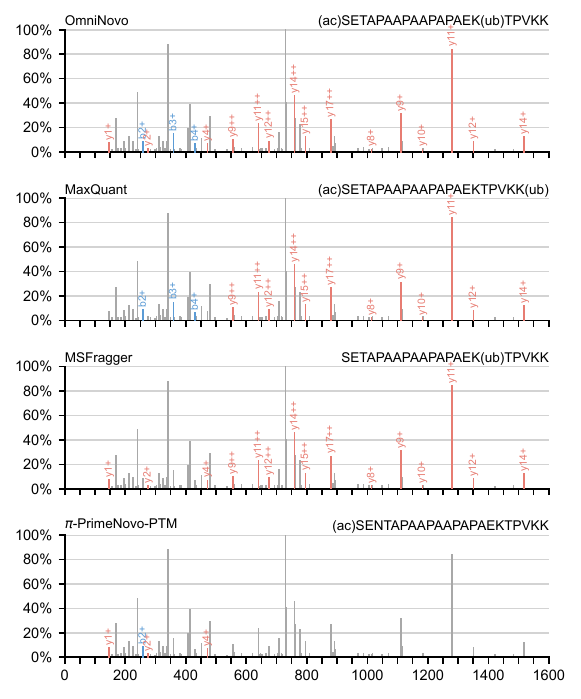}
        \caption{Case 6}
        \label{fig:sup_ub_case_6}
    \end{subfigure}
    \vspace{1em}

    \begin{subfigure}[b]{0.48\linewidth}
        \centering
        \includegraphics[width=\linewidth]{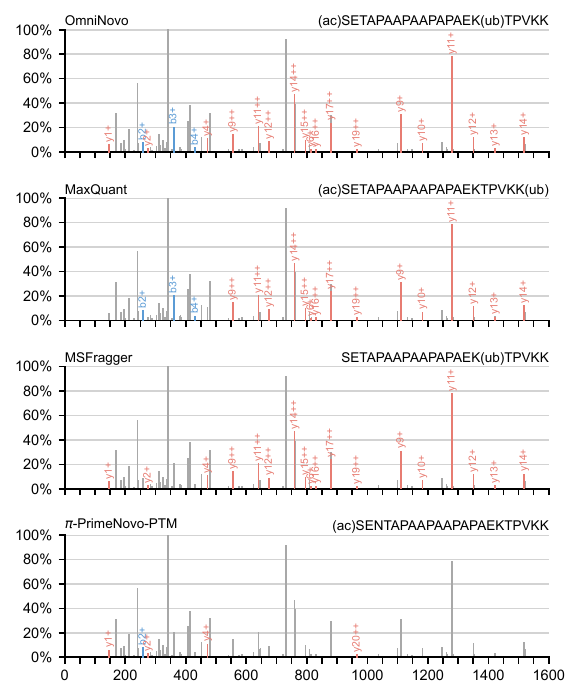}
        \caption{Case 7}
        \label{fig:sup_ub_case_7}
    \end{subfigure}\hfill
    \begin{subfigure}[b]{0.48\linewidth}
        \centering
        \includegraphics[width=\linewidth]{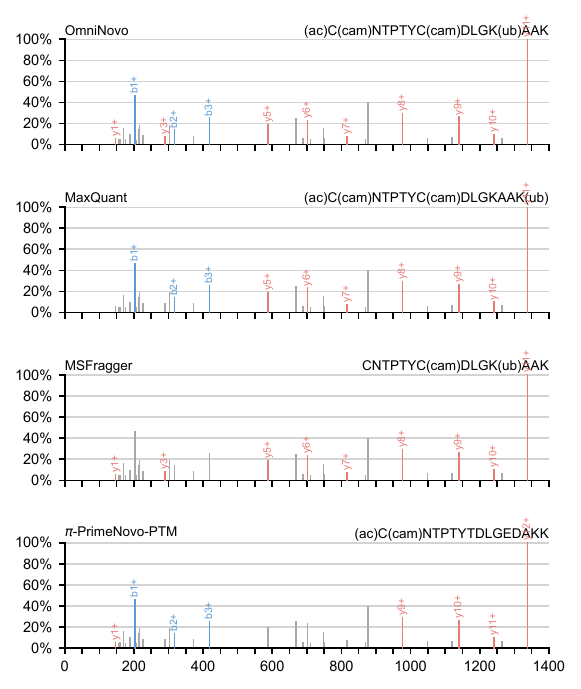}
        \caption{Case 8}
        \label{fig:sup_ub_case_8}
    \end{subfigure}
    \caption{PSM comparison for Ubiquitination (Cases 5--8).}
\end{figure}
\clearpage
\begin{figure}[htbp]
    \centering
    \begin{subfigure}[b]{0.48\linewidth}
        \centering
        \includegraphics[width=\linewidth]{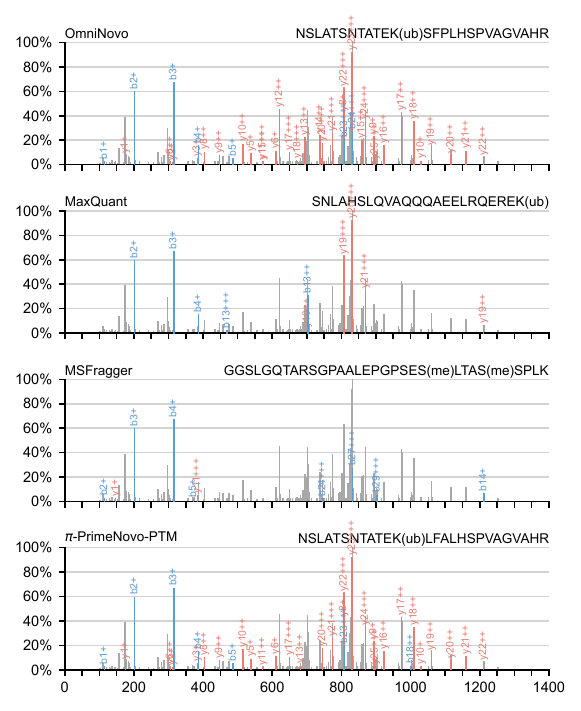}
        \caption{Case 9}
        \label{fig:sup_ub_case_9}
    \end{subfigure}\hfill
    \begin{subfigure}[b]{0.48\linewidth}
        \centering
        \includegraphics[width=\linewidth]{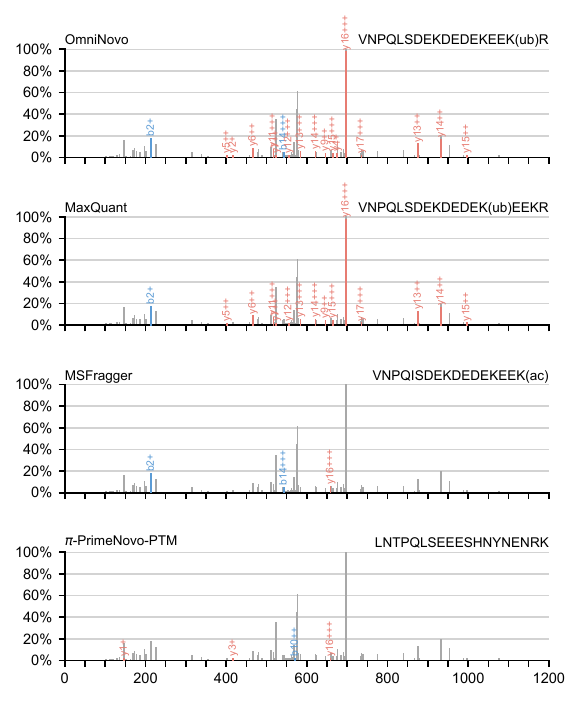}
        \caption{Case 10}
        \label{fig:sup_ub_case_10}
    \end{subfigure}
    \vspace{1em}

    \begin{subfigure}[b]{0.48\linewidth}
        \centering
        \includegraphics[width=\linewidth]{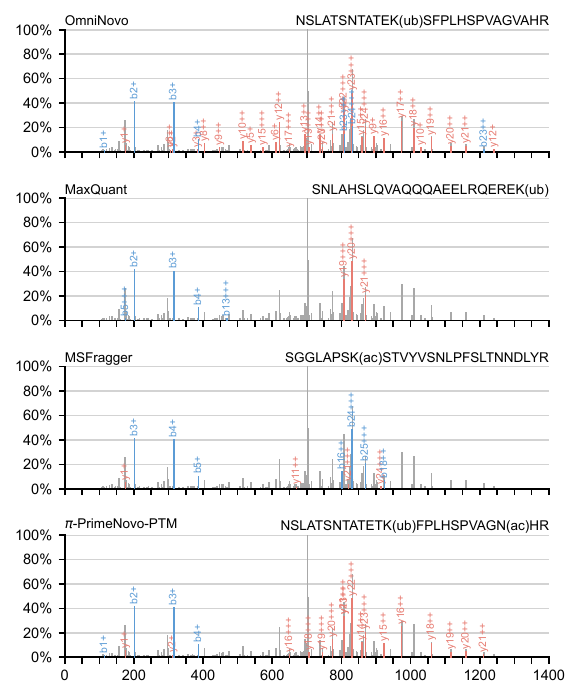}
        \caption{Case 11}
        \label{fig:sup_ub_case_11}
    \end{subfigure}\hfill
    \begin{subfigure}[b]{0.48\linewidth}
        \centering
        \includegraphics[width=\linewidth]{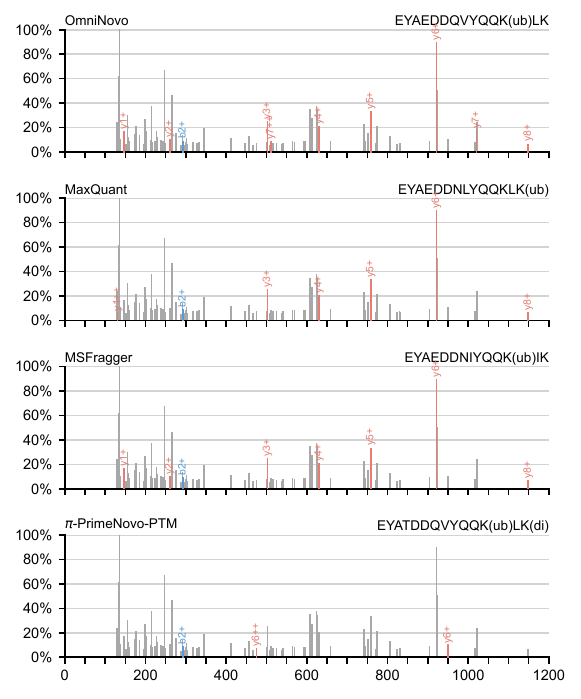}
        \caption{Case 12}
        \label{fig:sup_ub_case_12}
    \end{subfigure}
    \caption{PSM comparison for Ubiquitination (Cases 9--12).}
\end{figure}
\clearpage
\begin{figure}[htbp]
    \centering
    \begin{subfigure}[b]{0.48\linewidth}
        \centering
        \includegraphics[width=\linewidth]{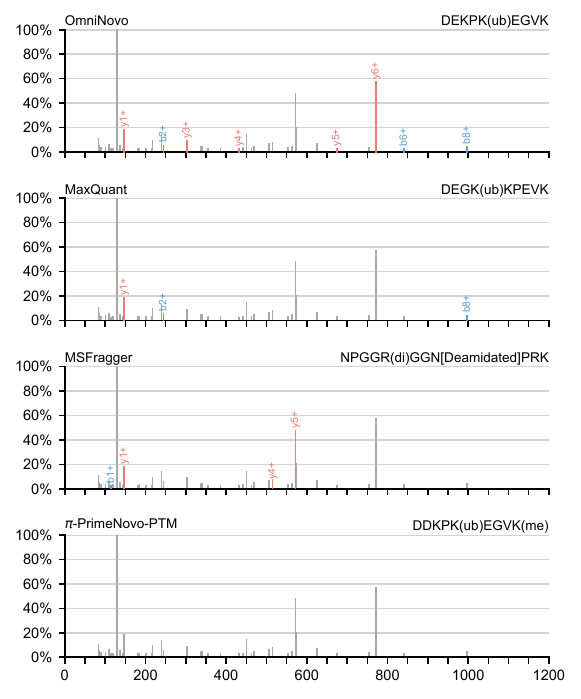}
        \caption{Case 13}
        \label{fig:sup_ub_case_13}
    \end{subfigure}\hfill
    \begin{subfigure}[b]{0.48\linewidth}
        \centering
        \includegraphics[width=\linewidth]{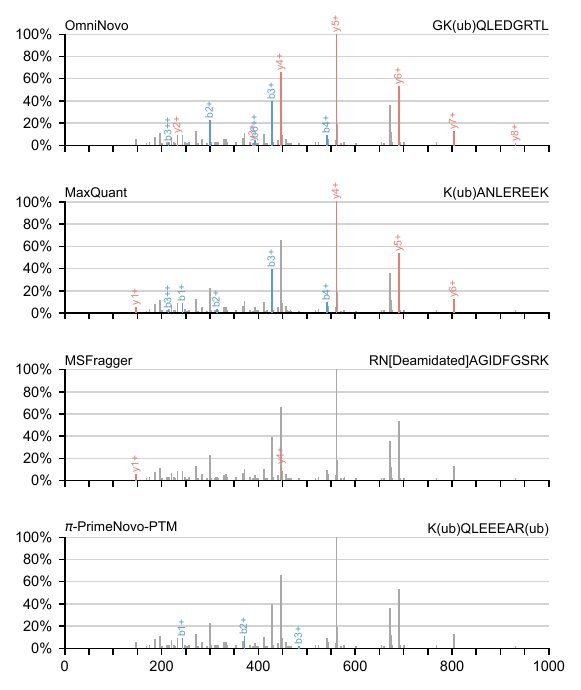}
        \caption{Case 14}
        \label{fig:sup_ub_case_14}
    \end{subfigure}
    \vspace{1em}

    \begin{subfigure}[b]{0.48\linewidth}
        \centering
        \includegraphics[width=\linewidth]{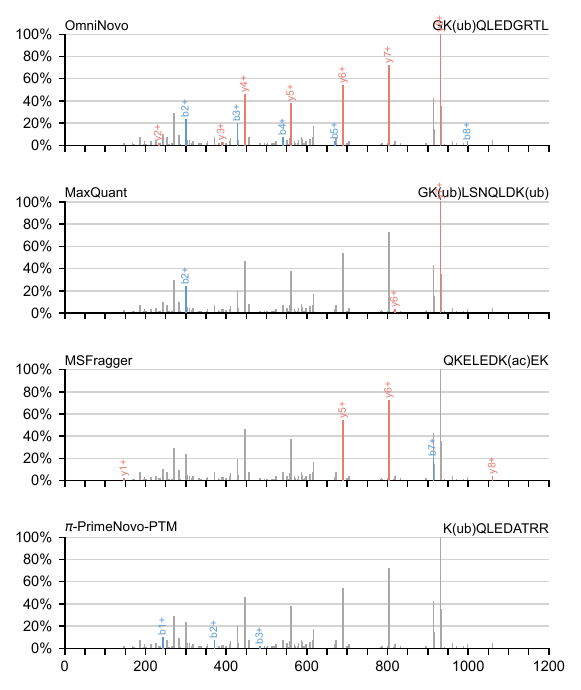}
        \caption{Case 15}
        \label{fig:sup_ub_case_15}
    \end{subfigure}\hfill
    \begin{subfigure}[b]{0.48\linewidth}
        \centering
        \includegraphics[width=\linewidth]{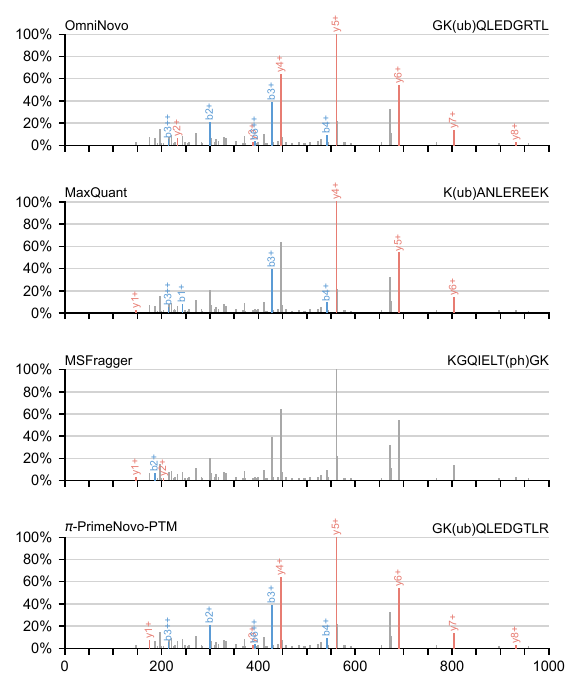}
        \caption{Case 16}
        \label{fig:sup_ub_case_16}
    \end{subfigure}
    \caption{PSM comparison for Ubiquitination (Cases 13--16).}
\end{figure}
\clearpage
\begin{figure}[htbp]
    \centering
    \begin{subfigure}[b]{0.48\linewidth}
        \centering
        \includegraphics[width=\linewidth]{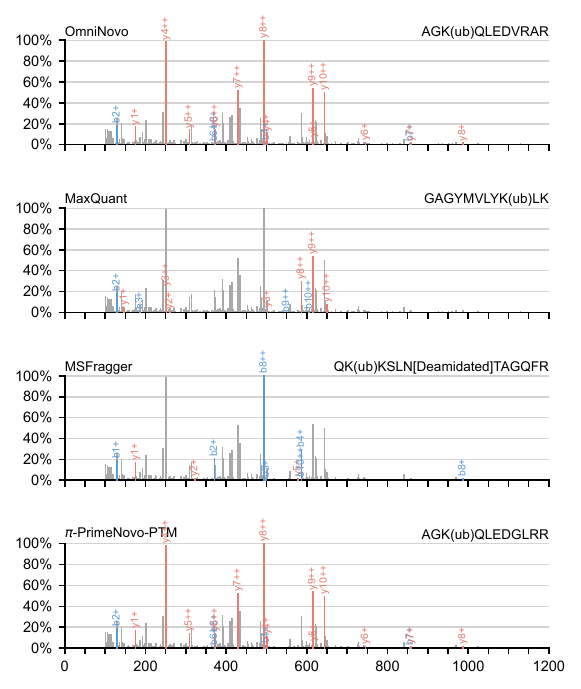}
        \caption{Case 17}
        \label{fig:sup_ub_case_17}
    \end{subfigure}\hfill
    \begin{subfigure}[b]{0.48\linewidth}
        \centering
        \includegraphics[width=\linewidth]{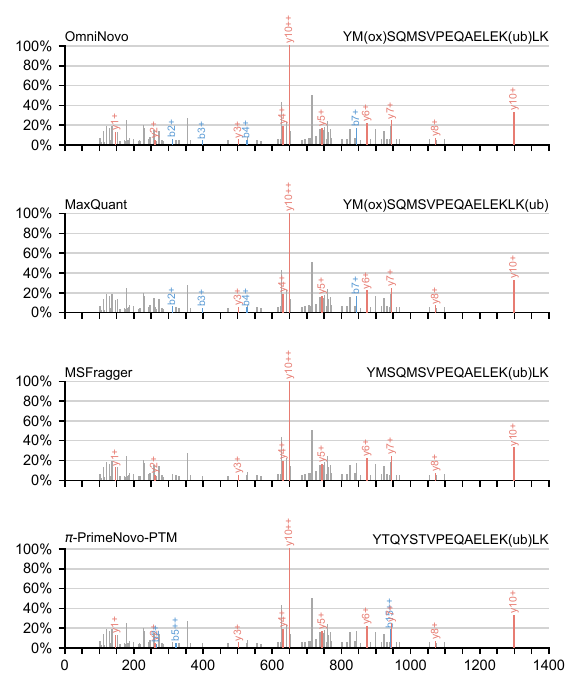}
        \caption{Case 18}
        \label{fig:sup_ub_case_18}
    \end{subfigure}
    \vspace{1em}

    \begin{subfigure}[b]{0.48\linewidth}
        \centering
        \includegraphics[width=\linewidth]{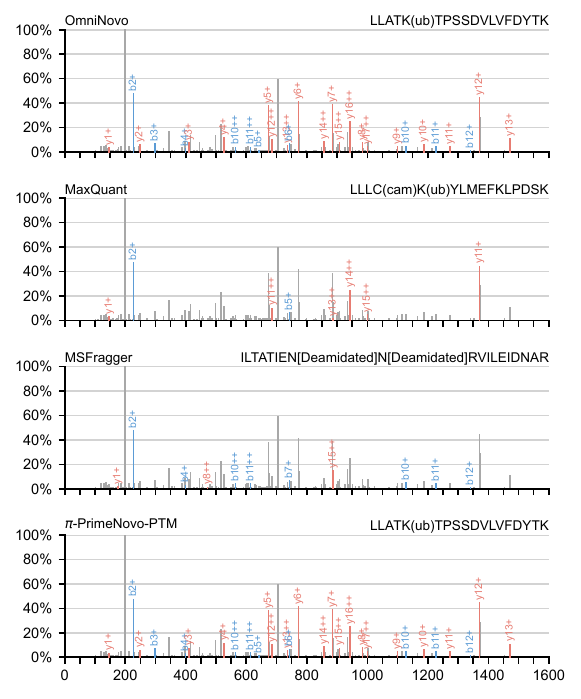}
        \caption{Case 19}
        \label{fig:sup_ub_case_19}
    \end{subfigure}\hfill
    \begin{subfigure}[b]{0.48\linewidth}
        \centering
        \includegraphics[width=\linewidth]{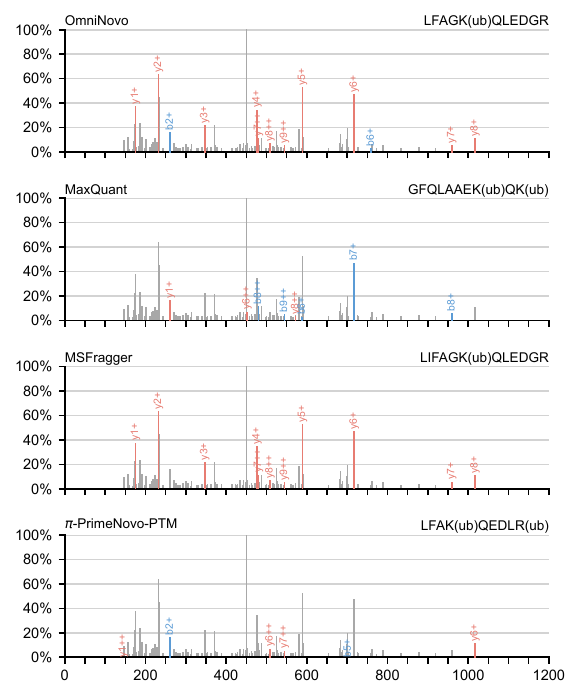}
        \caption{Case 20}
        \label{fig:sup_ub_case_20}
    \end{subfigure}
    \caption{PSM comparison for Ubiquitination (Cases 17--20).}
\end{figure}
\clearpage





\end{appendices}

\end{document}